\documentclass[traditabstract]{aa}

\newcommand{\hii}{H\,{\scriptsize II}}

\usepackage{epsfig,graphicx,color}
\usepackage{ctable}

\usepackage{txfonts}

\begin{document}

\bibliographystyle{aa} 
\title{Globules and Pillars in Cygnus X}
\subtitle{I. {\sl Herschel}\thanks{Herschel is an ESA space
observatory with science instruments provided by European-led
Principal Investigator consortia and with important participation from
NASA.} Far-infrared imaging of the Cyg~OB2 environment}
  \author{N. Schneider \inst{1,2,3} 
  \and S. Bontemps \inst{1,2} 
  \and F. Motte \inst{4,5}     
  \and A. Blazere \inst{1,2}
  \and Ph. Andr\'e \inst{4}
  \and L.D. Anderson \inst{6}
  \and D. Arzoumanian \inst{4,7}
  \and F. Comer\'on \inst{8} 
  \and P. Didelon \inst{4}
  \and J. Di Francesco \inst{9,10}
  \and A. Duarte-Cabral \inst{11}     
  \and M.G. Guarcello \inst{12,13}
  \and M. Hennemann \inst{4} 
  \and T. Hill \inst{14}
  \and V. K\"onyves \inst{4} 
  \and A. Marston \inst{15}  
  \and V. Minier \inst{4}    
  \and K.L.J. Rygl \inst{16}
  \and M. R\"ollig \inst{3} 
  \and A. Roy \inst{4}
  \and L. Spinoglio \inst{17}
  \and P. Tremblin \inst{18}   
  \and G.J. White \inst{19,20}
  \and N.J. Wright \inst{21}
}
 \institute{
  Univ. Bordeaux, LAB, UMR 5804, 33270, Floirac, France
  \and
  CNRS, LAB, UMR 5804, 33270, Floirac, France
  \and
  I. Physik. Institut, University of Cologne, Z\"ulpicher Str. 77, 50937 Cologne, Germany
  \and
  IRFU/SAp CEA/DSM, Laboratoire AIM CNRS - Universit\'e Paris 
  Diderot, 91191 Gif-sur-Yvette, France
  \and
  IPAG, University Grenoble Alpes, 38000 Grenoble, France
  \and 
  Dep. of Physics, West Virginia University, WV 26506, USA
  \and
  IAS, CNRS/Universit\'e Paris-Sud 11, 91405 Orsay, France
  \and
  ESO, Alonso de Cordova 3107, Vitacura, Santiago 19, Chile
  \and
  Dep. of Physics and Astronomy, University of Victoria, Victoria, BC, Canada
  \and
  NRCC, Victoria, BC, Canada
  \and
  Astrophysics Group, University of Exeter, EX4 4QL Exeter, UK 
  \and  
  INAF Osservatorio Astronomico di Palermo 90134 Palermo, Italy
  \and  
  Smithsonian Astrophysical Observatory, Cambridge, MA02138, USA
  \and 
  Joint ALMA Observatory, 3107 Alonso de Cordova, Vitacura, Santiago 19, Chile
  \and
  Herschel Science Centre, European Space Astronomy Centre (ESAC)/ESA, Villanueva de la Canada, Spain  
  \and 
  INAF-ORA, Via P.Gobetti 101, 40129 Bologna, Italy
  \and
  INAF-IAPS, Via Fosso del Cavaliere 100, 00133 Roma, Italy
  \and
  Maison de la Simulation, CEA-CNRS-INRIA-UPS-UVSQ, USR 3441, CEA Saclay, 91191 Gif-sur-Yvette, France
  \and
  Department of Physics \& Astronomy, The Open University, 
  Milton Keynes MK7 6AA, UK 
  \and 
  RALSpace, The Rutherford Appleton Laboratory, 
  Chilton, Didcot, Oxfordshire OX11, 0NL,UK
  \and 
  Centre for Astrophysics Research, University of Hertfordshire,  Hatfield, AL10 9AB, UK 
  }   
%
\mail{nschneid@ph1.uni-koeln.de}
\titlerunning{FAR-IR imaging of Cyg OB2}
\authorrunning{N. Schneider}

\date{Received February 18, 2016; accepted April 5, 2016}

\abstract { The radiative feedback of massive stars on molecular
  clouds creates pillars, globules and other features at the interface
  between the \hii\, region and molecular cloud.  Optical and
  near-infrared observations from the ground as well as with the {\sl
    Hubble} or {\sl Spitzer} satellites have revealed numerous
  examples of such cloud structures. We present here {\em Herschel}
  far-infrared observations between 70 $\mu$m and 500 $\mu$m of the
  immediate environment of the rich Cygnus OB2 association, performed
  within the HOBYS (Herschel imaging survey of OB Young Stellar
  objects) program. All of the observed irradiated
  structures were detected based on their appearance at 70 $\mu$m, and
  have been classified as pillars, globules, evaporating gasous
  globules (EGGs), proplyd-like objects, and condensations.  From the
  70 $\mu$m and 160 $\mu$m flux maps, we derive the local
  far-ultraviolet (FUV) field on the photon dominated surfaces. In
  parallel, we use a census of the O-stars to estimate the overall
  FUV-field, that is 10$^3$-10$^4$ G$_0$ (Habing field) close to the
  central OB cluster (within 10 pc) and decreases down to a few tens
  G$_0$, in a distance of 50 pc.  From a spectral energy distribution
  (SED) fit to the four longest {\sl Herschel} wavelengths, we
  determine column density and temperature maps and derive masses,
  volume densities and surface densities for these structures. We find
  that the morphological classification corresponds to distinct
  physical properties.  Pillars and globules are massive ($\sim$500
  M$_\odot$) and large (equivalent radius $r\sim$0.6 pc) structures,
  corresponding to what is defined as `clumps' for molecular
  clouds. EGGs and proplyd-like objects are smaller ($r \sim$0.1 and
  0.2 pc) and less massive ($\sim$10 and $\sim$30 M$_\odot$). Cloud
  condensations are small ($\sim$0.1 pc), have an average mass of 35
  M$_\odot$, are dense ($\sim$6$\times$10$^4$ cm$^{-3}$), and can thus
  be described as molecular cloud `cores'.  All pillars and globules
  are oriented toward the Cyg OB2 association center and have the
  longest estimated photoevaporation lifetimes, a few million years,
  while all other features should survive less than a million
  years. These lifetimes are consistent with that found in simulations
  of turbulent, UV-illuminated clouds. We propose a tentative
  evolutionary scheme in which pillars can evolve into globules, which
  in turn then evolve into EGGs, condensations and proplyd-like
  objects.}

\maketitle

\begin{figure*}[ht]
\begin{center} 
\hspace{-0.5cm}
\includegraphics [width=11cm, angle={0}]{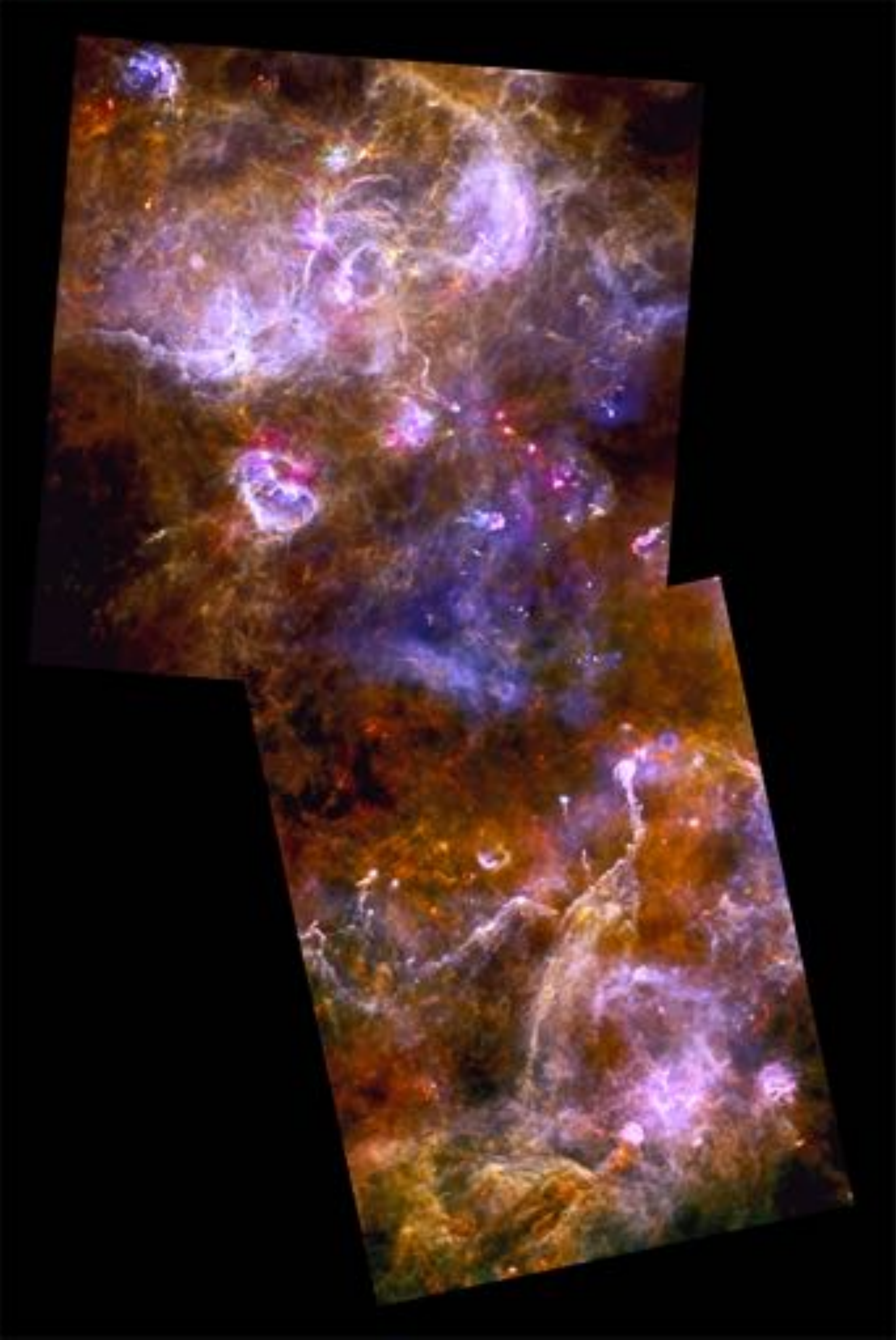}
\caption [] {Three-colour image (blue=70 $\mu$m, green=160 $\mu$m,
  red=500 $\mu$m, all maps are absolutely calibrated) of the Cygnus X
  complex (credit: ESA/PACS/SPIRE) oriented in RA/Dec, i.e., north to
  the top, east to the left.  The Cyg OB2 association is located
  slightly above the center of the figure. The image covers a range of
  RA(2000)$\sim$20$^h$30$^m$ to 20$^h$45$^m$ and
  Dec(2000)$\sim$38$^\circ$ and 43$^\circ$.  Blue colours indicate gas
  heated by high-mass stars of the cluster and white colours indicate
  smaller \hii\, regions that locally heat the gas and create
  bubble-like features. Colder extended emission mainly arising from
  the molecular cloud is shown in red. Numerous pillars and globules,
  in particular in the southern part of Cygnus X, point towards Cyg
  OB2.}
\label{cygnus}
\end{center} 
\end{figure*}

\section{Introduction} \label{intro} 
 
In the environment of high-mass stars, a rich diversity of large and
small dusty gas condensations are produced under the influence of
ionizing radiation. These structures have mainly been detected through
optical observations (see, e.g., Herbig \cite{herbig1974}, Schneps et
al. \cite{schneps1980}, and White et al.  \cite{white1997} for an
overview). Column-like features were named `elephant-trunks', and more
isolated globule-shaped objects were named `teardrop' or `cometary'
globules.  In particular {\sl Hubble} and {\sl Spitzer} revealed
manifold examples and the detailed structure of pillars in Galactic
\hii\, regions, like the famous Pillars of Creation in M16 or the
giant dust pillars in Carina.  Pillars have a column shape, are
attached to their native molecular cloud, and cover scales from
$\sim$0.5 pc up to a few pc in length. Globules (on similar size
scales) are isolated and have a characteristic head-tail
structure. Low-mass star formation can take place in pillars (Hester
et al. \cite{hester1996}, White et al. \cite{white1999}), and globules
(e.g., Sugitani et al.  \cite{sugitani2002}), but there are only a few
examples of high-mass stars in globules.  Intermediate-mass early
B-stars have been found in a globule in Cygnus (Schneider et
al. \cite{schneider2012a}, Djupvik et al., in prep.), which is part of
this study.

On smaller size scales ($<$0.5 pc), there is a whole collection of
different expressions for observed features: small globules (`EGGs',
evaporating gaseous globules) are interpreted as fragments of a cloud
(e.g., Smith et al.  \cite{smith2003}) which possibly form stars
(McCaughrean \& Andersen \cite{mccaughran2002}), while proplyds
(O'Dell et al. \cite{odell1993}) are evaporating circumstellar disks.
After the discovery of proplyds in the Orion Nebula (e.g., Laques \&
Vidal \cite{laques1979}) they were searched for in the vicinity of
other OB associations. Wright et al. (\cite{wright2012}) found
proplyd-like objects in the immediate environment of the Cyg OB2
association but doubted that these are all disk-features.  Another
group of tiny ($<$0.05 pc) isolated condensations were detected by
Gahm et al. (\cite{gahm2007}) in various Galactic \hii\, regions who
named them globulettes.

\begin{figure*}[ht]     
\begin{center}  
\includegraphics[angle=0,width=11cm]{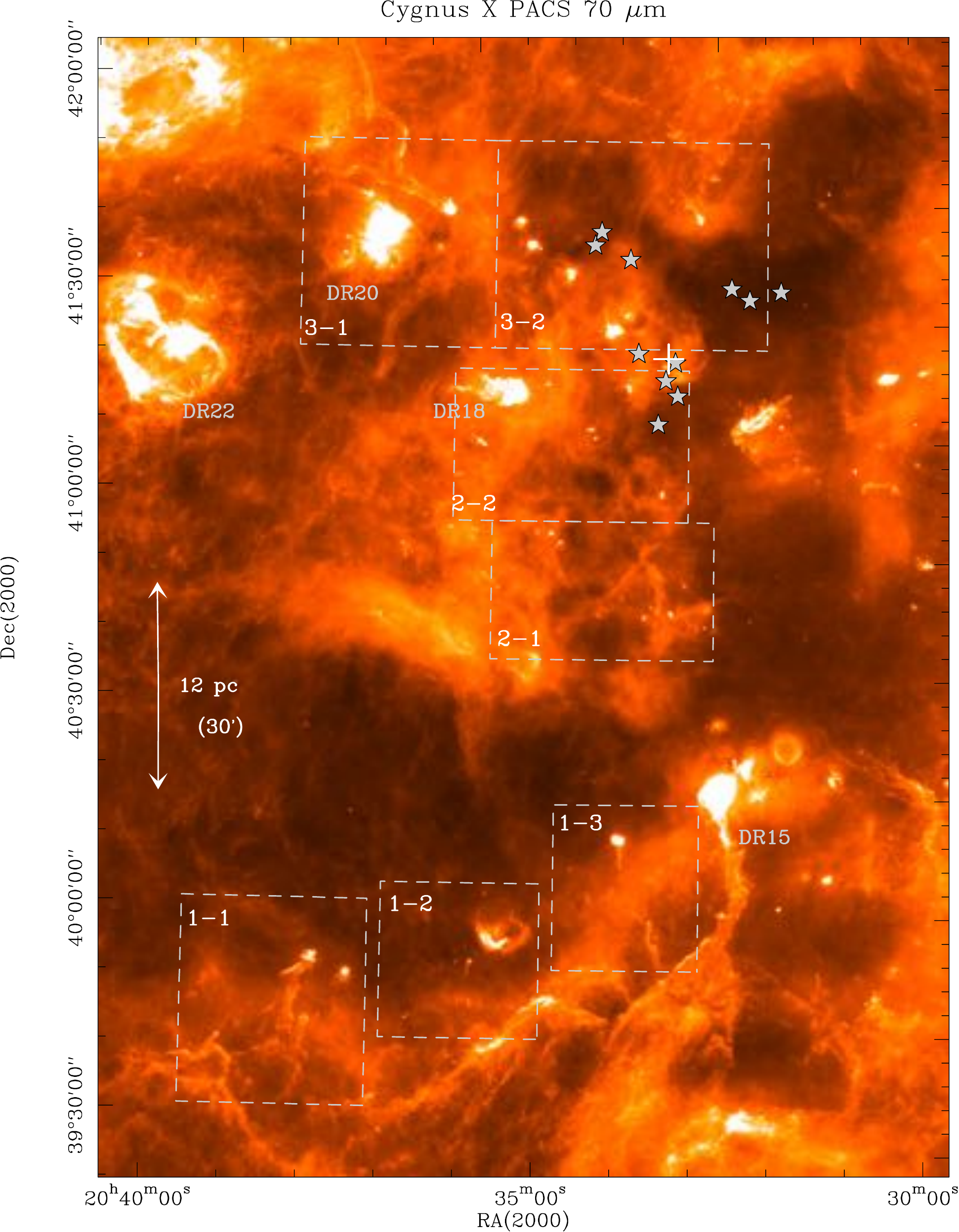} 
\end{center}  
\caption [] {{\sl Herschel} 70 $\mu$m PACS image (0 to 1000 MJy/sr) of
  the central region around Cyg OB2 (the most massive stars of the
  cluster are indicated by gray stars). The cross indicates the
  position of star Cyg OB2 \#8 at RA(2000)=20$^h$33$^m$16$^s$ and
  Dec(2000)=41$^\circ$18$'$45$''$, frequently used to define the
  center of Cyg OB2 (e.g., Wright et al. \cite{wright2015}).  The
  \hii\, regions DR15, 18, 20, and 22 are labelled and the regions
  shown and discussed in more detail are outlined with dashed lines
  and numbered.}
\label{cygnus_70} 
\end{figure*} 

With {\sl Herschel} far-infrared (FIR) imaging of high-mass
star-forming regions within the HOBYS\footnote{PIs: Motte, Zavagno,
  Bontemps; www.herschel.fr/cea/hobys} key program (Motte et al.
\cite{motte2010}), it is now possible to detect these features in the
FIR (provided that they are not too small) and also to determine
systematically their physical properties such as size, mass, and
temperature. At the same time, {\sl Herschel} observations can provide
a census of pre- and protostellar sources, and thus link the
properties of pillars and globules to star formation within them. So
far, {\sl Herschel} observations have shown that the majority of
low-mass stars form within filaments (e.g., Andr\'e et al.
\cite{andre2010}, 2014), and OB clusters where filaments merge (e.g.,
Schneider et al.  \cite{schneider2012b}, Hennemann et
al. \cite{hennemann2012}). It is not clear whether star formation in
pillars, globules, proplyds etc.  follows the same path. It is
possible that dense filaments and accretion flows are shaped by UV
radiation into the form of pillars, as was recently shown in numerical
simulations with stellar feedback processes (Dale et
al. \cite{dale2014}). These pillar-like features can then fragment
into smaller units that evolve under the influence of radiation into
globules, EGGs, condensations, proplyds, etc., depending on the intial
pre-existing density structure, and finally form stars.  The more
classical view is that large-scale compression of an expanding
\hii\ region on a molecular cloud surface creates pillars that also
evolve into globules, EGGs, condensations etc.  Various studies (e.g.,
Lefloch \& Lazareff \cite{lefloch1994}, Williams et
al. \cite{williams2001}, Miao et al. \cite{miao2006}, \cite{miao2009})
have shown that instabilities in the \hii\, region/molecular cloud
interface create bright-rimmed clouds and pillars that can detach to
form isolated globules.  The importance of the turbulent gas structure
was recognized by Gritschneder et al. (2009, 2010).  Tremblin et
al. (2012a,b) have shown that the turbulent density structure of
molecular clouds can lead to local curvatures of the dense shell
formed by the ionization compression, which may develop into pillars
that can subsequently detach from the cloud.  When the turbulent ram
pressure of the molecular gas is larger than the ionized-gas pressure,
globules can form.

In any case, both star-formation schemes outlined above produce
isolated stars and the difference between them lies mainly in its
primoridal phase. It is thus very difficult, perhaps impossible, to
find observational which would allow us to discriminate between the
two scenarios described.

Several HOBYS studies have already observed the \hii\,
region/molecular cloud interface, including pillars, globules, and
\hii\, bubbles, and how these are impacted by OB-clusters through
heating (e.g., Schneider et al.  \cite{schneider2010}, Hill et
al. \cite{hill2012}, Anderson et al.  \cite{anderson2012}, and Didelon
et al. \cite{didelon2015}), external compression (Zavagno et al.
\cite{zavagno2010}, Hill et al. \cite{hill2011}, Minier et
al. \cite{minier2013}, Tremblin et al.  \cite{tremblin2014}), and
ionization (Deharveng et al.  \cite{deharveng2012}, Tremblin et
al. \cite{tremblin2013}). In this paper, we focus on investigating the
immediate environment of the Cyg OB2 association in the so-called
Cygnus X region (see, e.g., Reipurth \& Schneider \cite{reipurth2008}
for an overview).  Cygnus X is a large $\sim$10~degree wide radio
emission feature (Piddington \& Minnett \cite{piddington1952})
composed of numerous individual \hii\ regions. A major problem is the
uncertainty in distances because we look down a spiral arm, with the
resulting confusion of regions as near as a few hundred parsec with
others at 1-2 kpc and even well beyond.  Kinematical distances in this
region are very unreliable for distances up to $\sim$4 kpc because of
the near-zero radial velocity gradient. For the objects studied in
this paper, however, we do not expect a significant confusion because
all spatially resolved objects clearly point towards Cyg OB2 and are
thus shaped by the radiation of the stars.
  
Using the observational {\sl Herschel} FIR-data (Sec.~\ref{obs}) and
the {\sl Herschel} derived column density, temperature, and FUV-field
maps (Sec.~\ref{uv} and \ref{ident}), we classify the various features
seen in the data (Sec. \ref{results}), calculate their lifetimes, and
discuss their possible evolution (Sec.~\ref{discuss}). A comparison of
the pre- and protostellar sources found within various structures will
be made in a subsequent paper.

 
\begin{figure*}[ht] 
\begin{center}  
\includegraphics[angle=0,width=9cm]{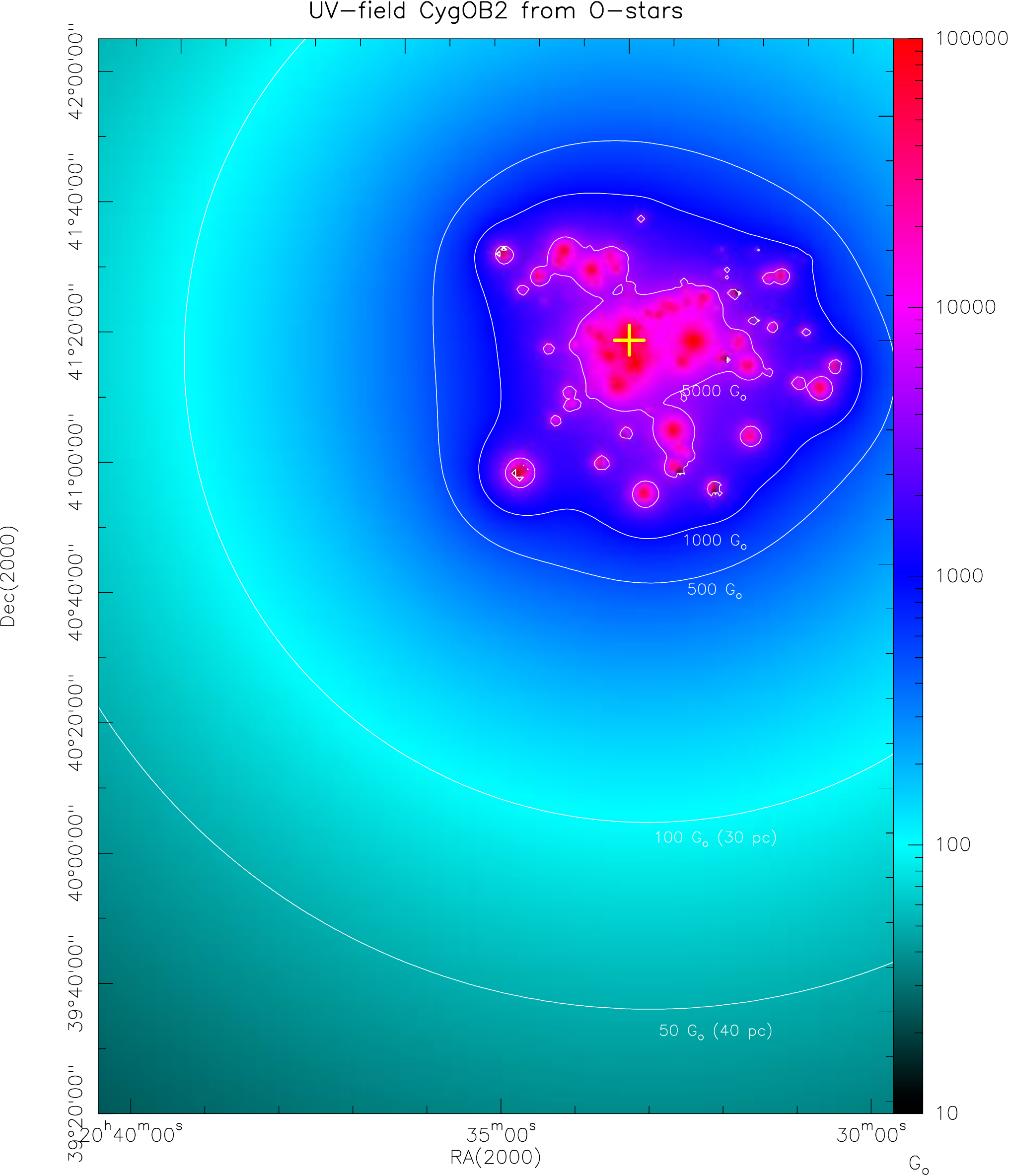}
\includegraphics[angle=0,width=9cm]{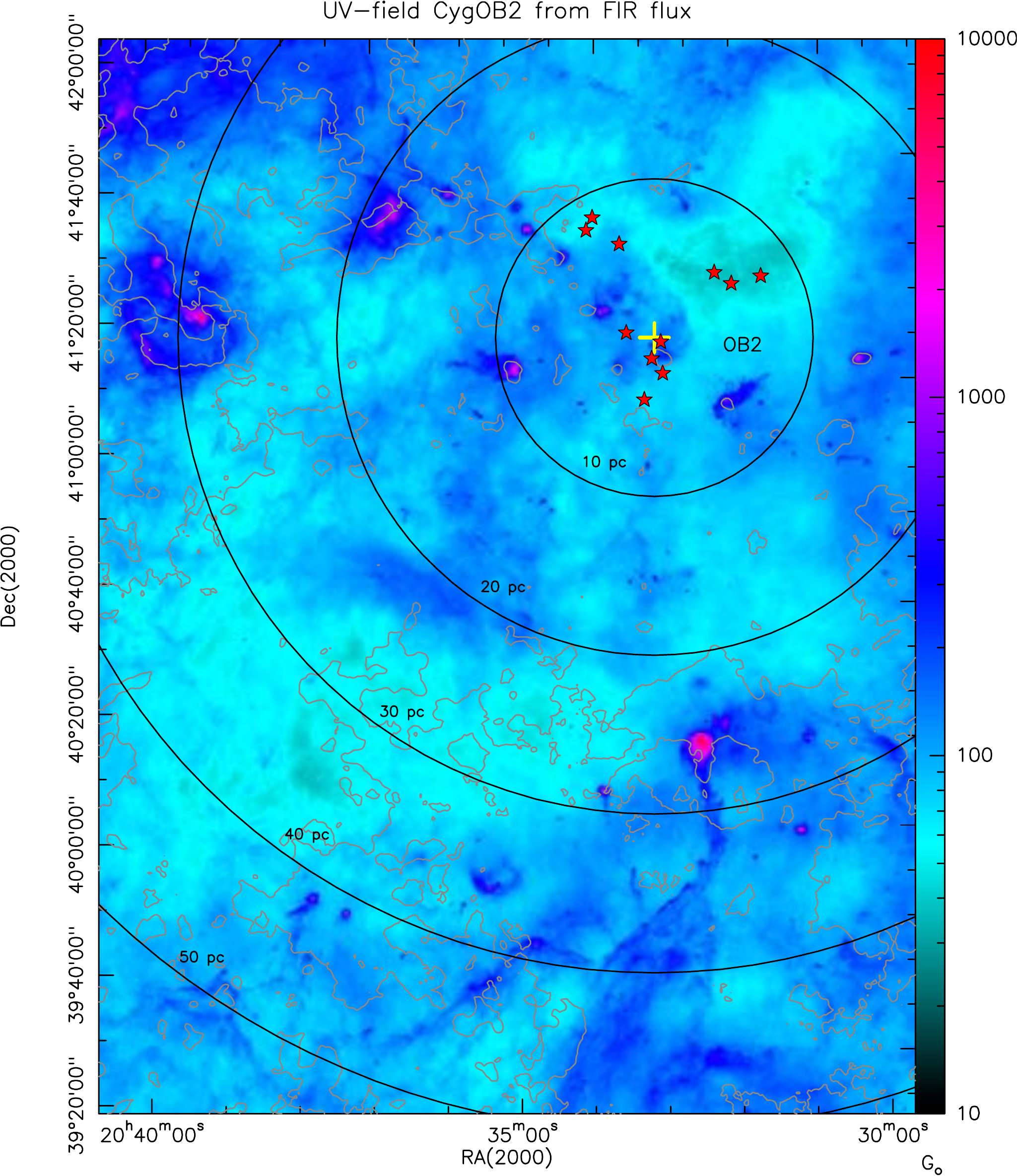}
\caption [] {{\bf Left:} UV-field in units of Habing (G$_0$)
  determined from a census of all O-stars given in Wright et
  al. (2015). {\bf Right:} UV-field in units of Habing (G$_0$) for the
  central region around Cyg OB2 (limits as in Fig.~2) determined from
  the {\sl Herschel} 70 $\mu$m and 160 $\mu$m fluxes. This image shows
  the impinging radiation. The most luminous O-stars of Cyg OB2
  (Wright et al. 2015) are indicated as red stars and the yellow cross
  indicates the approximate center of the association. Concentric
  circles indicate the projected distance to this center in the plane
  of the sky, assuming a distance of 1.4 kpc to Cyg OB2. Gray contours
  (2 K kms$^{-1}$ to 22 K kms$^{-1}$ in steps of 2 K kms$^{-1}$)
  display line integrated (-10 to 20 km s$^{-1}$) $^{13}$CO 1$\to$0
  emission at a resolution of 45$''$ (Schneider et al.
  \cite{schneider2011}).}
\label{uv-region1} 
\end{center}  
\end{figure*} 

\section{Observations, dust column density and dust temperature maps} \label{obs}  
 
Two fields covering the Cygnus X region were observed with the {\sl
  Herschel} satellite (Pilbratt et al. \cite{pilbratt2010}) in
parallel mode.  We present {\sl Herschel} imaging observations at 70
$\mu$m and 160 $\mu$m from PACS (Photoconductor Array Camera and
Spectrometer) (Poglitch et al. \cite{poglitsch2010}), and at 250
$\mu$m, 350 $\mu$m, and 500 $\mu$m from SPIRE (Spectral and
Photometric Imaging Receiver) (Griffin et al. \cite{griffin2010}).
{\bf Cygnus X South} was observed on May 24, 2010 (obsIDs 1342196917
and 1342196918 for the nominal and orthogonal directions,
respectively) and {\bf Cygnus X North} on December 18, 2010 (obsIDs
13422211307 and 1342211308). This paper uses the observations of a
part of both regions. The DR21 filament in Cygnus X North has already
been presented in Hennemann et al. (\cite{hennemann2012}, {\sl
  Herschel} imaging) and White et al. (\cite{white2010}, {\sl
  Herschel} spectroscopy).  For this paper, we employed a more recent
HIPE (Herschel Interactice Processing Environment) version (10.0.2751)
for the PACS and SPIRE data reduction in which we used modified
pipeline scripts. Data collected during the turnaround of the
satellite were included to insure a larger area was covered by the
observations.  The resulting Level1 contexts for each scan direction
were combined using the `naive' map maker in the destriper module.
The conversion of the maps into surface brightness (from Jy/beam into
MJy/sr) was made using the beam-areas obtained from measurements of
Neptune (March 2013, see SPIRE handbook). The map offsets of the
Level2 data were then determined using {\sl Planck} and {\sl IRAS}
observations (Bernard et al.  \cite{bernard2010}). The zero levels for
the SPIRE 250 $\mu$m, 350 $\mu$m, and 500 $\mu$m data were also
determined using the {\sc zeroPointCorrection} task in HIPE and found
to be consistent with offsets provided by J.-P. Bernard. PACS data
were reduced up to Level1 using HIPE 10.0.2751 and then v. 20 of the
Scanamorphos software package which performs baseline and drift
removal before regridding (Roussel \cite{roussel2013}). See Hennemann
et al. (\cite{hennemann2012}), Hill et al.  (\cite{hill2011}) and
K\"onyves et al. (\cite{vera2015}) for further details.  The angular
resolutions at 70 $\mu$m, 160 $\mu$m, 250 $\mu$m, 350 $\mu$m, and 500
$\mu$m, are $\sim$6$''$, $\sim$12$''$, $\sim$18$''$, $\sim$25$''$, and
$\sim$36$''$, respectively.

Figure~\ref{cygnus} shows a three-colour image of the whole Cygnus X
region observed as part of the HOBYS program. The picture reveals
impressively the way in which bright star-forming regions, emitting
mainly at 70 $\mu$m and 160 $\mu$m (blue/green), are nestled in the
larger cloud which is pervaded by a web of filaments. Colder gas is
well traced by dust emission at 500 $\mu$m (red).  At the very center,
the diffuse blue emission of PACS at 70 $\mu$m indicates heating by
the Cyg OB2 association (see Fig.~\ref{cygnus_70} to locate the
highest-mass stars). The \hii\, regions of Cygnus X (DR18, 20, and 22
in the northern region and DR15 in the southern region) stand out as
bright emission regions in the three-colour image. The majority of
pillars and globules in the figure, however, point towards the center
of Cyg OB2, the most important source of UV-illumination (see
Sec.~\ref{uv}). A zoom into the vicinity of Cyg OB2 is shown by a PACS
image at 70 $\mu$m (Fig.~\ref{cygnus_70}); individual maps at 70
$\mu$m to 500 $\mu$m of these cutouts are shown in appendix A. In
sections~\ref{results} and \ref{discuss}, we focus on the regions
indicated in this plot and discuss the properties of the observed
features.

Column density and dust temperature maps at an angular resolution of
36$''$ (all maps were smoothed to this lowest resolution of the 500
$\mu$m map) were made with a pixel-by-pixel SED fit from 160 $\mu$m to
500 $\mu$m, as described in, for example, Hill et
al. (\cite{hill2011}).  For the SED fit, we used a specific dust
opacity per unit mass (dust and gas) approximated by the power law
$\kappa_\lambda \, = \,0.1 \, (\lambda/300 \,{\rm \mu m})^\beta$
cm$^{2}$g$^{-1}$ with $\beta$=2 (Roy et al. \cite{roy2014}), and left
the dust temperature and column density as free parameters. We checked
the SED fit of each pixel and determined from the fitted surface
density the H$_2$ column density.  The fit assumes a constant
temperature for each pixel along the line of sight, an assumption that
is not fullfilled in regions with strong temperature gradients.  A
detailed study of the dust properties in Orion A (Roy et
al. \cite{roy2013}), however, showed that the single temperature model
provides a reasonable fit, but that the dust opacity varies with
column density by up to a factor of two. We thus estimate that the
final uncertainties of the column density map are between 20\% and
30\%.

\section{The FUV-field in Cygnus X } \label{uv} 

Cyg OB2 is one of the largest and most massive OB associations in the
Galaxy with a mass of around 3$\times$10$^4$ M$_\odot$ (Drew et al.
\cite{drew2008}) at a distance of only 1.4 kpc (Rygl et
al. \cite{rygl2012} from parallax observations).  It is the richest
aggregate in the Cygnus X region (e.g., Hanson \cite{hanson2003},
Comer\'on \& Pasquali \cite{comeron2012}), though the whole Cygnus X
region contains several OB-associations, and is clearly the most
important source of FUV-impact for the features we study here.  A
recent inventory of the central massive stars in Cyg OB2 compiles 52
O-stars and 3 Wolf-Rayet stars (Wright et al. \cite{wright2015}). They
are the most important sources of feedback (radiation, winds) though
the whole Cyg OB2 association is more extended (Comer\'on et al.
\cite{comeron2008}, Comer\'on \& Pasquali \cite{comeron2012}).  We
evaluate the FUV-field expressed as a Habing field\footnote{With G$_0$
  in units of the Habing field (Habing \cite{habing1968})
  2.7$\times$10$^{-3}$erg cm$^{-2}$ s$^{-1}$ and the relation
  G$_0$=1.7$\chi$ with the Draine field $\chi$.}, produced by the 52
O-stars listed in Wright et al. (\cite{wright2015}).  The ionising
fluxes (Wright et al., priv. comm.) were calculated taking into
account not just their spectral types, but also their exact
luminosities, and binary companions. The most recent stellar
atmosphere models were used, including the revised effective
temperature scale in Martins et al. (\cite{martins2005}), and the
luminosity shift that this leads to.  We assume a simple 1/r$^2$
decrease of the flux and project all stars in the plane of the sky,
ignoring possible attenuation by diffuse gas and blocking by molecular
clumps. Additional sources of illumination, for example the more
widespread O- and B-star population of Cyg OB2, are also not
considered in this method. The resulting FUV-field is shown in
Fig.~\ref{uv-region1} (left panel).  Obviously, the field is strongest
in the immediate environment of the stars, that is, the central 10 pc
circle, reaching values of up to a few 10$^5$ G$_0$ and drops to
$\sim$50 G$_0$ in 40 pc distance.

In a second approach, we estimate the FUV-field from the total FIR
intensity ($I_{FIR}$), assuming that the radiation from the massive
stars heating the dust is re-radiated mainly at FIR wavelengths. With
{\sl Herschel}, we evaluate the FUV-field by adding the intensities at
70 $\mu$m and 160 $\mu$m (at an angular resolution of 12$''$) to
I$_{FIR}$ [10$^{-17}$ erg cm$^{-2}$ s$^{-1}$ sr$^{-1}$] (see Kramer et
al. \cite{kramer2008}, Roccatagliata et al. \cite{rocca2013} for the
methodology):
\begin{equation}
F_{fuv} [G_0] \,\, = \,\, (4\, \pi I_{FIR} \, 1000)/1.6
\end{equation}
Note that the use of 160 $\mu$m emission as a tracer of the FUV-field
can be disputable in regions dominated by cooler gas (which is not the
case for Cygnus X) because flux at this wavelength can also come from
cold thermal dust emission from the molecular cloud (the wavelength
range 160 $\mu$m to 500 $\mu$m is used for the SED fit to determine
the column density). For Cygnus X, the 160 $\mu$m emission contributed
approximately 20\% to the total UV-field, displayed in
Fig.~\ref{uv-region1} (right), that ranges between a few hundred G$_0$
to more than 10$^4$ G$_0$.

This method emphasizes the PDR surfaces that stand out prominently.
These are illuminated by a UV-field of at least a few hundred G$_0$.
Internal \hii\, regions can also contribute to the local FUV-field as
in the case of the bright globule ({\tt g1}) in Cygnus X (see below)
that contains early B-stars. Here, the average UV-field across the
globule head is 550 G$_0$ (see Table~1), but increases up to 7500
G$_0$ in a 12$''$ beam at the position of the B-star. Note, however,
that the UV-field can only be calculated on the surfaces of molecular
condensations (clumps, filaments, pillars, globules, etc.).  In the
ionized phase, the UV-field is, of course, present but can not be
evaluated with this method.  Instead, the field is estimated as
outlined above, that is, from the photon flux of the stars
(Fig.~\ref{uv-region1}, right). The maps are thus not comparable but
complementary, and a combination of both maps probably represents the
overall FUV-field in Cygnus X best.

\begin{figure*}[ht]     \label{1-1}
\begin{center}  
\includegraphics[angle=0,width=7cm]{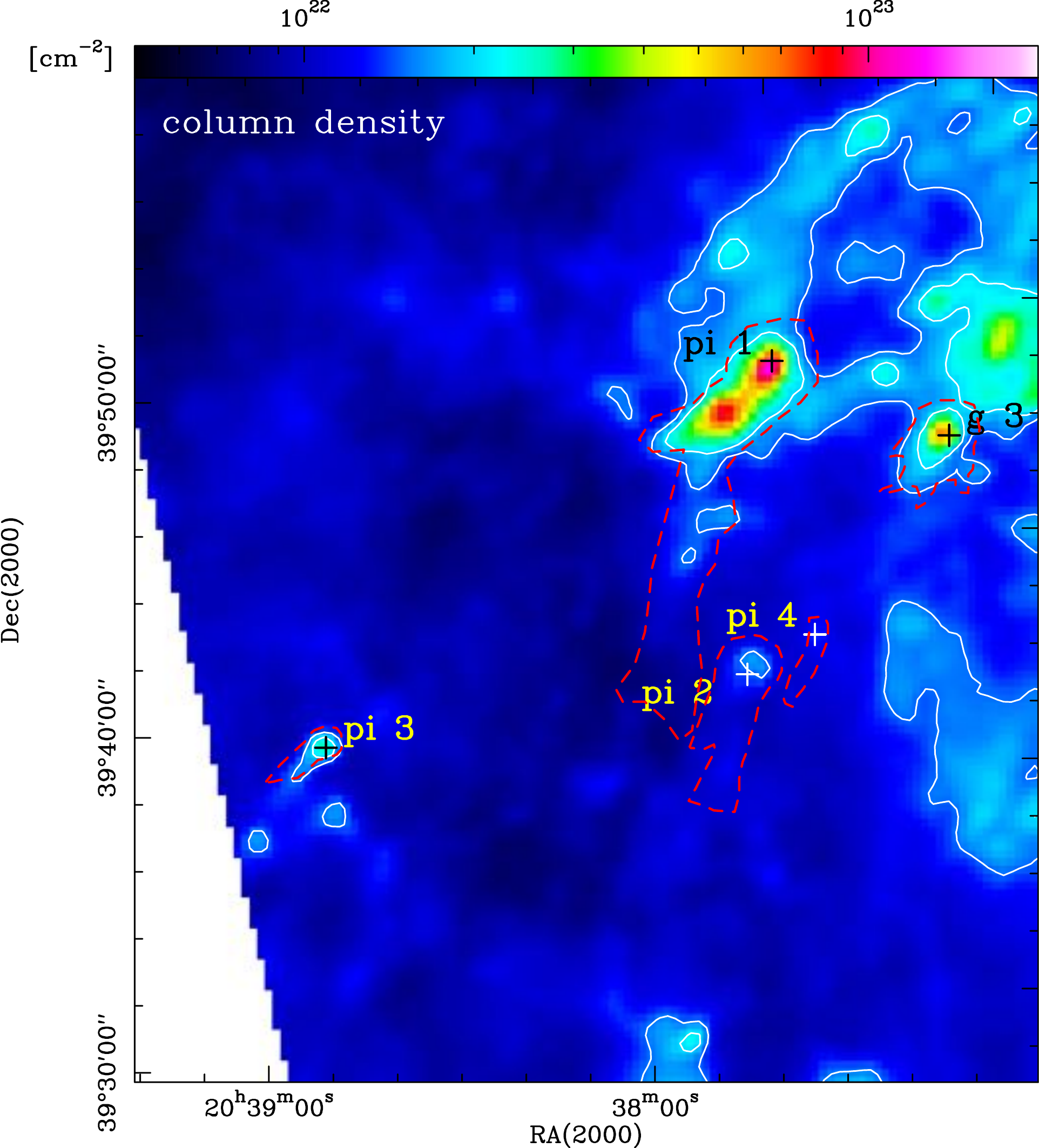} 
\includegraphics[angle=0,width=7cm]{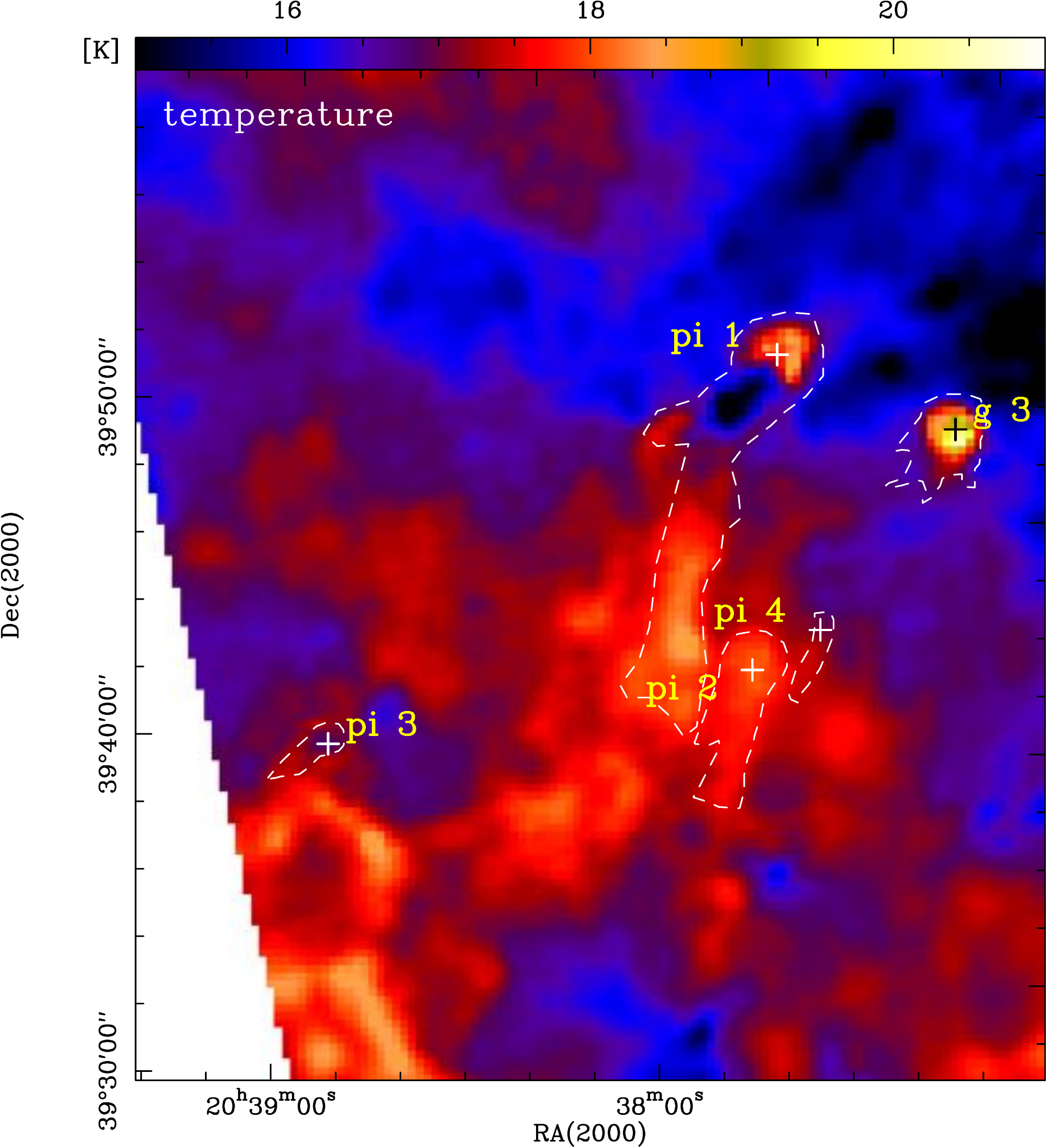} 
\end{center}  
\caption [] {Region 1-1: column density (left) and dust temperature
  (right). Pillars are indicated by `pi', globules by `g',
  proplyds/proplyd-like by `pr', condensations by `c', and EGGs by
  `e'. The red dashed (white dashed in the right panel) contours
  outline the source shape based on the 70 $\mu$m map (derived from
  the 400 MJy/sr threshold). The column density contours correspond to
  the levels 1.5 and 2.0 10$^{22}$ cm$^{-2}$. }
\end{figure*} 

\begin{figure*}[ht]  \label{1-2} 
\begin{center}
\includegraphics[angle=0,width=7cm]{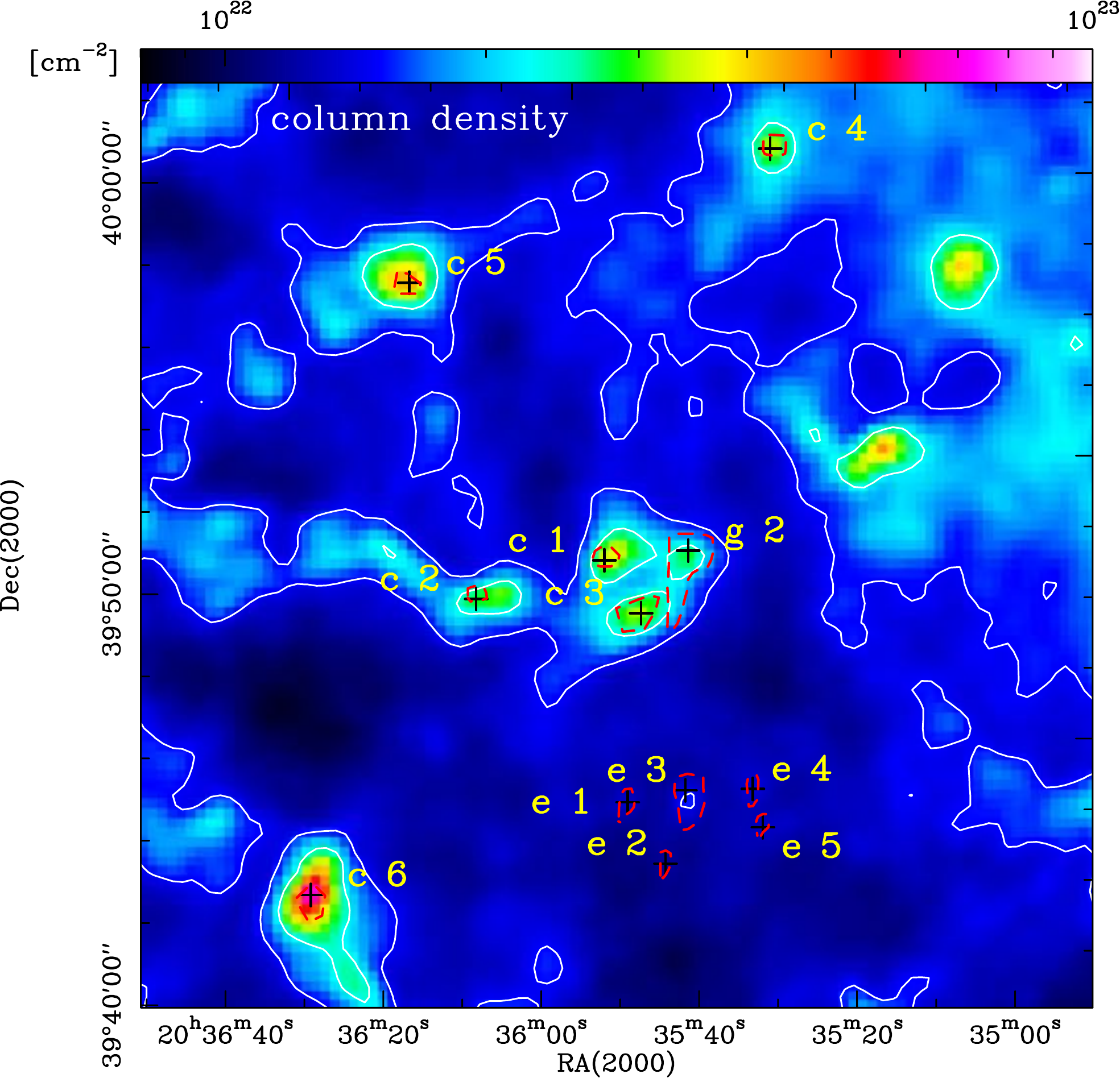} 
\includegraphics[angle=0,width=7cm]{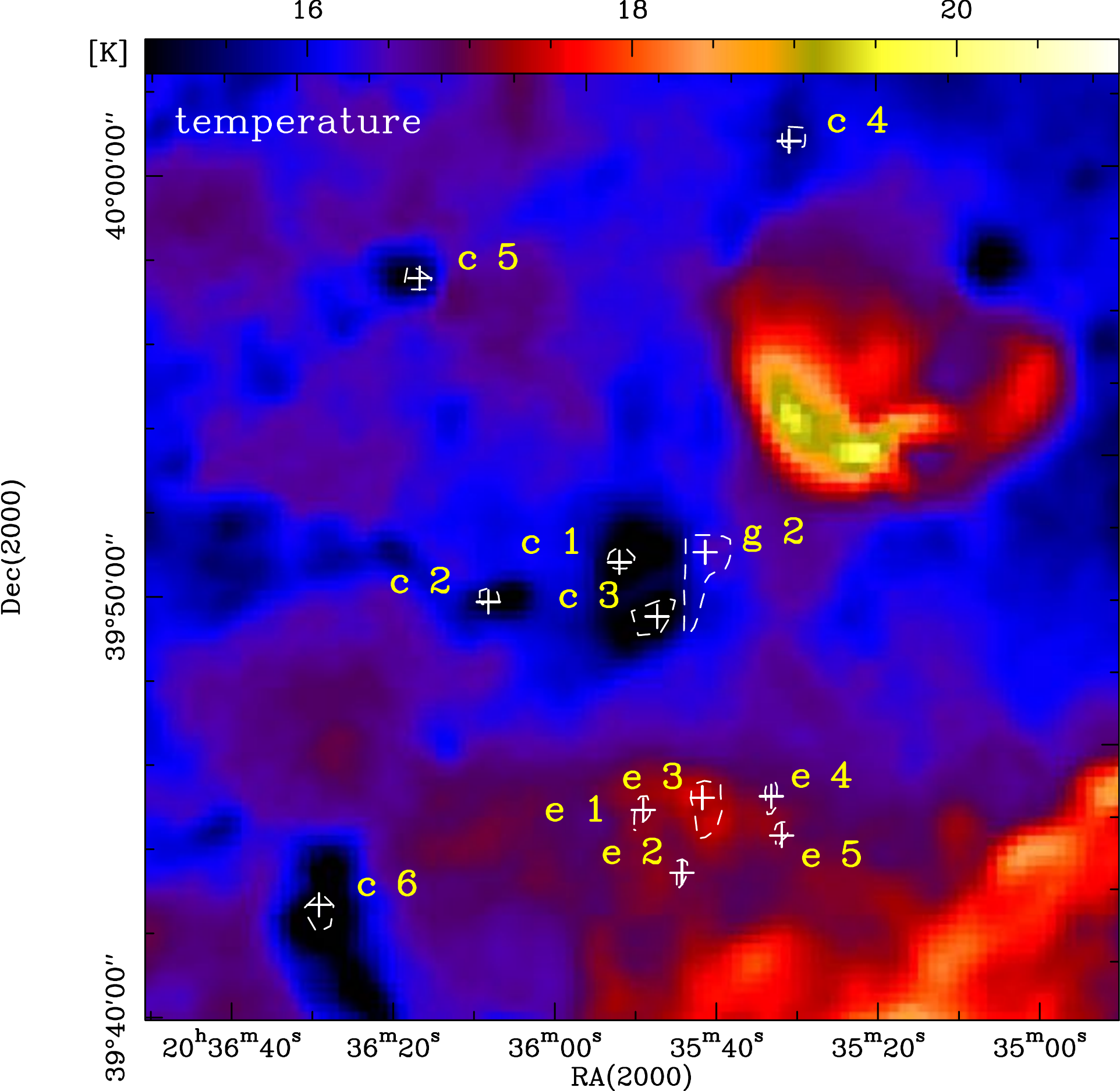} 
\end{center}  
\caption [] {Region 1-2: as in Fig.~4. The column density contours 
correspond to the levels 1.5 and 2.3 10$^{22}$ cm$^{-2}$. }   
 \end{figure*} 

\begin{figure*}[ht]     \label{1-3}  
\begin{center}
\includegraphics[angle=0,width=6.5cm]{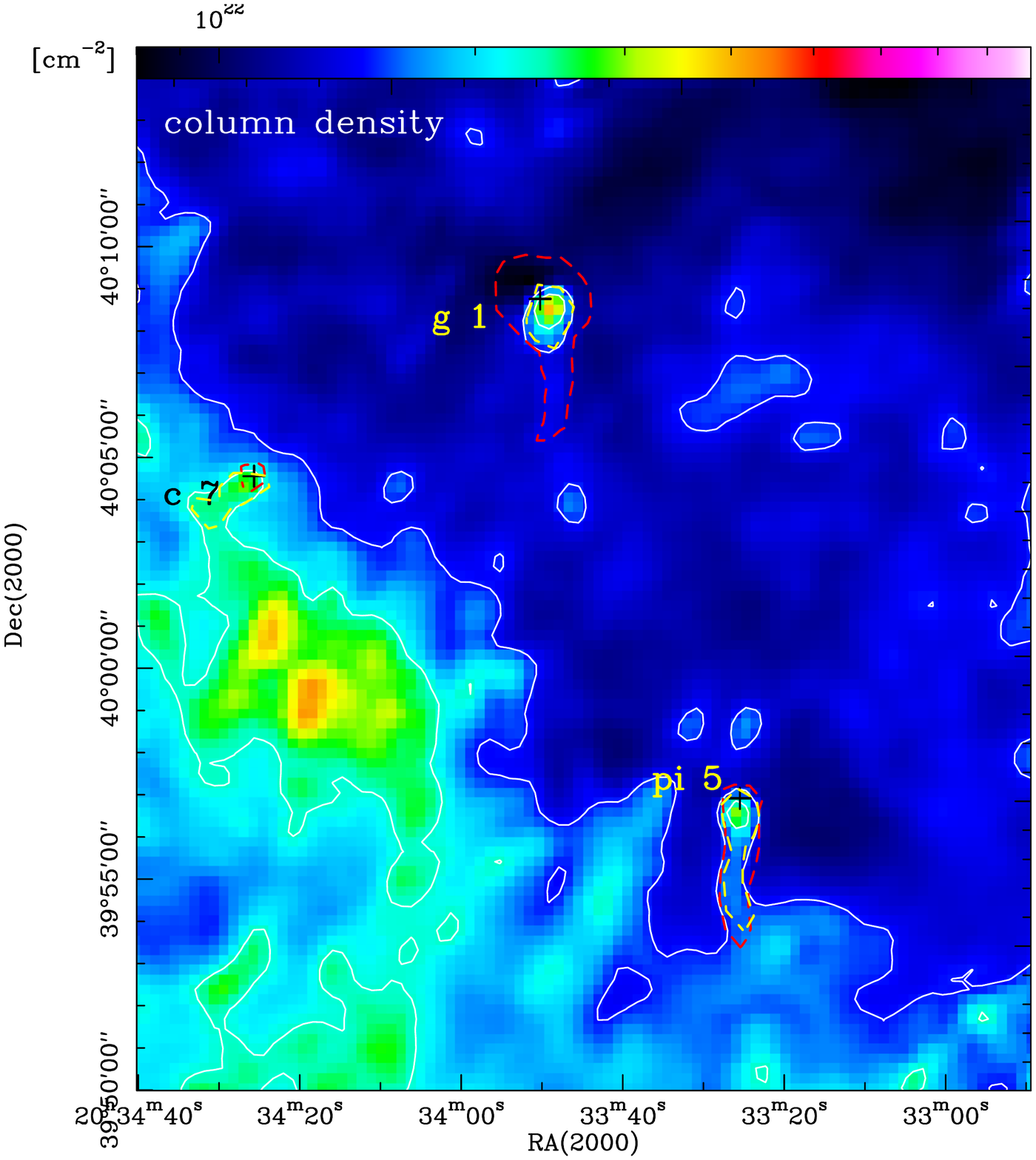} 
\includegraphics[angle=0,width=6.5cm]{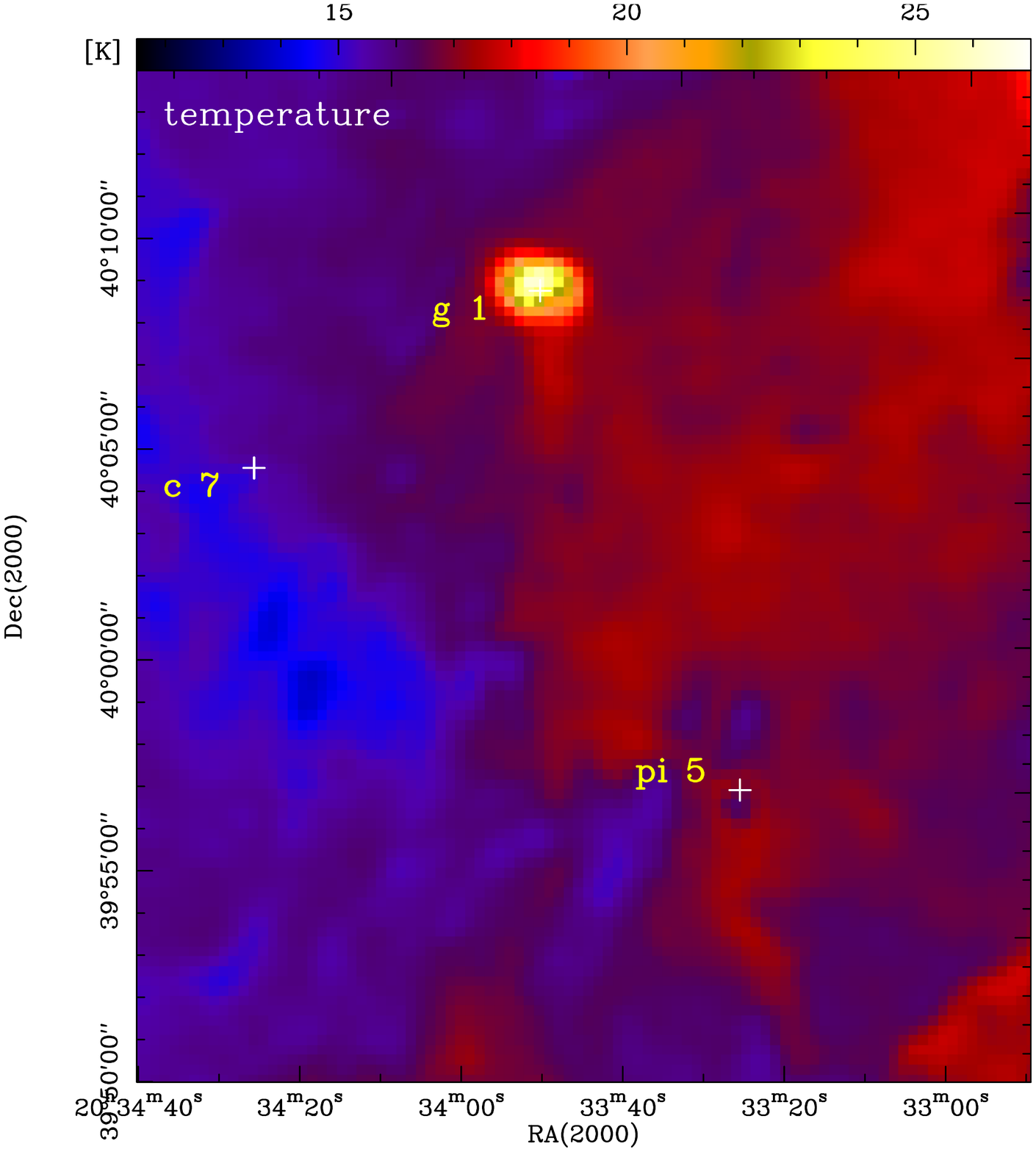} 
\end{center}  
\caption [] {Region 1-3: as in Fig.~4. The column density contours 
correspond to the levels 1.5 and 2.3 10$^{22}$ cm$^{-2}$.  }   
\end{figure*} 

\section{Identification of objects} \label{ident}

From the five {\sl Herschel} wavelength bands, the 70 $\mu$m map
(Fig.~\ref{cygnus_70} and Appendix A) is the best suited to trace
UV-illuminated features because the emission at 70 $\mu$m arises from
warm dust and has the highest angular resolution (6$''$).  The maps at
shorter wavelengths (70 $\mu$m, 160 $\mu$m, and 250 $\mu$m) all start
at 0 MJy/sr and show that with increasing wavelength, the amount of
emission from colder dust increases and is visible as a diffuse
background.

The 70 $\mu$m maps are dominated by emission from UV-heated dust, and
so pillars, globules, and other features clearly stand out from a
lower background, as is indicated by the contour at 400 MJy/sr (i.e.,
the sharp transition from red to green in the map). We thus used this
contour to trace UV-illuminated features in the 70 $\mu$m maps and not
the maps at longer wavelengths that are dominated by cold dust, mixed
with mostly molecular gas. For display reasons, we set the lower
limits to 150 MJy/sr and 70 MJy/sr for the 350 $\mu$m and 500 $\mu$m
maps, respectively. The 500 $\mu$m maps thus show only the densest
parts of the objects, that is, the heads of pillars or globules which
are also traced in the column density maps. It is only with the
temperature maps, that it is possible to distinguish whether the gas
is warm or cold.

For display reasons, we divided the map shown in Fig.~\ref{cygnus_70}
into subregions covering many observed features. The resulting maps
reveal many objects with different shapes and sizes in the vicinity of
the Cyg OB2 cluster. As explained above, we take the contour level 400
MJy/sr (shown in the individual 70 $\mu$m maps of each subregion; see
Appendix A) as a threshold to define the borders of all objects.  This
contour typically corresponds to a column density of
1--2$\sim$10$^{22}$ cm$^{-2}$. We note, however, that the shape of the
objects defined in the 70 $\mu$m maps do not always correspond
one-to-one to the column density contour. For example, clumps within
molecular clouds are prominent column density peaks but do not show up
in the 70 $\mu$m map. We classified elongated, column-like structures
that are still attached to the molecular cloud as {\bf pillars},
isolated head-tail features as {\bf globules} when they are large
(typically $>$1$'$) and prominent and as {\bf EGGs} when they are
small ($<$1$'$) and faint. Centrally condensed objects without tails
were called {\bf condensations}.  The term {\bf proplyd-like} was used
for the sources given in Wright et al. (\cite{wright2012}) as well as
to some new ones, which probably fit into this scheme, detected in the
{\sl Herschel} images.  All other features in the map that we did not
classify have a more or less arbitrary shape and are most often bright
edges of \hii\, regions. Our classification is complete, within the
observed area, though we do not cover the whole Cygnus X region. Our
main objective, however, is to derive typical values and discuss
differences between the features representing different object classes
in the various subregions.

The contours and labels of the objects identified in this way are
shown in the column density and temperature maps of each subregion
(see Figs.~4 to 10). Obviously, for some sources, the shape outlined
by the 70 $\mu$m contour does not fully correspond to its shape in the
column density or temperature map. For example, Pillar 1 in region 1-1
(Fig.~\ref{1-1}) consists of a dense head with two peaks in column
density of which one is cold ($\sim$14 K) and the other warm ($\sim$19
K), while the base of the pillar does not show up in the column
density map. With this caveat in mind, we then determine the physical
properties (column density, temperature, density, mass, length and
width or radius, etc.) of the objects (Table~1 lists these
quantities). In the following sections, we qualitatively describe the
maps and objects and quantitatively discuss their various properties.

\begin{figure*}[ht]       \label{3-2}    
\begin{center}  
\includegraphics[angle=0,width=7cm]{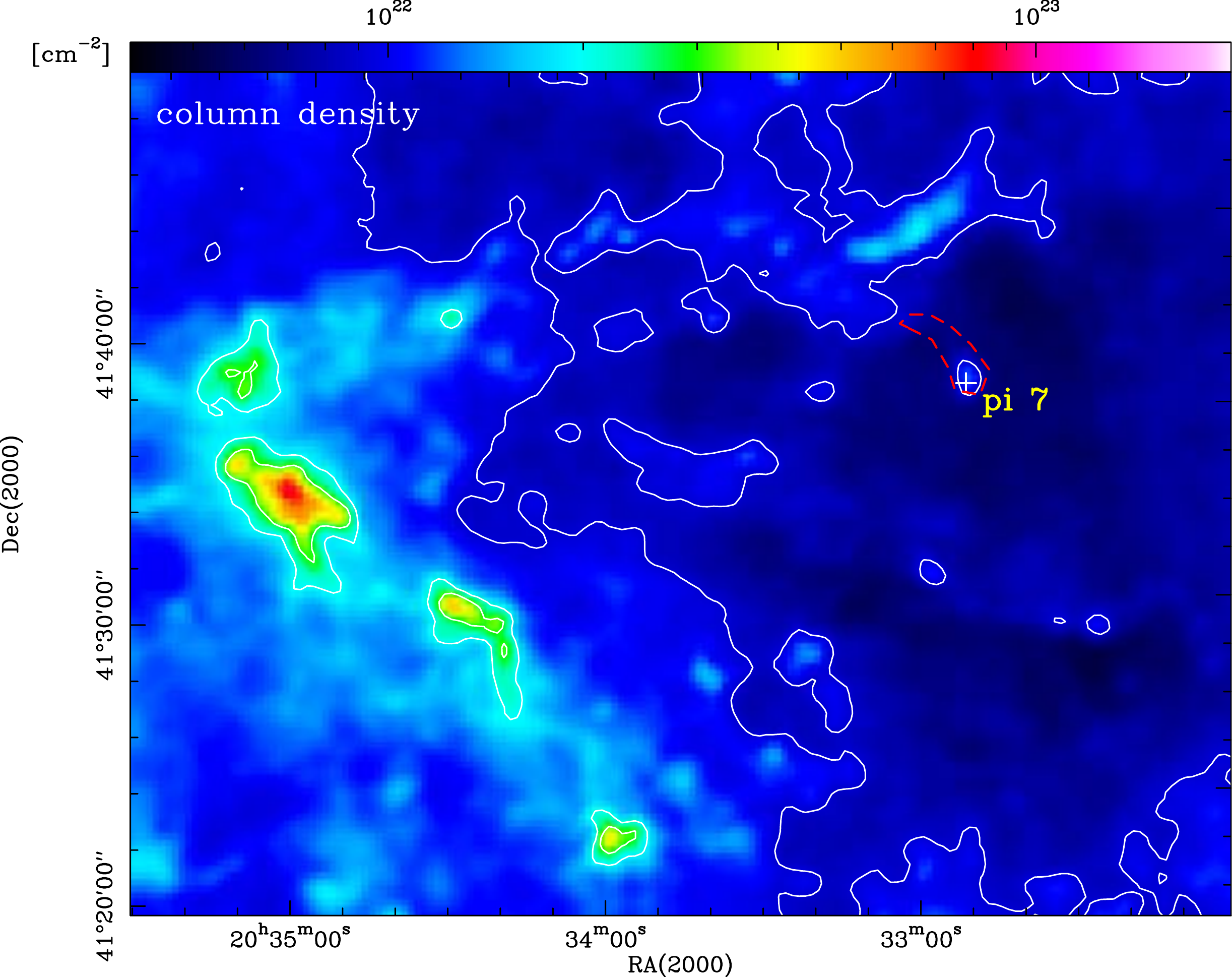} 
\includegraphics[angle=0,width=7cm]{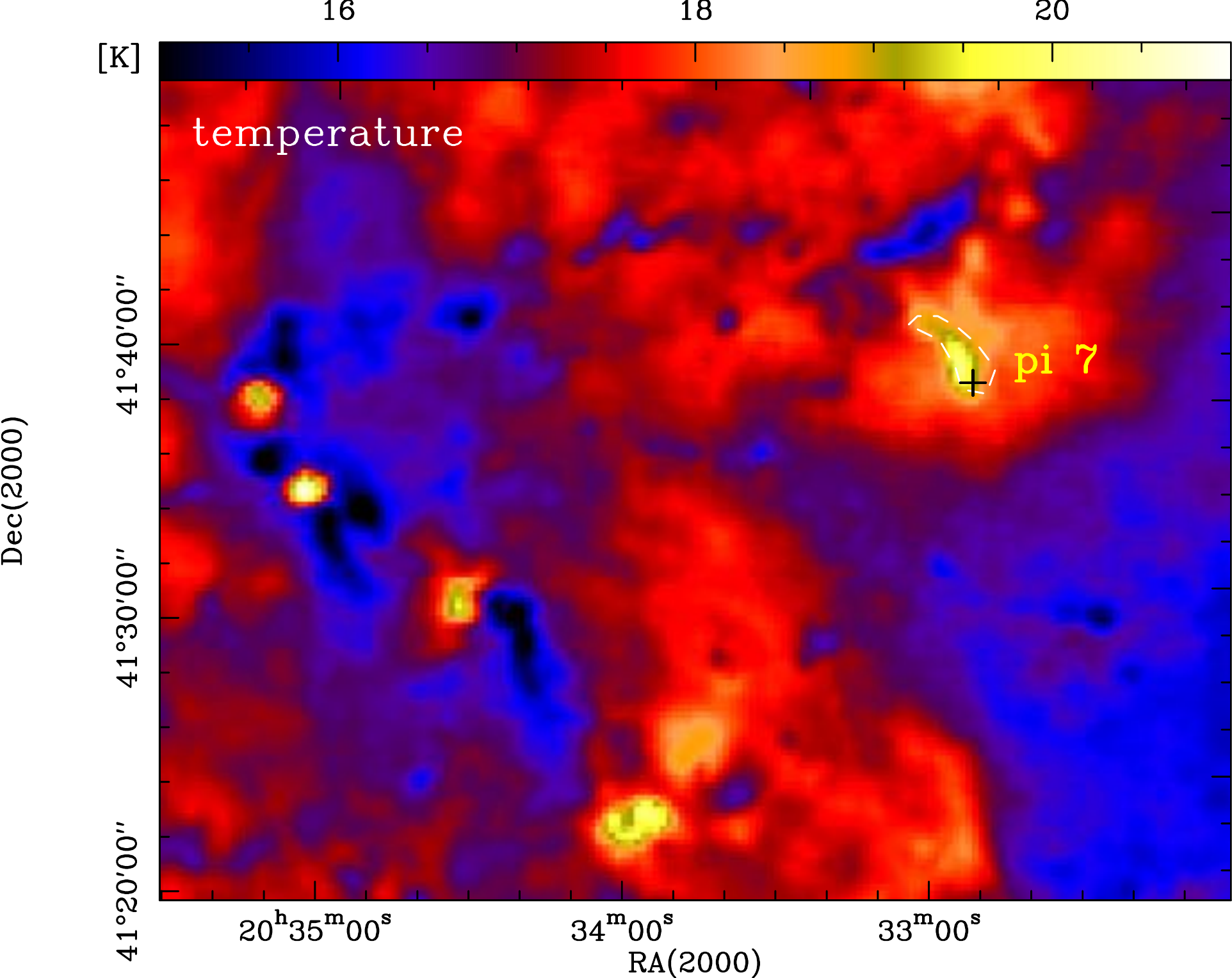} 
\end{center}  
\caption [] {Region 3-2: as in Fig.~4. The column density contours 
correspond to the levels 1.0 and 1.7 10$^{22}$ cm$^{-2}$.}   
\end{figure*}

\begin{table*}[ht] 
 \centering 
   \begin{tabular}{lcccccccccc} 
\hline 
\hline 
Source  &$\langle N(H_2) \rangle$&$M$&$\langle n(H_2) \rangle$&$\langle$T$\rangle$&T$_{min}$&T$_{max}$&$r$&$l\times w$&$\Sigma$&$\langle$flux$\rangle$ \\ 
        & [10$^{21}$ cm$^{-2}$]&[M$_\odot$]&[10$^{3}$ cm$^{-3}$]&[K]&[K]&[K]&[pc]&[pc$\times$pc]&[M$_\odot$/pc$^2$]&[Go] \\  
        &   (1)  & (2)  &  (3) & (4)  & (5) & (6)   & (7)  & (8) & (9) & (10) \\ 
\hline 
\multicolumn{3}{l}{ {\bf Pillars}} &  \\   
1       &  21.0  & 1680 &  2.9 & 17.3 & 14.8 & 18.7 & 1.17 & 2.86 $\times$ 0.46 & 390 & 198 \\  
2       &  11.5  & 282  &  3.1 & 17.7 & 17.4 & 18.2 & 0.65 & 1.22 $\times$ 0.31 & 213 & 147 \\
3       &  16.1  & 83   &  8.8 & 17.1 & 16.7 & 17.3 & 0.30 & 0.61 $\times$ 0.17 & 298 & 177 \\
4       &  11.5  & 50   &  6.9 & 17.3 & 16.8 & 17.4 & 0.27 & 0.63 $\times$ 0.13 & 213 & 122 \\
5       &  17.2  & 156  &  7.1 & 17.1 & 16.2 & 17.4 & 0.39 & 0.87 $\times$ 0.18 & 319 & 175 \\
6       &  21.0  & 1403 &  3.2 & 18.0 & 17.3 & 20.2 & 1.07 & 2.38 $\times$ 0.43 & 390 & 295 \\
7       &  7.9   & 86   &  3.0 & 19.1 & 18.1 & 20.1 & 0.43 & 0.88 $\times$ 0.24 & 147 & 191 \\
{\bf mean}&{\bf 15.2$\pm$1.9}&{\bf 534$\pm$263}&{\bf 5.0$\pm$0.9}&{\bf 17.7$\pm$0.3}&{\bf 16.8$\pm$0.4}&{\bf 18.5$\pm$0.5}
&{\bf 0.61$\pm$0.14}& & {\bf 281$\pm$35}&{\bf 186$\pm$21} \\
\hline
\multicolumn{3}{l}{ {\bf Globules}} &  \\  
1       &  14.2  & 238  & 4.3  & 19.7 & 17.2 & 26.3 & 0.54 & & 263 & 549 \\
2       &  22.1  & 108  & 23.8 & 16.1 & 15.7 & 16.5 & 0.29 & & 410 & 156 \\
3       &  30.6  & 180  & 15.8 & 17.5 & 15.6 & 20.2 & 0.31 & & 568 & 294 \\ 
4       &  15.0  & 1356 & 2.0  & 19.7 & 18.6 & 23.7 & 1.24 & & 279 & 428 \\ 
{\bf mean}&{\bf 20.5$\pm$3.8}&{\bf 470$\pm$296}&{\bf 12$\pm$5}&{\bf 18.3$\pm$0.9}&{\bf 16.8$\pm$0.7}&{\bf 21.7$\pm$2.1}
&{\bf 0.60$\pm$0.22}& &{\bf 380$\pm$71}&{\bf 357$\pm$85} \\
\hline
\multicolumn{3}{l}{ {\bf EGGs}}&  \\  
1       & 12.9   & 10.8 & 18.6 & 17.1 &      &      & 0.11 & & 239 & 137 \\
2       & 11.7   & 5.9  & 22.8 & 17.0 &      &      & 0.08 & & 217 & 113 \\
3       & 13.5   & 38.6 & 10.1 & 17.3 &      &      & 0.22 & & 251 & 153 \\
4       & 13.5   & 4.5  & 34.4 & 16.9 &      &      & 0.06 & & 251 & 137 \\
5       & 12.8   & 6.4  & 25.0 & 17.1 &      &      & 0.08 & & 238 & 127 \\
{\bf mean}&{\bf 12.9$\pm$0.3}&{\bf 13$\pm$6}&{\bf 22$\pm$4}&{\bf 17.1$\pm$0.1}& & &{\bf 0.11$\pm$0.03}& &{\bf 239$\pm$6}
&{\bf 133$\pm$7} \\
\hline
\multicolumn{3}{l}{ {\bf Proplyd-like}} &  \\  
1       &  10.9  & 11.0 & 13.5 & 17.1 &      &      & 0.12 & 0.12 $\times$ 0.08 & 202 & 160 \\  
2       &  10.5  &  5.3 & 20.4 & 17.8 &      &      & 0.08 & 0.31 $\times$ 0.11 & 194 & 211 \\  
3       &  16.2  & 48.8 & 11.7 & 17.8 &      &      & 0.22 & 0.55 $\times$ 0.29 & 300 & 394 \\  
4       &  13.1  & 24.2 & 12.3 & 17.6 &      &      & 0.17 & 0.49 $\times$ 0.16 & 243 & 315 \\  
5       &  17.4  & 69.9 & 10.9 & 17.2 &      &      & 0.26 & 0.36 $\times$ 0.20 & 322 & 340 \\ 
5a      &  13.8  & 37.0 & 10.6 & 17.2 &      &      & 0.21 &                    & 256 & 192 \\
5b      &  12.7  & 23.5 & 12.0 & 17.1 &      &      & 0.17 &                    & 236 & 157 \\
5c      &  12.7  & 17.0 & 14.1 & 17.0 &      &      & 0.15 &                    & 235 & 171 \\
5d      &  12.3  & 12.4 & 16.0 & 17.2 &      &      & 0.12 &                    & 229 & 179 \\
6       &  21.1  & 84.9 & 13.2 & 16.9 &      &      & 0.26 & 0.37 $\times$ 0.13 & 392 & 296 \\  
7       &  12.8  & 42.8 &  8.8 & 18.3 &      &      & 0.24 & 0.37 $\times$ 0.11 & 237 & 349 \\  
8       &  13.7  & 50.5 &  9.0 & 17.2 &      &      & 0.25 & 0.21 $\times$ 0.07 & 254 & 314 \\  
9       &  10.5  &  3.5 & 26.8 & 17.4 &      &      & 0.06 & 0.09 $\times$ 0.06 & 195 & 138 \\  
10      &  14.6  & 26.9 & 13.7 & 17.7 &      &      & 0.17 & 0.40 $\times$ 0.13 & 270 & 255 \\ 
11      &  26.0  & 34.9 & 28.9 & 17.7 &      &      & 0.15 &                    & 483 & 394 \\
12      &  10.2  &  8.5 & 14.7 & 17.3 &      &      & 0.11 &                    & 189 & 128 \\
13      &  11.8  & 11.9 & 15.4 & 17.4 &      &      & 0.12 &                    & 219 & 144 \\
{\bf mean}&{\bf 14.1$\pm$1.0}&{\bf 31$\pm$6}&{\bf 15$\pm$1}&{\bf 17.4$\pm$0.09}& & &{\bf 0.17$\pm$0.02}& & {\bf 262$\pm$19}
&{\bf 243$\pm$23} \\
\hline
\multicolumn{3}{l}{ {\bf Condensations}} &  \\  
1       &  32.6  & 21.9 & 53.5 & 14.4 &      &      & 0.10 & & 605 & 110 \\
2       &  26.3  &  8.8 & 67.0 & 15.1 &      &      & 0.06 & & 488 & 123 \\
3       &  29.8  & 49.8 & 29.4 & 15.3 &      &      & 0.16 & & 552 & 122 \\
4       &  32.4  & 21.7 & 53.1 & 15.5 &      &      & 0.10 & & 601 & 168 \\
5       &  42.4  & 21.3 & 82.7 & 15.1 &      &      & 0.08 & & 787 & 156 \\
6       &  59.3  & 69.5 & 70.9 & 15.1 &      &      & 0.14 & & 1100 & 272 \\
7       &  25.0  & 30.0 & 30.0 & 15.3 &      &      & 0.14 & & 465 & 131 \\
{\bf mean}&{\bf 35.4$\pm$4.5}&{\bf 35$\pm$5}&{\bf 55$\pm$8}&{\bf 15.1$\pm$0.1}& & &{\bf 0.11$\pm$0.01}&&{\bf 657$\pm$84}
&{\bf 155$\pm$21} \\
\hline
\end{tabular} 
   \caption{Physical properties of pillars, globules, EGGS, proplyd-like, and condensations.
     The last line of each section gives the average (mean)
     values for each class of sources in bold. } 
\tablefoot{
(1) Average column density derived within the 70 $\mu$m contour level. \\   
(2) Mass derived from column density map within the contours of the 70 $\mu$m data. \\ 
(3) Average density from the mass $M$, assuming a spherical shape with an equivalent radius $r$. \\   
(4) Average temperature (average across the area covered by the 70 $\mu$m contour). \\
(5) and (6) Minimum and maximum temperature. \\ 
(7) Equivalent radius ($r=\sqrt(area/\pi)$), deconvolved with the beam (6$''$ for 70 $\mu$m that corresponds to 
0.04 pc for a distance of 1.4 kpc). \\
(8) Length and width for pillars (this study) and proplyd-like (sizes from Wright et al. \cite{wright2012}). \\
(9) Surface density. \\
(10) Average UV-flux in units of Habing field. } 
\label{values} 
\end{table*} 

\section{Results} \label{results} 
\subsection{Description of maps} 

{\bf Regions~1-1, 1-2, and 1-3} (Figs.~4, 5, 6) contain five pillars
and two globules and some examples of EGGs (five in total gathered in
a `swarm', named {\tt e1} to {\tt e5}), best visible in the 70 $\mu$m
{\sl Herschel} map in the Appendix. They have the same orientation
toward north as globule {\tt g2} suggesting that they are being
influenced by Cyg OB2. The condensations ({\tt c1} to {\tt c3}) in
subregion 1-2 are also faint but are more spherical and show up mainly
at 500 $\mu$m.

\begin{figure*}[ht]    \label{2-1}  
\begin{center}  
\includegraphics[angle=0,width=8cm]{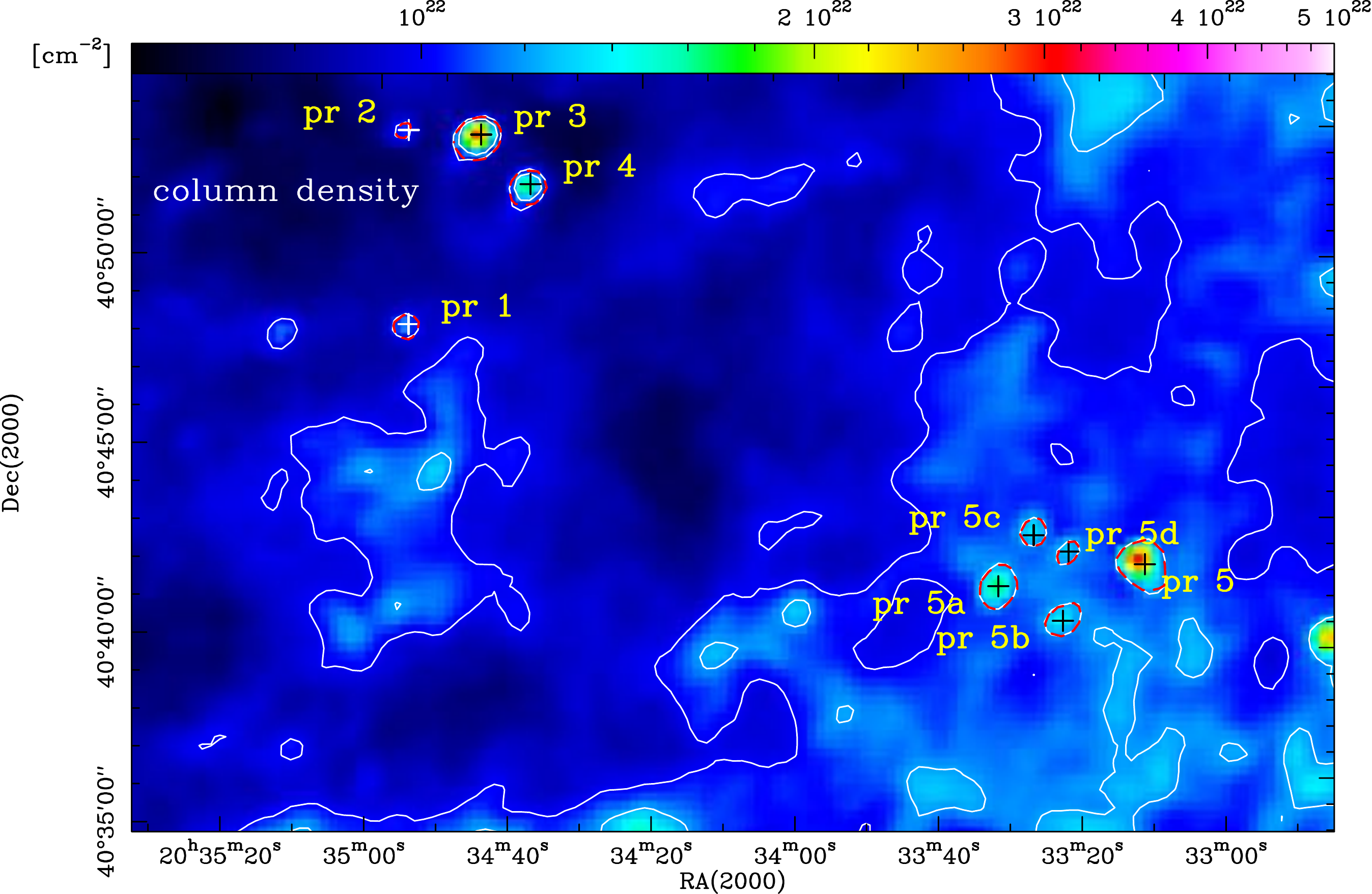} 
\includegraphics[angle=0,width=8cm]{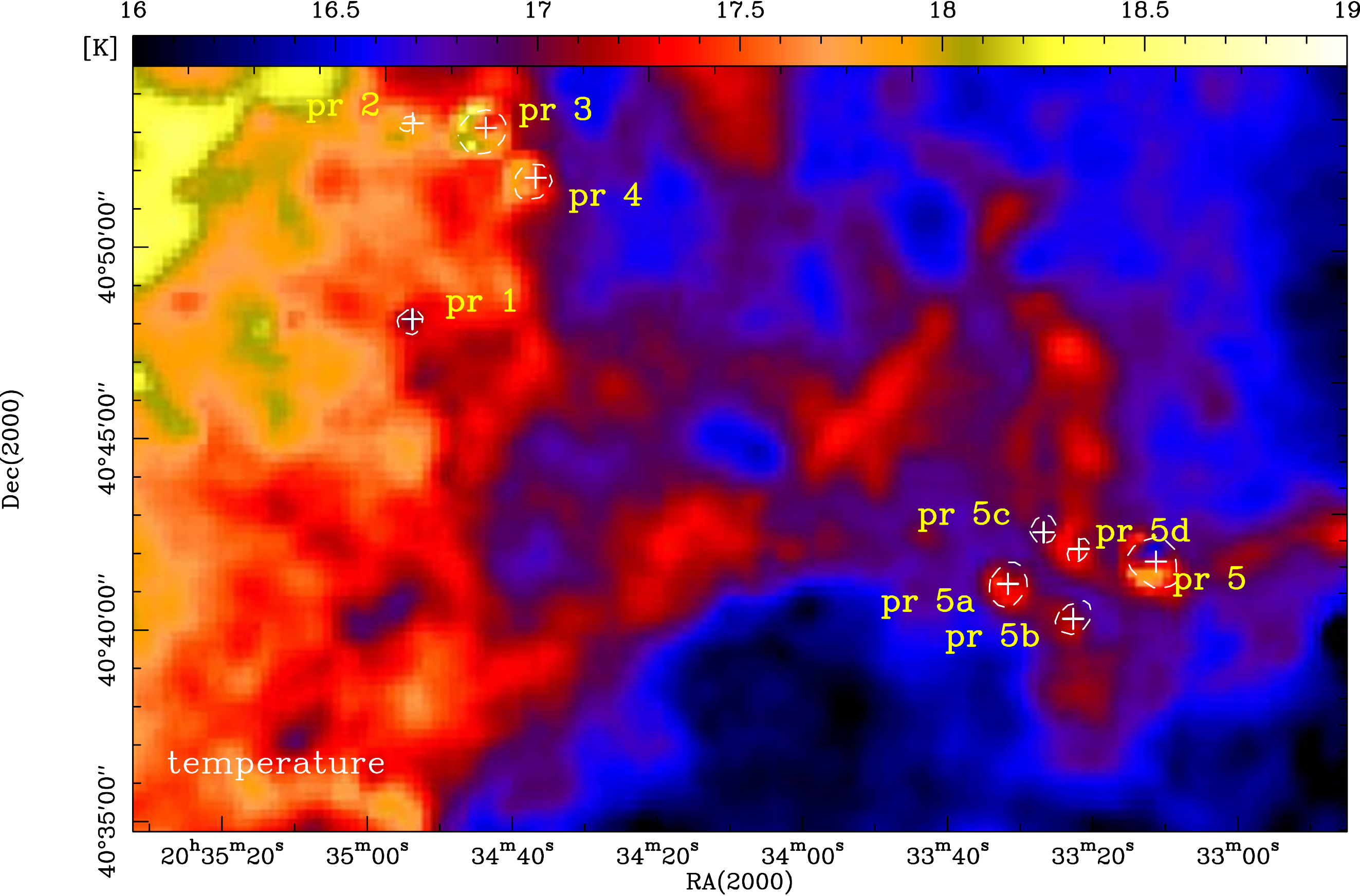} 
\end{center}  
\caption [] {Region 2-1: as in Fig.~4. The column density contours 
correspond to the levels 1.0 and 1.2 10$^{22}$ cm$^{-2}$.  }   
\end{figure*} 

\begin{figure*}[ht]     \label{2-2} 
\begin{center}  
\includegraphics[angle=0,width=8cm]{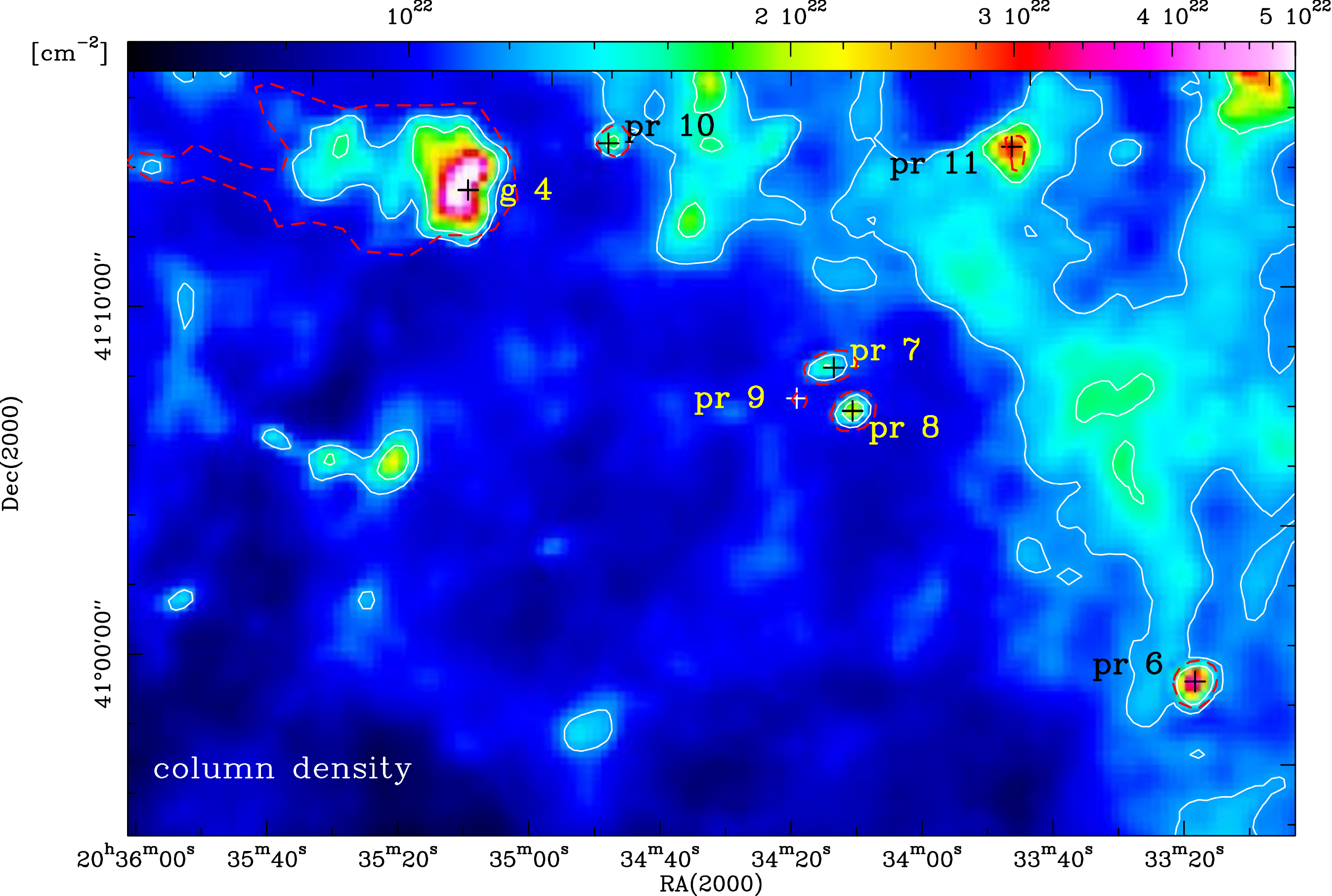} 
\includegraphics[angle=0,width=8cm]{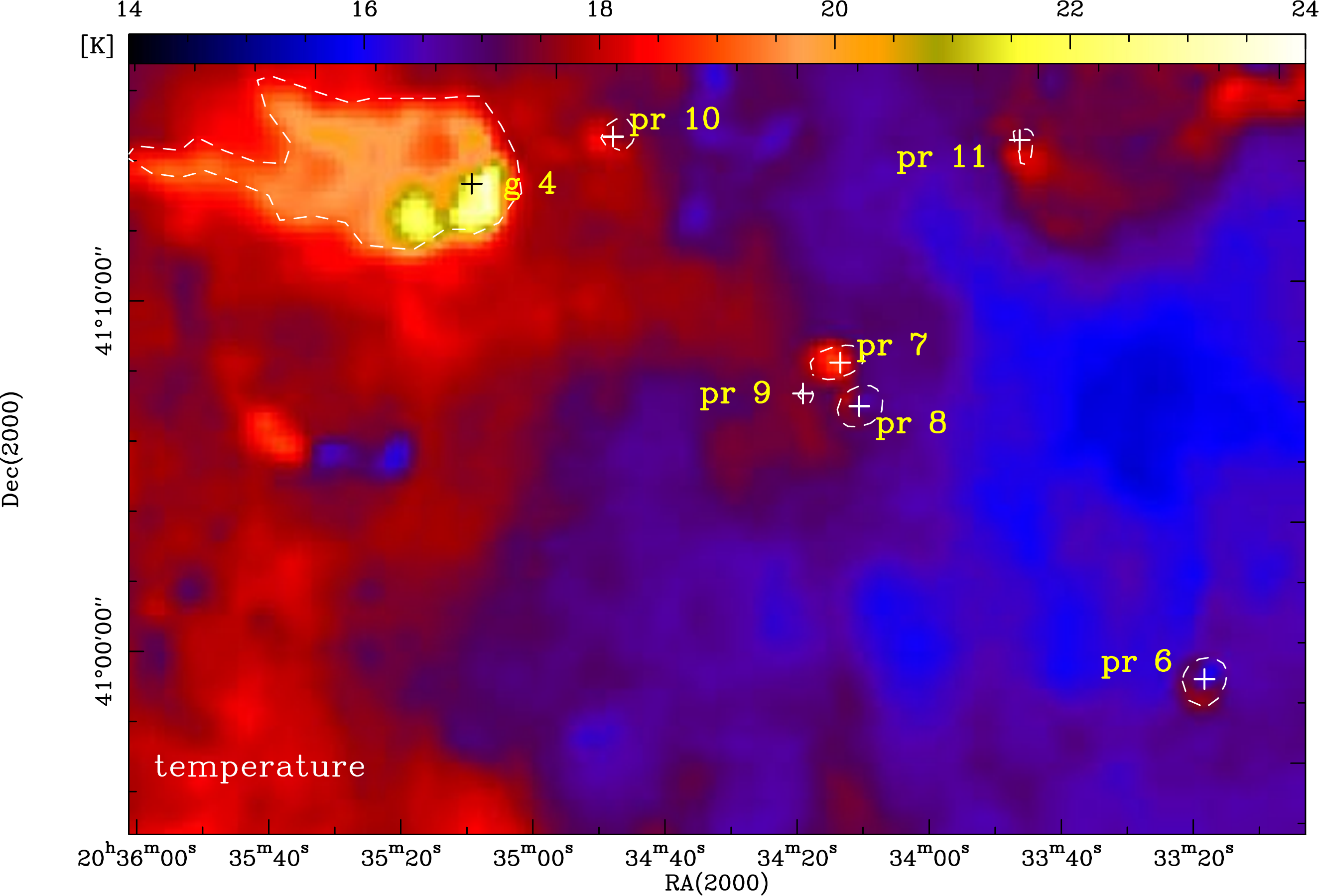} 
\end{center}  
\caption [] {Region 2-2: as in Fig.~4. The column density contours
  correspond to the levels 1.2 and 1.6 10$^{22}$ cm$^{-2}$.}
\end{figure*} 


\noindent {\bf Regions 2-1 and 2-2} (Figs.~7, 8) contain the ten Cyg
OB2 proplyd-like objects reported by Wright et al.  (\cite{wright2012})
that were detected in H$\alpha$ and {\sl Spitzer} images.  Two sources
look spherical ({\tt pr1, pr8}) in the {\sl Herschel} images, even in
the highest angular resolution 70 $\mu$m map, and can thus be partly
unresolved, while the others have more complex head/tail shapes with
bright ionization fronts that are oriented towards the center of Cyg
OB2. However, all proplyd-like objects show an elongated structure
pointing toward the center of Cyg OB2 in high resolution optical and
near-IR images (Wright et al. \cite{wright2012}).  Close to proplyd
\#5 (Wright et al. 2012), we detected some additional faint features
that we name {\tt pr5a-d}. The size scale of all these objects is
typically around 0.1 pc, in agreement with what Wright et al. found.
These authors are not certain about the nature of these features,
suggesting that they are photo-evaporating protostars and naming them
proplyd-like objects. Guarcello et al. (\cite{guarcello2013}) classified
some of them as embedded stars with disks, and a recent spectroscopic
study (Guarcello et al. \cite{guarcello2014}) showed that at least \#5
and \#7 are actively accreting protostars and that the disk of \#7 is
photoevaporating.  Globule {\tt g4} coincides with the \hii\, region
DR18, which hosts a strong IR-source (IRAS20343+4129,
MSX6G080.3624+00.4213) at near-, mid-, and far-IR wavelengths.
Comer\'on \& Torra (\cite{comeron1999}) observed DR18 at IR and
optical wavelengths and found an arc-shaped nebula, externally
illuminated by a nearby B0.5V star. The globular shape is most likely
caused by the object's proximity to the Cyg OB2 cluster (see Fig.~2),
though this source is not a typical example of a globule (like {\tt
  g1}) because it is significantly more massive and extended.

\begin{figure*}[ht]     \label{3-1}   
\begin{center}  
\includegraphics[angle=0,width=8cm]{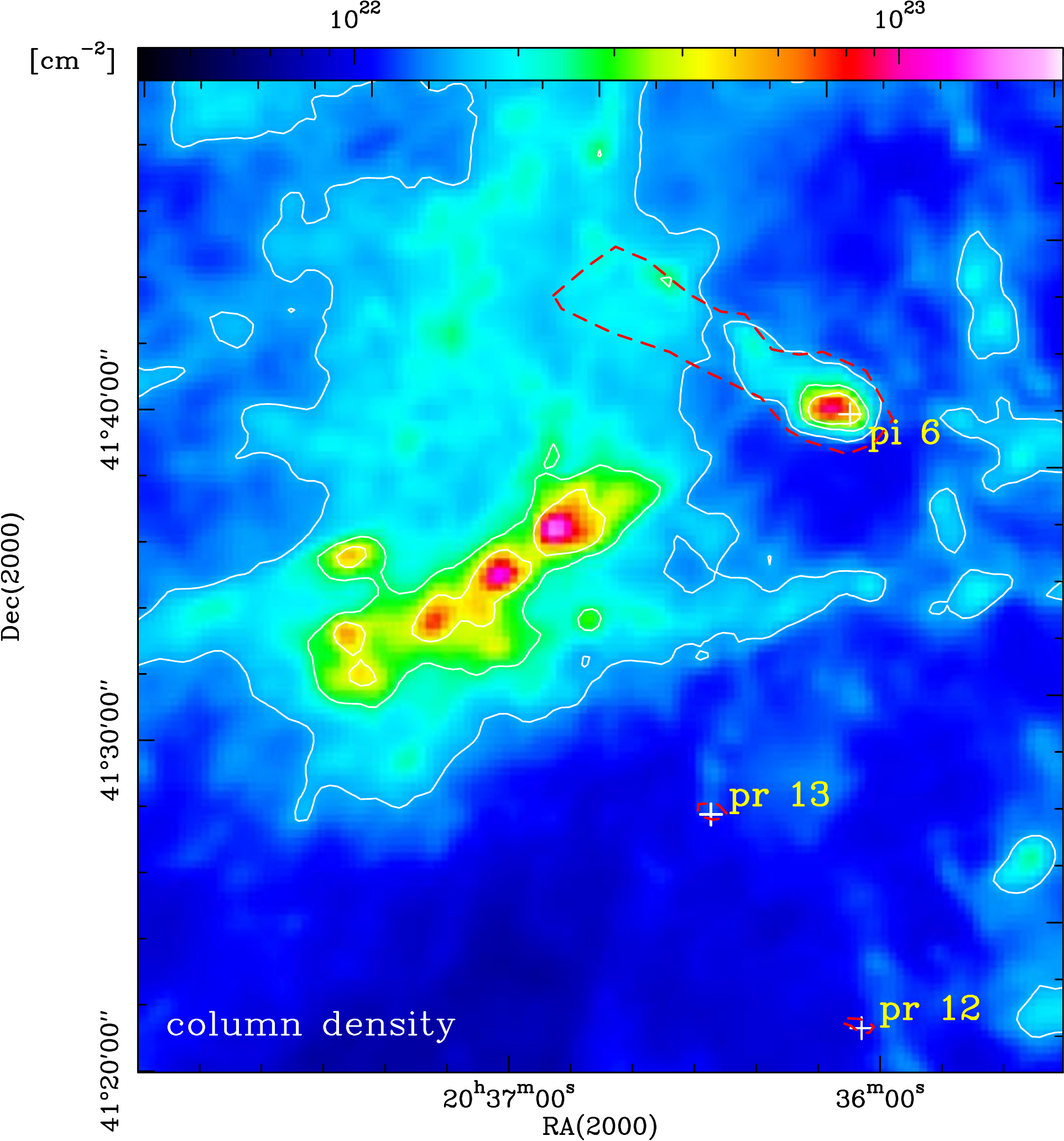} 
\includegraphics[angle=0,width=8cm]{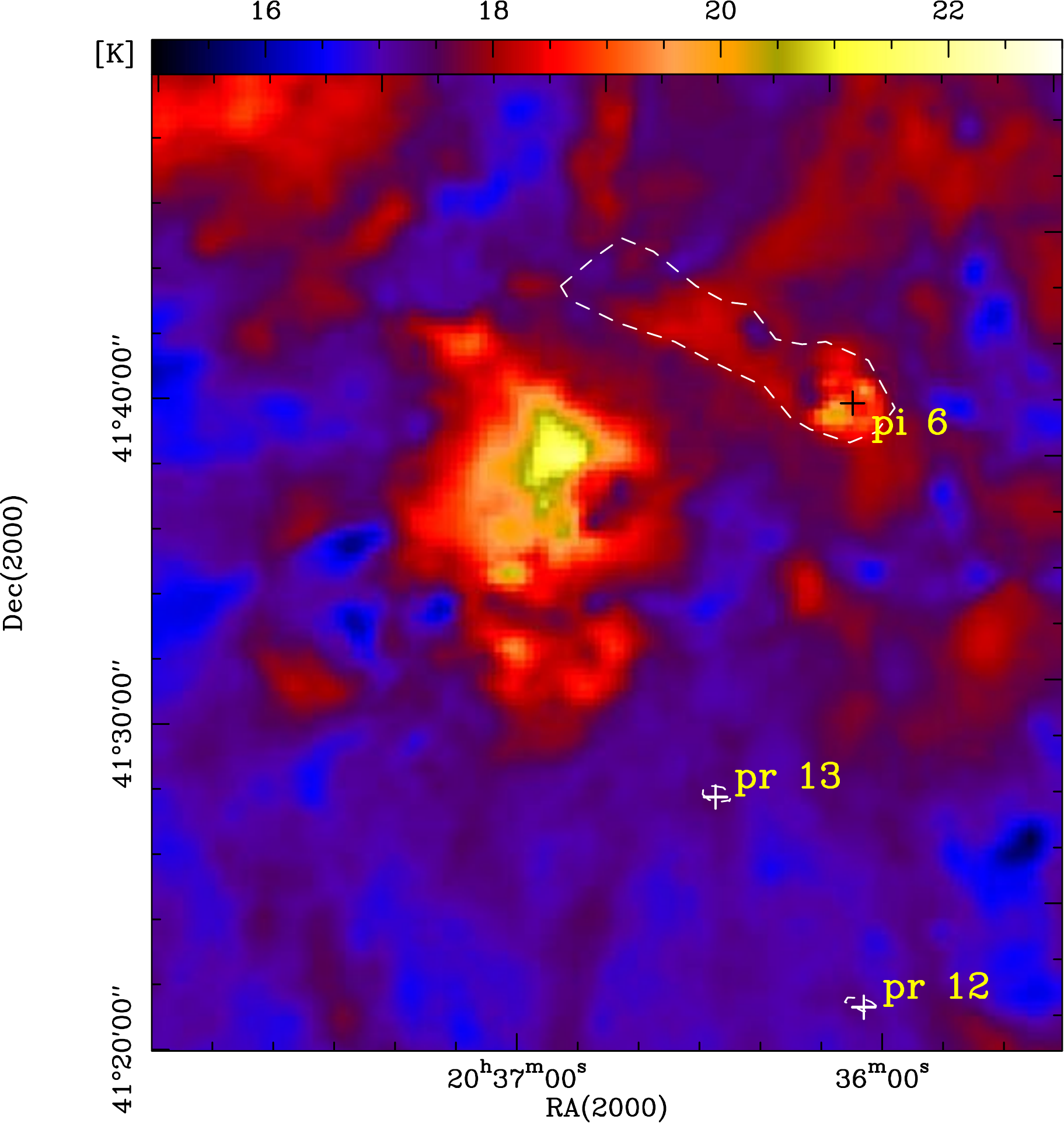} 
\end{center}  
\caption [] {Region 3-1: as in Fig.~4. The column density contours 
correspond to the levels 1.5 and 4.0 10$^{22}$ cm$^{-2}$.}   
\end{figure*}

\noindent {\bf Region 3-1} (Fig.~9) corresponds to the star-forming
region DR20 (e.g., Odenwald et al. \cite{odenwald1990}) and includes
one of the largest and most prominent pillars ({\tt pi6}) in the Cyg
OB2 environment. The pillar clearly points towards the center of
OB2. \\
{\bf Region 3-2} (Fig.~10) contains another pillar ({\tt pi7})
and three cold, irregularly shaped gas clumps.  These clumps are
probably not influenced by the close-by Cyg OB2 cluster because they
have no preferred direction, in contrast to {\tt pi7} which points
clearly towards the cluster center. Although projection effects can
not be excluded, it is more likely that the clumps are massive
structures still embedded in a more extended molecular cloud, and
which are probably forming small clusters because we observe internal
heating (see temperature map).

\subsection{Physical properties} \label{prop}

Table~1 gives an overview of the column density $N$(H$_2$), mass $M$,
density $n$, temperature $T$, equivalent radius $r$, length $l$ and
width $w$ for the elongated features, surface density $\Sigma$, and
UV-flux F$_{fuv}$ (derived from the {\it Herschel} data), we
determined for each object from the {\sl Herschel} observations. The
length and width of the pillars were directly measured on the 70
$\mu$m map.  The column density maps were used to calculate average
column density, density, and mass within the areas defined by the 70
$\mu$m contour at 400 MJy/sr that we used to identify different
features (see Sec.~\ref{ident}).  Mass was determined following
Eq. (2), using the column density $N$(H$_2$) in cm$^{-2}$, distance
$d$ in parsec, and area $A$ in square degrees (see, e.g., Schneider et
al. \cite{schneider1998})
\begin{eqnarray}
M & = & 6.6 \,\times \, 10^{-24} \, N(H_2) \, d^2 \, A \,\,\, .
\end{eqnarray}
We estimate that the column density, mass, and density values are correct to
within a factor of $\sim$2. The uncertainties mainly arise from the
assumed dust opacity, and the possible variation in temperature along
the line-of-sight not accounted for in the SED fitting. We consider
the error on the distance to be low because the distance of Cyg OB2
has been very accurately determined through parallax (Rygl et
al. \cite{rygl2012}). Since all objects we discuss here are directly
affected by this cluster they must be at more or less the same
distance. The minimum, maximum and average temperatures given in the
Table were derived from the temperature maps.  Since proplyd-like objects,
EGGs, and condensations were barely resolved, minimum and
maximum values were omitted.  We did not include any irregular
features in the Table because they do not form an individual class of
objects, but are mostly \hii\, regions. The last line in each source
category lists the average (mean) value.  The main results extracted from the
table are: \\

\noindent{\bf Size and density} \\
\noindent EGGs, proplyd-like objects, and condensations are small
(typically 0.1 pc to 0.25 pc). Their average densities are
2$\times$10$^4$ cm$^{-3}$, 1.5$\times$10$^4$ cm$^{-3}$, and
5.5$\times$10$^4$ cm$^{-3}$, respectively. Comparing them to general
molecular cloud features, condensations correspond to the dense cores
typically found in submm-continuum imaging of high-mass star-forming
regions (e.g., Motte et al.  \cite{motte2007} for the Cygnus X
region), but they are significantly larger and more massive than the
dense cores found with {\sl Herschel} in the nearby clouds of the
Gould Belt (e.g., K\"onyves et al. \cite{vera2015}).  The proplyd-like
objects are much larger than the ones found in Orion A ($\sim$ a
factor of 10, O'Dell et al. 1993), and also larger than those detected
in the Carina nebula ($\sim$a factor of 2, Smith et
al. \cite{smith2003}). Pillars and globules show a density gradient
from high values $n>$10$^4$ cm$^{-3}$ for the head, and $n <$10$^4$
cm$^{-3}$ for the tail. The length of pillars varies between 0.6 pc
and 3 pc. \\

\noindent{\bf Temperature} \\
\noindent The temperature range of all objects is between 14 K and 26
K (average temperatures).  The lowest temperatures are found for
condensations ($\langle $T$\rangle \sim$ 15 K), and highest for
globules ($\langle$T$\rangle \sim$18 K). All other objects typically
have a temperature around 17 K.  Pillars and globules show a
temperature gradient from their warm UV-heated (external and possibly
internal) head to a colder tail. \\

\noindent{\bf Mass} \\
\noindent The highest mass objects (typically a few hundred M$_\odot$)
are found amongst pillars and globules, mainly due to their large
extents (e.g., equivalent radii larger than 0.5 pc) since their
average densities ($\sim$5000 cm$^{-3}$) are low compared to the other
features.  We note, however, that the mass range can be quite varied,
for example $\sim$50 M$_\odot$ for {\tt pi4} but $\sim$1700 M$_\odot$
for {\tt pi1}. Their physical properties thus correspond to those
typically derived for {\bf clumps} inside molecular clouds (e.g.,
Schneider et al. \cite{schneider2006}, Lo et al. \cite{lo2009}).
Condensations and proplyd-like objects have masses of typically
several tens of M$_\odot$ and EGGs have masses lower than 10
M$_\odot$. All objects are thus potential star-forming sites with
large enough gas reservoirs to be so. \\

\noindent{\bf Surface density} \\
\noindent Surface densities are highest for condensations and
globules, i.e., $\sim$660 M$_\odot$/pc$^2$ and $\sim$380
M$_\odot$/pc$^2$, respectively.  All other objects have $\langle
\Sigma \rangle$ around 240--280 M$_\odot$/pc$^2$.\\

\noindent{\bf UV-flux} \\
\noindent The UV-flux does not reflect an internal property of the
detected features, it depends on its location and if it is only
externally illuminated or contains an internal source.  Obviously,
high implied UV fluxes are found for all objects (mainly the
proplyd-like objects), close to the central Cyg OB2 cluster. Two
objects, globules {\tt g1} and {\tt g4}, presumably have internal
heating sources which explains their rather high implied average UV
fluxes (549 G$_\circ$ and 428 G$_\circ$), respectively.

\begin{table}[htbp]
\centering 
\begin{tabular}{lcccc|c} 
\hline 
\hline  
Source  & $d_{OB}$ & t$_{photo}$   & t$_{expos}$   & t$_{total}$   & t$_{ff}$ \\
        & [pc]     & [$10^{6}$ yr] & [$10^{6}$ yr] & [$10^{6}$ yr] & [$10^{6}$ yr] \\
        & (1)      & (2)           &  (3)          &  (4)          &    (5) \\
\hline
\multicolumn{5}{l}{ {\bf Pillars} \,\, $\langle n(H_2) \rangle$=5$\times$10$^3$ cm$^{-3}$, $\langle r \rangle$ = 0.61 pc  }  &  \\
1       & 43.4     & 8.38          & 0.92          & 9.30         & 0.14 \\  
2       & 46.2     & 3.77          & 0.64          & 4.41          & 0.15 \\
3       & 50.4     & 3.76          & 0.23          & 3.99          & 0.25 \\
4       & 44.8     & 2.30          & 0.78          & 3.08          & 0.22 \\
5       & 30.8     & 2.84          & 2.15          & 4.99          & 0.23 \\
6       & 19.6     & 3.63          & 3.24          & 6.87          & 0.15 \\
7       & 10.5     & 0.47          & 4.13          & 4.60          & 0.15 \\
\hline
\multicolumn{5}{l}{ {\bf Globules} \,\, $\langle n(H_2) \rangle$=12$\times$10$^3$ cm$^{-3}$, $\langle r \rangle$ = 0.60 pc  }  &  \\
1       & 26.6     & 2.35          & 2.56          & 4.91          & 0.16 \\
2       & 36.4     & 5.10          & 1.60          & 6.70          & 0.41 \\
3       & 42.0     & 6.19          & 1.05          & 7.24          & 0.34 \\ 
4       & 11.5     & 1.66          & 4.03          & 5.69          & 0.12 \\ 
\hline
\multicolumn{5}{l}{ {\bf Condensations} \,\, $\langle n(H_2) \rangle$=55$\times$10$^3$ cm$^{-3}$, $\langle r \rangle$ = 0.11 pc  }  &  \\
1       & 37.8     & 3.58          & 1.46          & 5.04          & 0.62 \\
2       & 37.8     & 2.54          & 1.46          & 4.00          & 0.69 \\
3       & 37.8     & 4.00          & 1.46          & 5.46          & 0.46 \\
4       & 32.2     & 3.02          & 2.01          & 5.03          & 0.62 \\
5       & 33.8     & 3.92          & 1.86          & 5.78          & 0.77 \\
6       & 39.7     & 7.71          & 1.28          & 8.99          & 0.71 \\
7       & 29.4     & 2.44          & 2.29          & 4.73          & 0.46\\
\hline
\multicolumn{5}{l}{ {\bf EGGs} \,\, $\langle n(H_2) \rangle$=22$\times$10$^3$ cm$^{-3}$, $\langle r \rangle$ = 0.11 pc  }  &  \\
1       & 37.8     & 1.48          & 1.47          & 2.95          & 0.37 \\
2       & 38.1     & 1.22          & 1.44          & 2.66          & 0.40 \\
3       & 38.6     & 2.04          & 1.39          & 3.43          & 0.27 \\
4       & 37.1     & 1.28          & 1.54          & 2.82          & 0.50 \\
5       & 37.8     & 1.32          & 1.47          & 2.79          & 0.42 \\
\hline
\multicolumn{5}{l}{ {\bf Proplyd-like} \,\, $\langle n(H_2) \rangle$=15$\times$10$^3$ cm$^{-3}$, $\langle r \rangle$ = 0.17 pc  }  &  \\
1       & 14.0     & 0.47          & 3.79          & 4.26          & 0.31 \\  
2       & 13.0     & 0.37          & 3.88          & 4.25          & 0.38 \\  
3       & 11.5     & 0.76          & 4.03          & 4.79          & 0.29 \\  
4       & 12.0     & 0.57          & 3.98          & 4.55          & 0.30 \\  
5       & 13.0     & 0.99          & 3.88          & 4.87          & 0.28 \\ 
5a      & 13.2     & 0.72          & 3.86          & 4.58          & 0.28 \\
5b      & 13.4     & 0.62          & 3.85          & 4.47          & 0.29 \\
5c      & 13.0     & 0.56          & 3.88          & 4.44          & 0.32 \\
5d      & 12.7     & 0.49          & 3.91          & 4.40          & 0.34 \\
6       &  5.9     & 0.55          & 4.58          & 5.13          & 0.31 \\  
7       &  6.4     & 0.34          & 4.52          & 4.86          & 0.25 \\  
8       &  6.1     & 0.36          & 4.56          & 4.92          & 0.25 \\  
9       &  6.9     & 0.18          & 4.48          & 4.66          & 0.44 \\  
10      &  9.3     & 0.49          & 4.24          & 4.73          & 0.31 \\ 
11      & 37.8     & 1.47          & 1.46          & 2.93          & 0.45 \\
12      & 37.8     & 0.61          & 1.46          & 2.07          & 0.32 \\
13      & 32.2     & 0.11          & 2.01          & 2.12          & 0.33 \\
\end{tabular} 
\caption{Photoevaporation-, total-, and free-fall lifetimes of observed objects. } 
\label{life} 
\tablefoot{
(1) Distance to the center of Cyg OB2. \\   
(2) Lifetime of object until complete destruction by photoevaporation. \\   
(3) Time the object has been exposed to UV radiation of the \hii\, region. \\   
(4) Total lifetime (t$_{photo}$ + t$_{expos}$). \\
(5) Free-fall time (Eq. 6).} 
\end{table}

\section{Analysis and discussion} \label{discuss} 

The original `radiative driven implosion' scenario (Bertoldi
\cite{bertoldi1989}, Lefloch \& Lazareff \cite{lefloch1994}, Miao et
al. \cite{miao2009}) for the formation of pillars and globules
involves UV-radiation illuminating a pre-existing clumpy molecular
cloud and photoevaporating the lower density gas, leaving only the
densest cores, which may collapse to form stars. Though recent
hydrodynamic simulations including radiation (Gritschneder et al.
\cite{gritschneder2009}, 2010; Tremblin et al. \cite{tremblin2012a},
\cite{tremblin2012b}) successfully model pillars and globules
emphasizing the importance of turbulence, we will focus here on
studying the lifetime of UV-illuminated features using the more
classical approach.

\subsection{Lifetimes} \label{times}

The model for photoevaporation of dense clumps we use here is based on that of 
Johnstone et al. \cite{johnstone1998}. It calculates the mass loss
rate $\dot{M}$ considering the external photon flux $\phi_{49}$,  the
number of ionizing photons from O-stars per second (in units of
10$^{49}$), that impinges on a clump with radius $r_{14}$ (in units of
10$^{14}$ cm) at the distance to the central source $d_{17}$ (in units
of 10$^{17}$ cm):
\begin{eqnarray}
\dot{M} & = & 9.5 \times 10^{-9} \, \phi_{49}^{1/2} \,  r_{14}^{3/2}  \, d_{17}^{-1} \,\, [M_{\odot}/yr]
.\end{eqnarray}
For the photon flux, we adopted as a lower limit the total flux from the
most luminous O-stars clearly identified as members of Cyg OB2 in
Hanson (\cite{hanson2003}) and Wright et al. (\cite{wright2015}),
and ignored possible extinction. The total photon
flux of 10$^{50.3}$ s$^{-1}$ was determined by their spectral type
using the published values given in Sternberg et al.
\cite{sternberg2003}. Assuming a constant density but a variable radius, Eq. (3)
can be solved analytically and we obtain for the time (t$_{photo}$) that
it takes to completely photoevaporate the object
\begin{eqnarray}
t_{photo} & = & 2.1 \times 10^{20} \, \phi_{49}^{-1/2} \, (2\pi\, m_H/3)^{1/2} \, n_3^{1/2}  \, d_{17} \, M^{1/2}\,\, [10^6 yr]
,\end{eqnarray}
with the number density $n_3$ in [10$^3$ cm$^{-3}$] and the mass $M$
[M$_{\odot}$] from Table~1, and the mass of hydrogen $m_H$ [g].  In
this simple approach, we neglected external compression which would
increase t$_{photo}$ because it reduces the size (i.e., radius) of the
object which in turn lowers the mass loss rate. We note also that the
values for the distances are only approximations because there is no
clearly defined center of the Cyg OB2 association, and there are
line-of-sight effects.  The time of exposure to UV
radiation for the object is estimated by t$_{expos}$ = d$_{front}$/$v_{exp}$ km/s
with the expansion velocity of $v_{exp}$ of the \hii\, region.
In this paper we use a value of 10 km s$^{-1}$ for our calculations which is
a typical value of expanding \hii\, regions (e.g., Williams et
al. \cite{williams2001}).

This value is approximately in accord with that determined by
comparing the relative velocity of the average radial stellar velocity
of Cyg OB2 of -28 km s$^{-1}$ (Simbad data base) and the molecular
cloud, that is --2 to -11 km s$^{-1}$ (Schneider et
al. \cite{schneider2006}). Another way to approximate $v_{exp}$ is to
use $v_{exp} \, = R_b/t_{OB2}$ with the extent of the bubble created
by the \hii\, region $(R_b$ ), and the age of the Cyg OB2 association
($t_{OB2}$ ). The average radius of the bubble is $\sim$40 pc, derived
from, for example, MSX-images (Schneider et al. \cite{schneider2006})
or the Canadian Galactic Plane Survey at 1.4 GHz (Reipurth \&
Schneider \cite{reipurth2008}).  The cluster age is under discussion,
with values ranging from 3--4 Myr (e.g., Comer\'on \& Pasquali
\cite{comeron2012}) and 5--7 Myr (e.g., Drew et al. \cite{drew2008},
Wright et al. \cite{wright2015}).  We thus determine an upper limit of
$\sim$17 km s$^{-1}$ for $v_{exp}$ and a lower limit of $\sim$7 km
s$^{-1}$.  Considering the spread in values from the different
methods, we take 10 km s$^{-1}$ as a conservative value and estimate
that the lifetime calculations are correct within a factor of 1.5--2.

The photoevaportation lifetimes can be compared to the free-fall time 
for isothermal, gravitational collapse, as 
\begin{eqnarray}
t_{ff} & = & \sqrt(3/(2 \, \pi \, G \, \rho),  
\end{eqnarray}
in which $G$ is the gravitational constant and $\rho = n \, m_H$ the average volume
density of the structure studied. 

The calculated lifetime values are listed in Table~\ref{life}.  The
most massive and extended objects -- the pillars and globules --
survive photoevaporation the longest, up to 8$\times$10$^6$ yr. There
is one exception, {\tt pi7}, which has t$_{photo}$ of only
5$\times$10$^5$ yr.  Condensations also have a relatively long
photoevaporation lifetime ($\sim$2 to 8$\times$10$^6$ yr).  EGGs have
shorter t$_{photo}$ of 1--2$\times$10$^6$ yr. For proplyd-like objects
the photoevaporation lifetime is in the order of only a few 10$^5$
yr. This timescale is consistent with what was estimated for the
circumstellar disks in Orion (Johnstone et al.  \cite{johnstone1998})
though it is not clear if all objects are indeed photoevaporating
disks.  The majority of the proplyd-like objects ($\sim$70\%) actually
already have stars within them, as evidenced by optical (Guarcello et
al. \cite{guarcello2012}) and near-IR (Wright et
al. \cite{wright2012}) point sources detected within them and the
spectroscopy of two of their central stars (Guarcello et
al. \cite{guarcello2014}), see also Sec.~\ref{results}.
Interestingly, Gahm et al. ({\cite{gahm2007}) found lifetimes of about
  4$\times$10$^6$ yr for the globulettes in the Rosette Nebula, and
  IC1805, which have sizes smaller than 0.05 pc and average densities
  of 3--100$\times$10$^3$ cm$^{-3}$.  These longer photo evaporation
  times are likely due to much weaker FUV-field in these regions (both
  have far fewer O stars than Cyg OB2).  Given our angular resolution,
  our census is not sensitive to these objects.

The ratio $t_{photo}/t_{expos}$ is an indicator of the evolutionary
state of the structure.  Pillars, globules, and condensations have an
average ratio of $\sim$5, $\sim$2, and $\sim$3, respectively,
suggesting that they are mostly in an early state of their
evolution. EGGs typically have a ratio of one, which may imply that they
are already half way through of being dissociated. Finally,
proplyd-like objects have very small ratios of $\sim$0.05 and could be
close to disappearing.

Taking objects' current densities (Table~1), we estimate free-fall
lifetimes\footnote{We note that $t_{ff}$ is a lower limit, the
  collapse time increases if the structure has support by turbulence,
  pressure, rotation, or magnetic fields.}  typically of a few 10$^5$
yr.  Since $t_{ff}$ is proportional to $1/\sqrt(n)$, the densest
objects - condensations - have the shortest collapse times, around
6$\times$10$^5$ yr. The condensations, but also pillars, EGGs, and
globules, are all potentially star-forming with $t_{ff} < t_{photo}$
and $t_{ff} \ll t_{total}$.  Only proplyd-like objects have comparable
lifetimes for photoevaporation and gravitational collapse. However,
most of them, at least the original 10 from Wright et
al. \cite{wright2012}, have central stars observed in the optical or
near-IR bands. We suspect that the large envelopes surrounding these
stars will be eroded and destroyed within a short period of time which
might influence how much more mass these stars can accrete, thus
determining their final stellar masses.

\subsection{An evolutionary sequence ?} \label{sequence} 

Many detached structures (globules, proplyds, etc.) are observed in
the Cygnus X region. The presence of such structures is also a
signature found by Tremblin et al. (\cite{tremblin2012b}), when the
initial turbulence in the molecular cloud dominates over the
compression of the ionized gas. Hence, the turbulent pressure is
likely to be as important in Cygnus X as the ionized gas pressure.
Interestingly, Gritschneder et al. (\cite{gritschneder2009}) found
dense pillar heads in their simulations and less material at the
bases, which is a morphology seen for all pillars in Cygnus X. This
behavior is also observed in the controlled simulations of Tremblin et
al. (\cite{tremblin2012a}, Fig. 5) and indicates that the structures
are relatively old (see below), with pillars stretched until their
heads are completely separated from the rest of the cloud.

The controlled runs in Tremblin et al. (2012a) represent
a scenario of the interaction between an ionization front and a medium
of constant density with a modulated interface geometry or a flat
interface with density enhancements. They show that pillars with a
large width/height ratio (w/h) will take a very long time (t) to
grow. The narrowest pillar with an initial w/h= 0.5 (see Fig. 6 of
Tremblin et al. (2012a) reached a length of 1.5 pc with
w/h = 0.16 in 5$\times$10$^5$ yr. In Cygnus X, the ratio R = w/h
(Table 1) is on average 0.22, varying between 0.16 and 0.28, and the
length of the pillars is between 0.6 pc and 3 pc with an average of
1.35 pc. If these values are compared to the ones shown in Tremblin et
al. (2012a), see also Kinnear et al. (\cite{kinnear2014}, 
\cite{kinnear2015}), the pillars observed in Cygnus X are likely to be
quite well evolved. Only the curve with the smallest w/h ratio with
large pillar sizes ($>$1 pc) is comparable to our observed values and
hence indicates an evolutionary state of t$>$5$\times$10$^5$ yr. This
lower limit of t is consistent with the exposure time t$_{expos}$ we
calculated in Sec.~7.1 which ranges between 2$\times$10$^5$ yr and
4$\times$10$^6$ yr.

Putting together the physical properties determined from the {\sl
  Herschel} observations (Sec.~\ref{prop}), the lifetime calculations
(Sec.~\ref{times}), and the results from comparison with simulations
(see above), we propose a tentative evolutionary scheme for the
observed features.

Pillars and globules are the most massive objects we see and they
correspond in their physical properties (average size, density, and
mass of $\sim$1 pc, 5--10 10$^{3}$ cm$^{-3}$, and $\sim$500 M$_\odot$,
respectively) to typical molecular cloud {\bf clumps}. Condensations,
and to a lesser extent EGGs, are likely molecular cloud {\bf dense
  cores} with typical size and density of 0.1 pc and 2--6 10$^4$
cm$^{-3}$, respectively. In this scheme, proplyd-like objects
correspond to less dense (10$^3$ cm$^{-3}$) {\bf cores}.  The main
difference between these objects is that clumps and cores form as a
part of a molecular cloud and are embedded in the clouds (e.g., cores
within a filament), whereas globules, EGGs, condensations, and
proplyd-like objects are isolated features. The resemblence of their
physical properties, however, may suggest a filamentary origin as a
possible scenario (see Introduction).

The remaining molecular envelopes of proplyd-like objects are less
massive than those in the EGGs, probably because some of the mass has
gone into the forming stars.  Pillars form a special category because
they are still attached to the cloud. This fact points towards a
scenario in which pillars can be the eroded leftovers of a
pre-existing clump or filament (Dale et al. \cite{dale2015}). Tremblin
et al. (2012a,b) show that pillars can form either in a density- or
surface-modulated region.

We find that pillars have the longest timescales for photoevaporation,
mainly because they are massive and large, and have the smallest times
of UV-exposure.  We speculate that a pillar evolves into a globule and
then on into a condensation. The major difference between globule and
condensation is that the latter has no head-tail structure and is a
factor of four denser and smaller. Larger globules might form multiple
stars (see the example of globule {\tt g1} where several B-stars were
found), possibly leading to multiple proplyd-like objects grouped
together. This would also explain why the proplyd-like objects are
grouped together a few at a time in little clumps.  A condensation may
be an evolved globule in which most of the lower density gas has
photoevaporated away, leaving only a dense (5.5$\times$10$^4$
cm$^{-3}$) and cold (T$\sim$15 K) core. We further speculate that EGGs
and proplyd-like objects could be also leftovers of initially larger
globules.  As their name already indicates, EGGs, that is evaporating
gaseous globules, have globular shapes. They are, however, less dense
(2$\times$10$^4$ cm$^{-3}$) and warm (T$\sim$17 K) and thus prone to
disappear fast, in contrast to condensations that are potential sites
of star/cluster formation. Proplyd-like objects may share the same
fate as EGGs but they are on average more massive and extended and so
likely to survive the photoevaporating impact of Cyg OB2 for longer.
Given that some of them show signatures of star-formation activity,
they could be the leftovers of condensations. If proplyd-like objects
follow EGGs and if the latter have time to form stars, then it is not
unreasonable to think that the former may have formed stars but still
be surrounded by a portion of an envelope that is still finishing
evaporating.

\section{Summary} \label{summary}

We used {\sl Herschel} FIR imaging observations of the Cyg OB2 region,
performed within the HOBYS keyprogram, to detect and characterize
features that are formed in the interface region between \hii\, region
and molecular cloud. Using a 400 MJy/sr flux threshold in the 70
$\mu$m map, we define pillars, globules, evaporating gaseous globules
(EGGs), proplyd-like objects, and condensations.  From SED fits to
the 160--500 $\mu$m {\sl Herschel} wavelengths, we determine column
density and temperature maps, and derive masses, volume densities, and
surface densities for these structures.  From the 70 $\mu$m and 160
$\mu$m flux maps, we estimate an average FUV-field of typically a few
hundred Habing on the photon dominated surfaces.  We find that
the initial morphological classification indeed corresponds to
distinct objects with different physical properties.

\noindent $\bullet$ Pillars are the largest structures (equivalent
mean average radius $\langle r \rangle\sim$0.6 pc). They have an average density
of $\langle n \rangle \sim$5$\times$10$^3$ cm$^{-3}$, and an average
temperature $\langle T \rangle$ of 18 K. They often show temperature
gradients along their longer axis. Their masses range between 50 M$_\odot$ and
1680 M$_\odot$, with an average mass $\langle M \rangle\sim$500
M$_\odot$. \\
\noindent $\bullet$  
Globules are also large ($\langle r \rangle \sim$0.6 pc) but they have a more defined
head-tail structure with a denser `head' 
than pillars with $\langle n \rangle \sim$1.2$\times$10$^4$
cm$^{-3}$. Their average mass and temperature are $\sim$500 M$_\odot$
and $\sim$18 K, respectively, similar to pillars. \\ 
\noindent $\bullet$  
EGGs and proplyd-like objects are smaller ($r\sim$0.1 and 0.2 pc) and less massive
($\sim$10 M$_\odot$ and $\sim$30 M$_\odot$, respectively), but they have a high average density of
2.2$\times$10$^4$ cm$^{-3}$ and 1.5$\times$10$^4$ cm$^{-3}$, respectively. They both have an
average temperature of 17 K. 
\\
\noindent $\bullet$ Condensations are small ($\sim$0.1 pc), have an
average mass of 35 M$_\odot$, and are the densest structures we found
in our sample with $\langle n \rangle\sim$5.5$\times$10$^4$
cm$^{-3}$.

In summary, pillars and globules are irradiated structures which
correspond to a subset of what is described as `clumps' in molecular
clouds while irradiated condensations correspond to a subset of
massive dense molecular cloud `cores'.  All pillars, globules and
proplyd-like objects show a clear orientation toward the center of the
Cyg OB2 association. They are thus directly influenced by the
radiation of the stars and we used a census of them to estimate the
lifetimes of all observed features using a model for photoevaporating
dense clumps (Johnstone et al. \cite{johnstone1998}).  Pillars and
globules have the longest estimated photoevaporation lifetimes, a few
million years, while all other features most likely survive less than
a million years. These lifetimes are consistent with what was found in
simulations of turbulent, UV-illuminated clouds (Tremblin et
al. 2012a,b). We propose a
tentative evolutionary scheme in which pillars can evolve into
globules, which in turn then evolve into EGGs, condensations and
proplyd-like objects.
  
\begin{acknowledgements} 
SPIRE has been developed by a consortium of institutes led by Cardiff
University (UK) and including Univ. Lethbridge (Canada); NAOC (China);
CEA, LAM (France); IFSI, Univ. Padua (Italy); IAC (Spain); Stockholm
Observatory (Sweden); Imperial College London, RAL, UCL-MSSL, UKATC,
Univ. Sussex (UK); and Caltech, JPL, NHSC, Univ. Colorado (USA). This
development has been supported by national funding agencies: CSA
(Canada); NAOC (China); CEA, CNES, CNRS (France); ASI (Italy); MCINN
(Spain); SNSB (Sweden); STFC (UK); and NASA (USA).  PACS has been
developed by a consortium of institutes led by MPE (Germany) and
including UVIE (Austria); KU Leuven, CSL, IMEC (Belgium); CEA, LAM
(France); MPIA (Germany); INAF-IFSI/OAA/OAP/OAT, LENS, SISSA (Italy);
IAC (Spain). This development has been supported by the funding
agencies BMVIT (Austria), ESA-PRODEX (Belgium), CEA/CNES (France), DLR
(Germany), ASI/INAF (Italy), and CICYT/MCYT (Spain). Part of this work
was supported by the ANR-11-BS56-010 project ``STARFICH'' and the ERC
advanced Grant no. 291294 ``ORISTARS''.  N.S. acknowledges support
from the DFG-priority program 1573 (ISM-SPP), through project number
Os 177/2-1 and 177/2-2. NJW acknowledges an RAS Research Fellowship.
We thank the referee, G. Gahm for his competent comments that improved
the clarity of the paper.
\end{acknowledgements}

 
\begin{appendix} 
 
\section{Herschel images at 70 to 500 $\mu$m of all regions}

\begin{figure*}[ht]     
\begin{center}  
\includegraphics[angle=0,width=7.5cm]{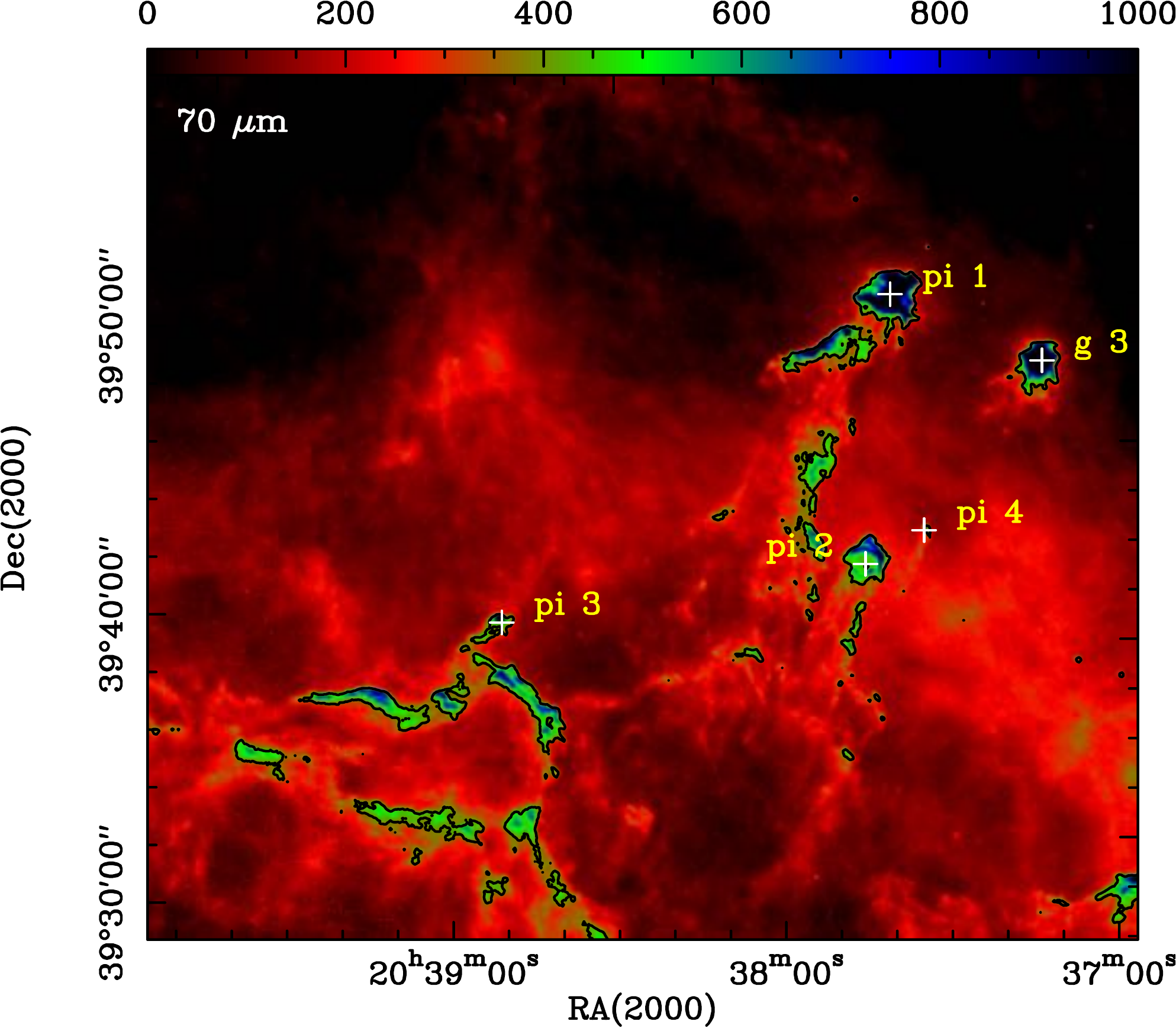} 
\includegraphics[angle=0,width=7.5cm]{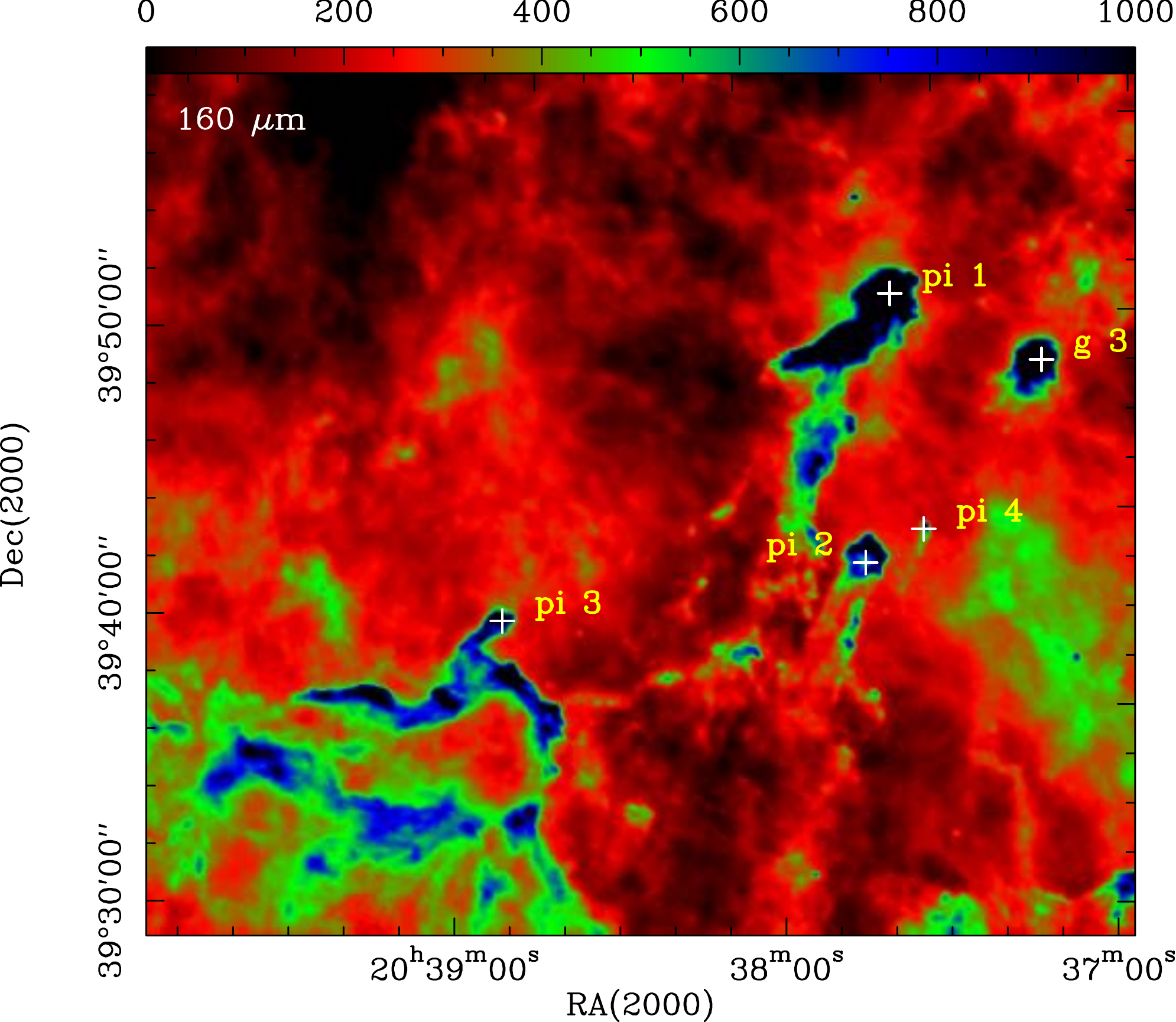} 
\includegraphics[angle=0,width=7.5cm]{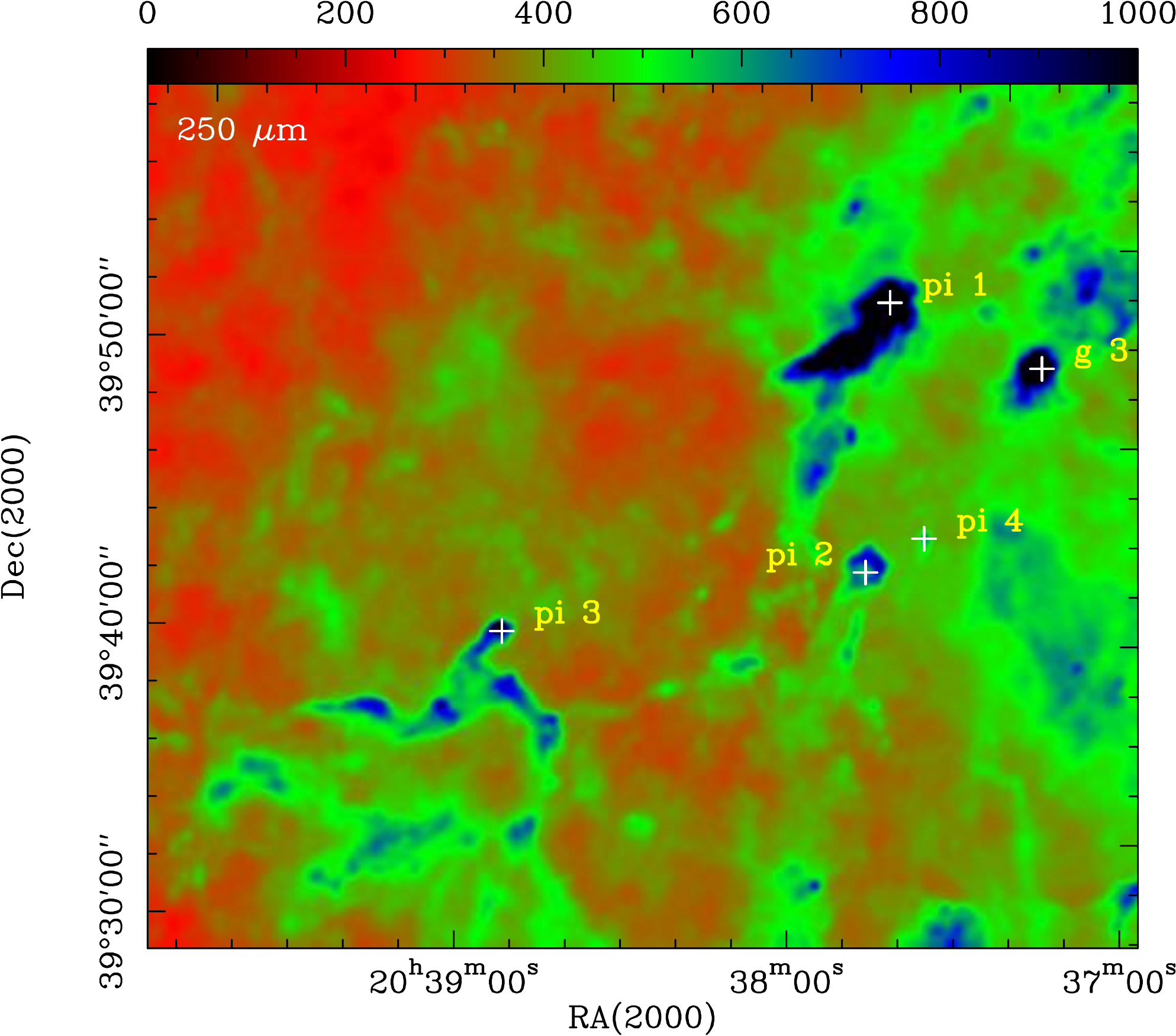} 
\includegraphics[angle=0,width=7.5cm]{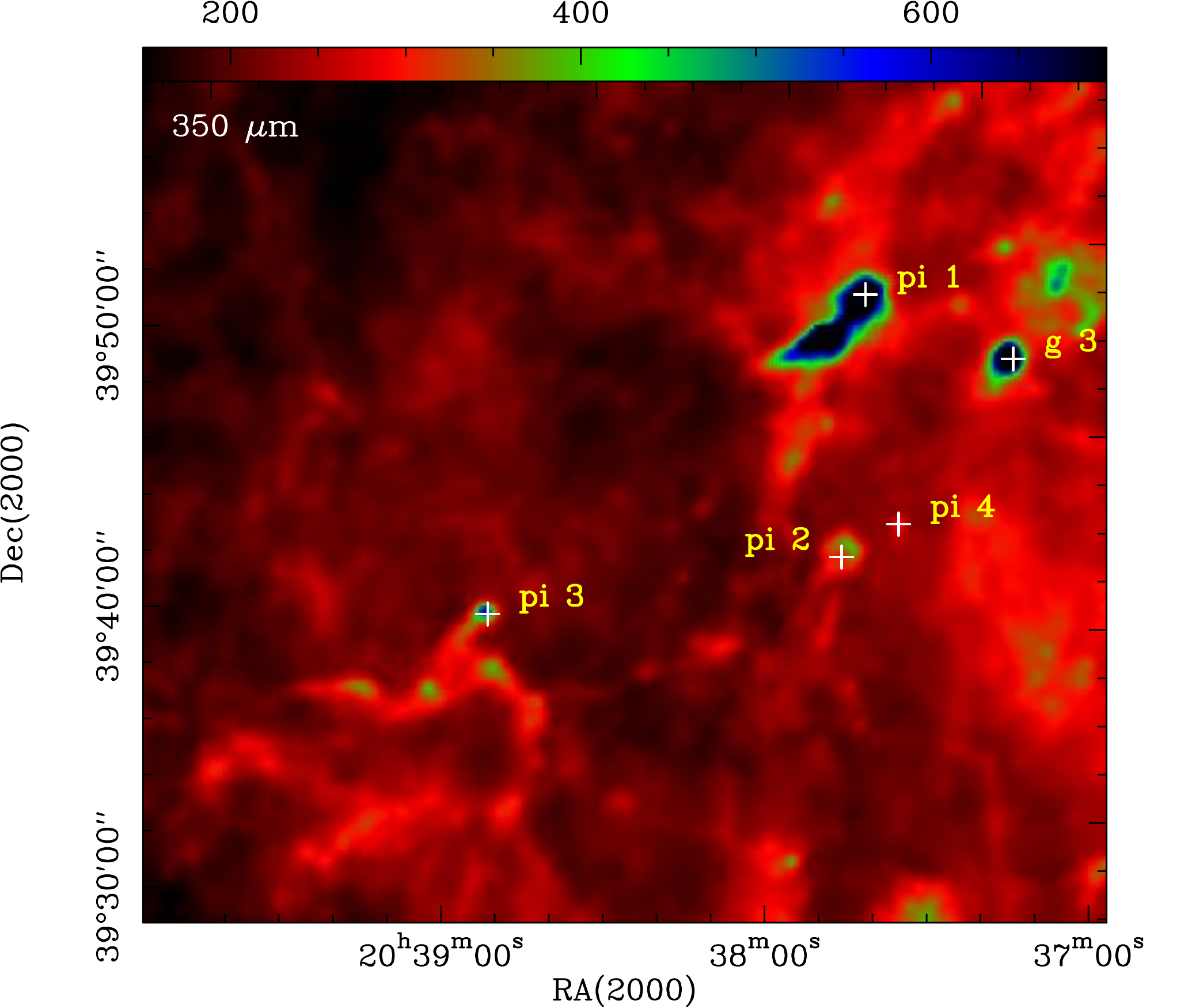} 
\includegraphics[angle=0,width=7.5cm]{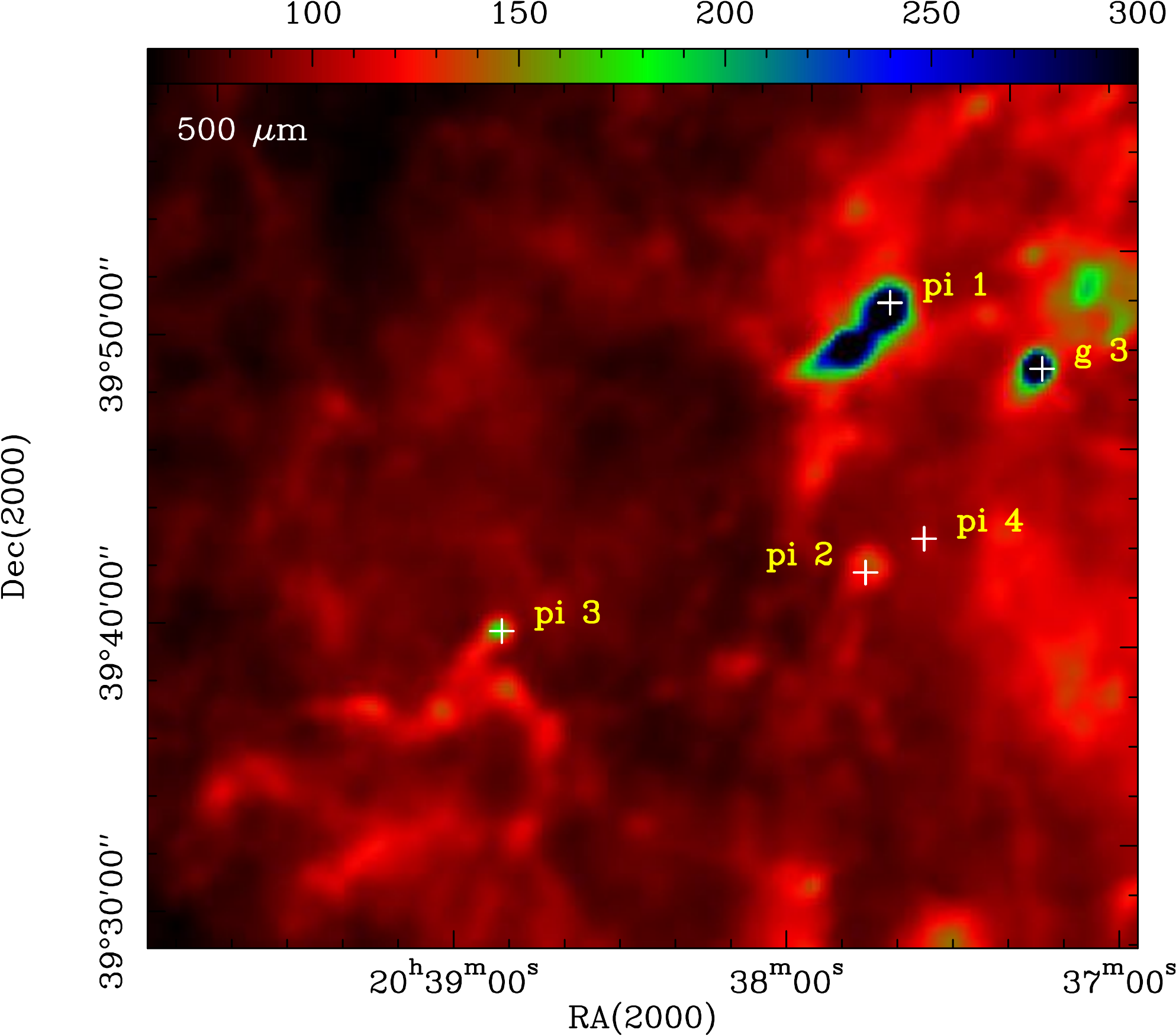} 
\end{center}  
\caption [] {PACS (70 and 160 $\mu$m) and SPIRE (250, 350, 500 $\mu$m)
  images of region 1-1. All maps are in units of MJy/sr and the
  features classified as pillars (`pi'), globules (`g'), condensations
  (`c') and EGGS (`e') are indicated by yellow crosses in the plots.
  The black contour in the 70 $\mu$m map outlines the 400 MJy/sr level
  that was used to approximately define the shapes of the various
  features.}
\end{figure*} 

\begin{figure*}[ht]     
\begin{center}  
\includegraphics[angle=0,width=7.5cm]{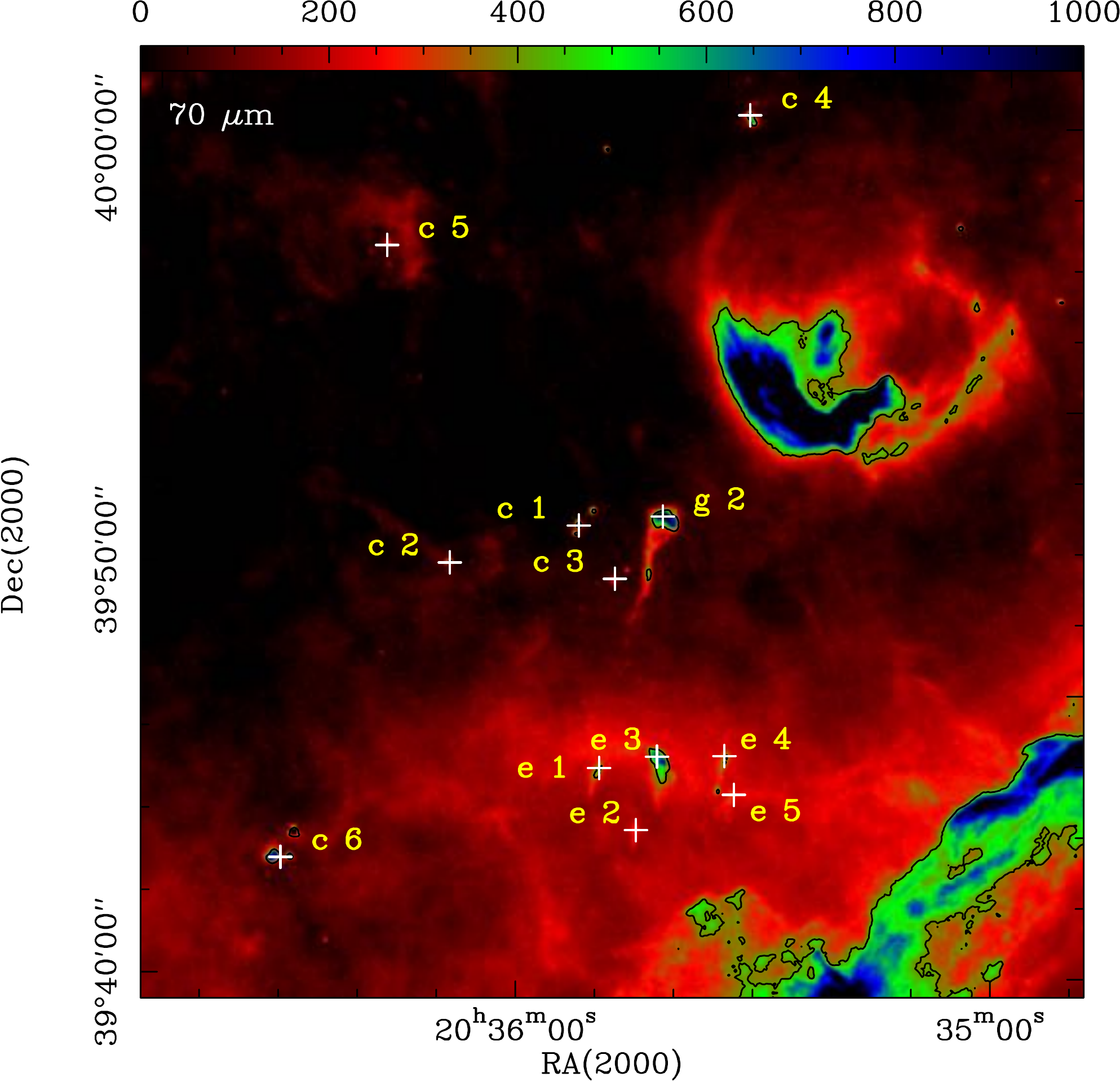} 
\includegraphics[angle=0,width=7.5cm]{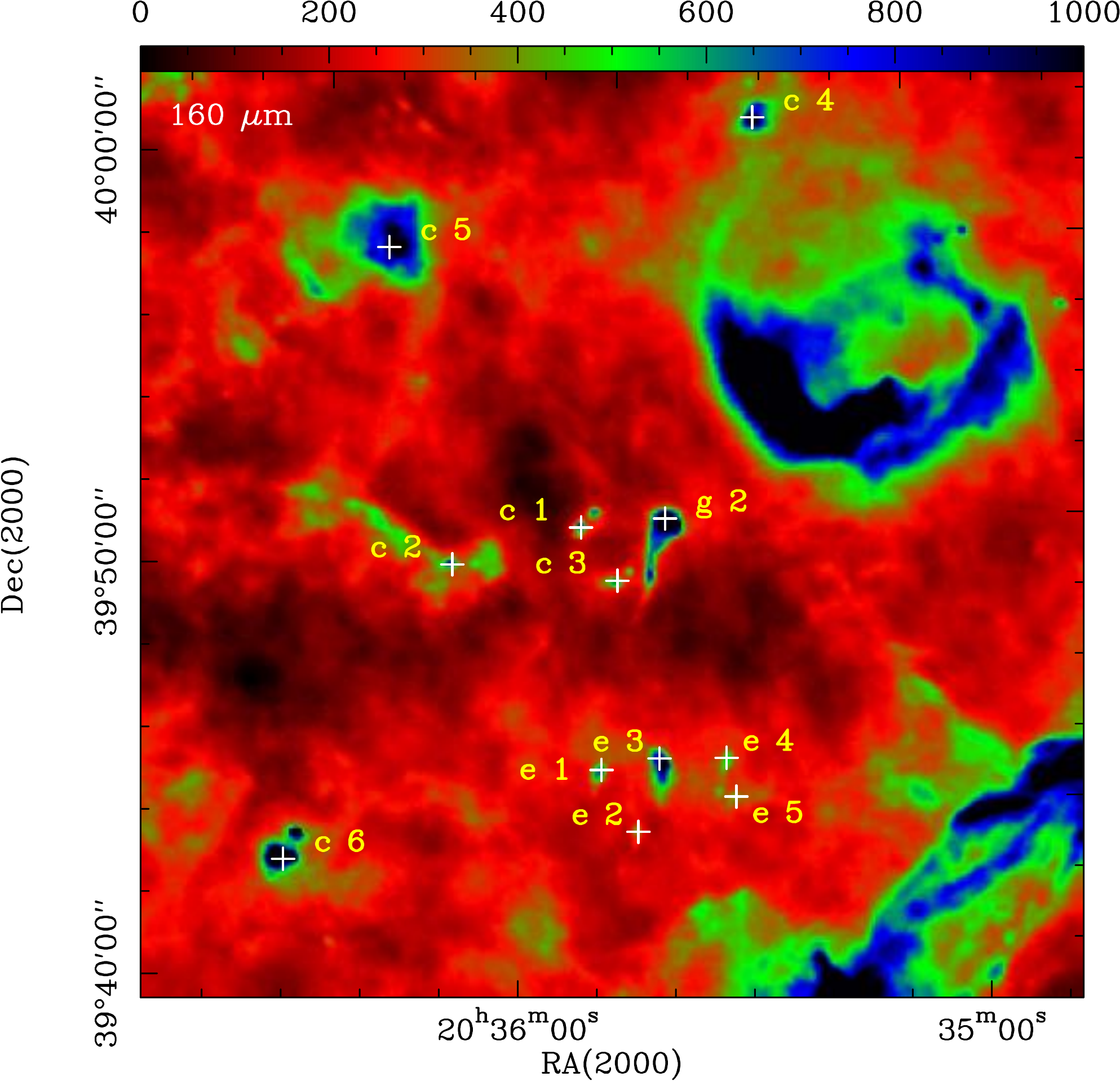} 
\includegraphics[angle=0,width=7.5cm]{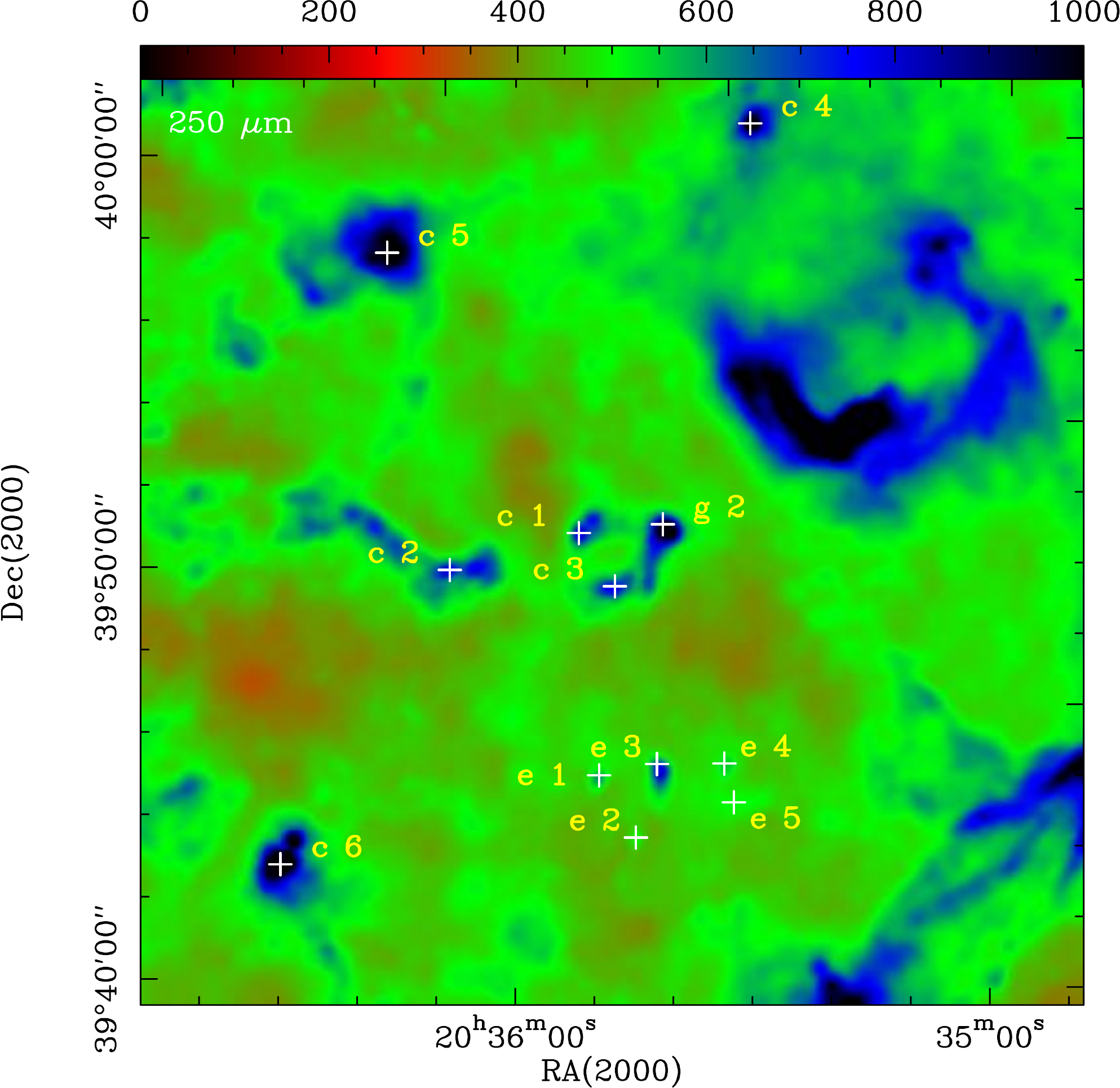} 
\includegraphics[angle=0,width=7.5cm]{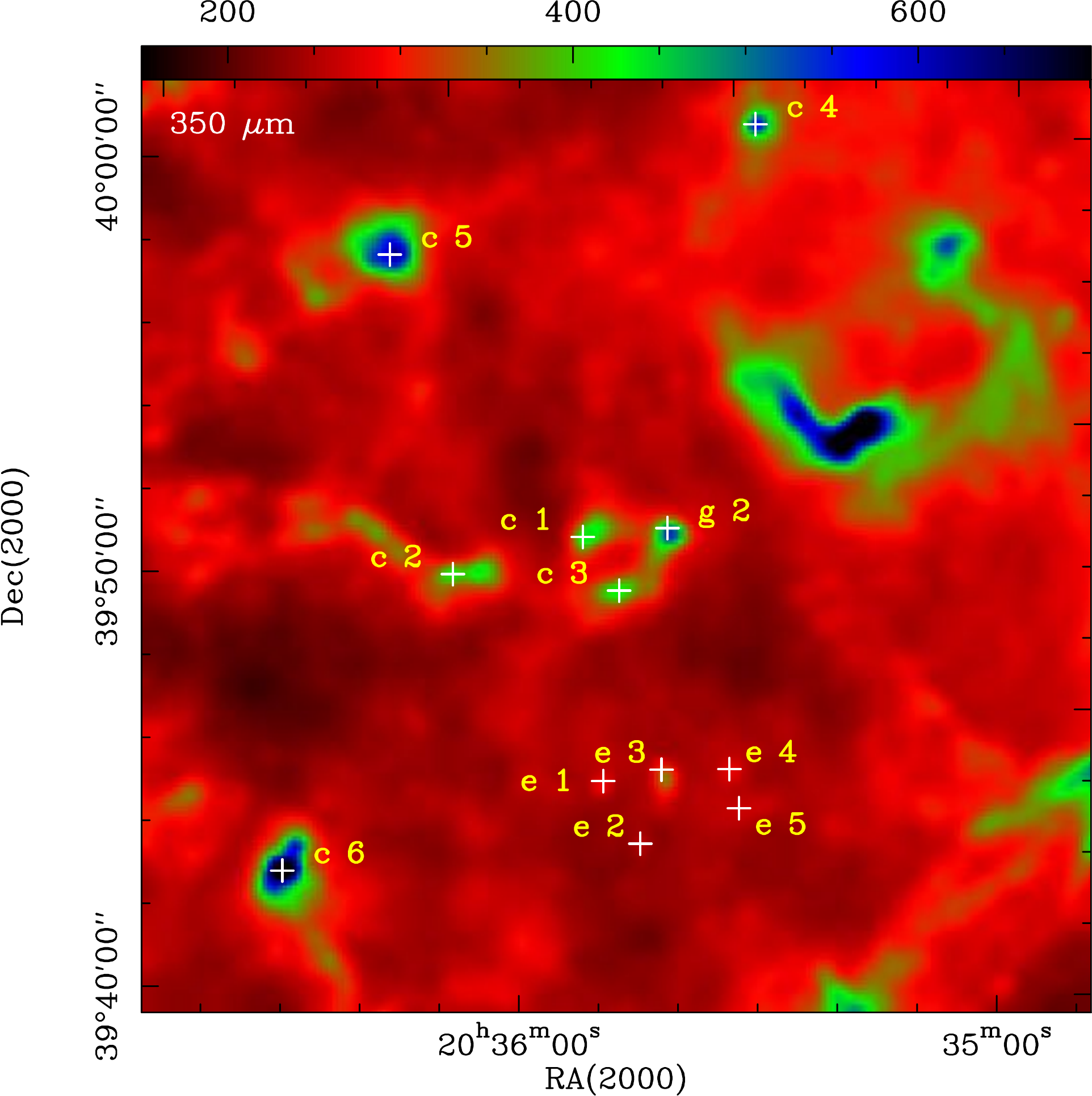} 
\includegraphics[angle=0,width=7.5cm]{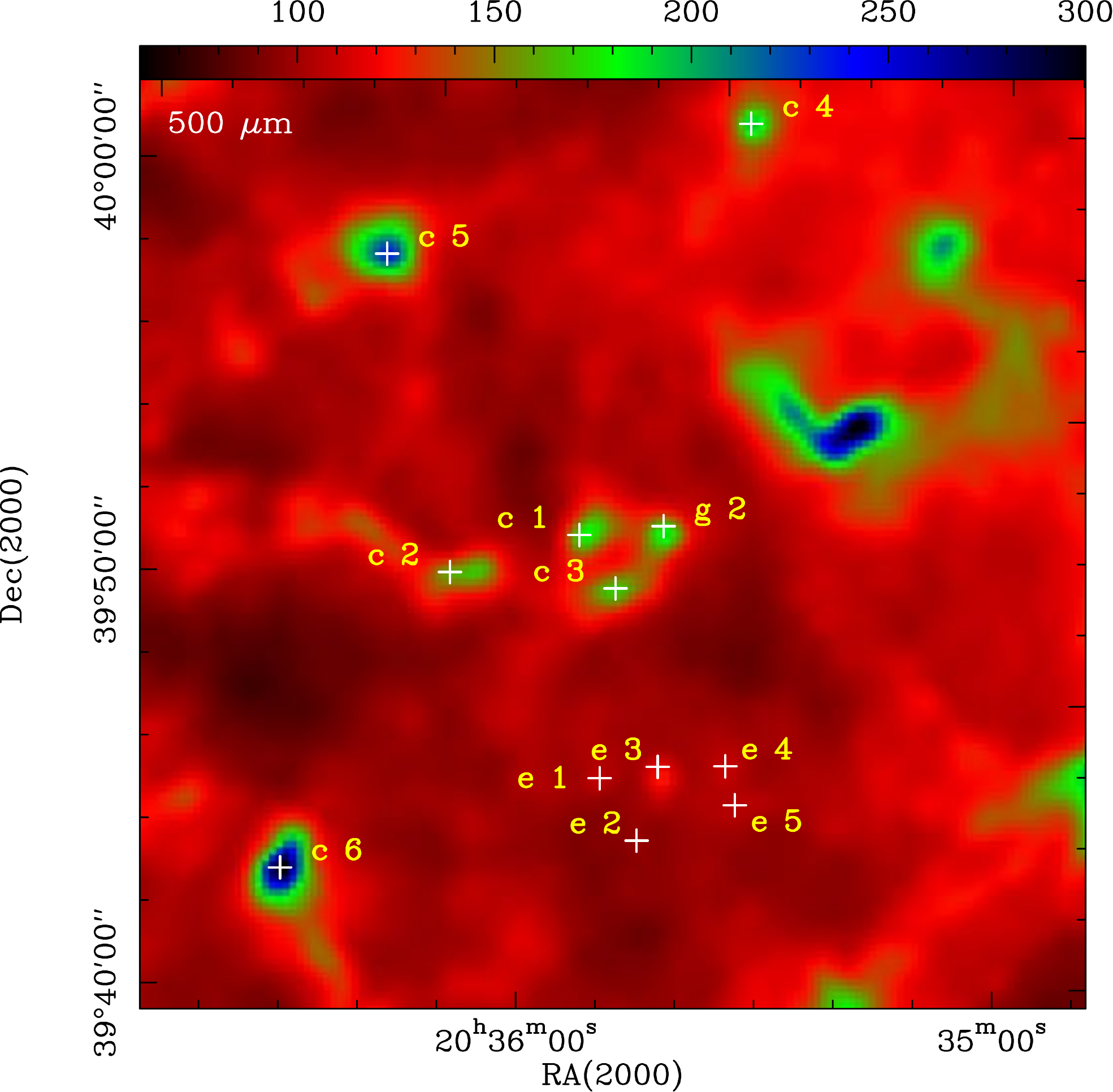} 
\end{center}  
\caption [] {PACS (70 and 160 $\mu$m) and SPIRE (250, 350, 500 $\mu$m) images 
of region 1-2.}   
\end{figure*} 

\begin{figure*}[ht]     
\begin{center}  
\includegraphics[angle=0,width=7.5cm]{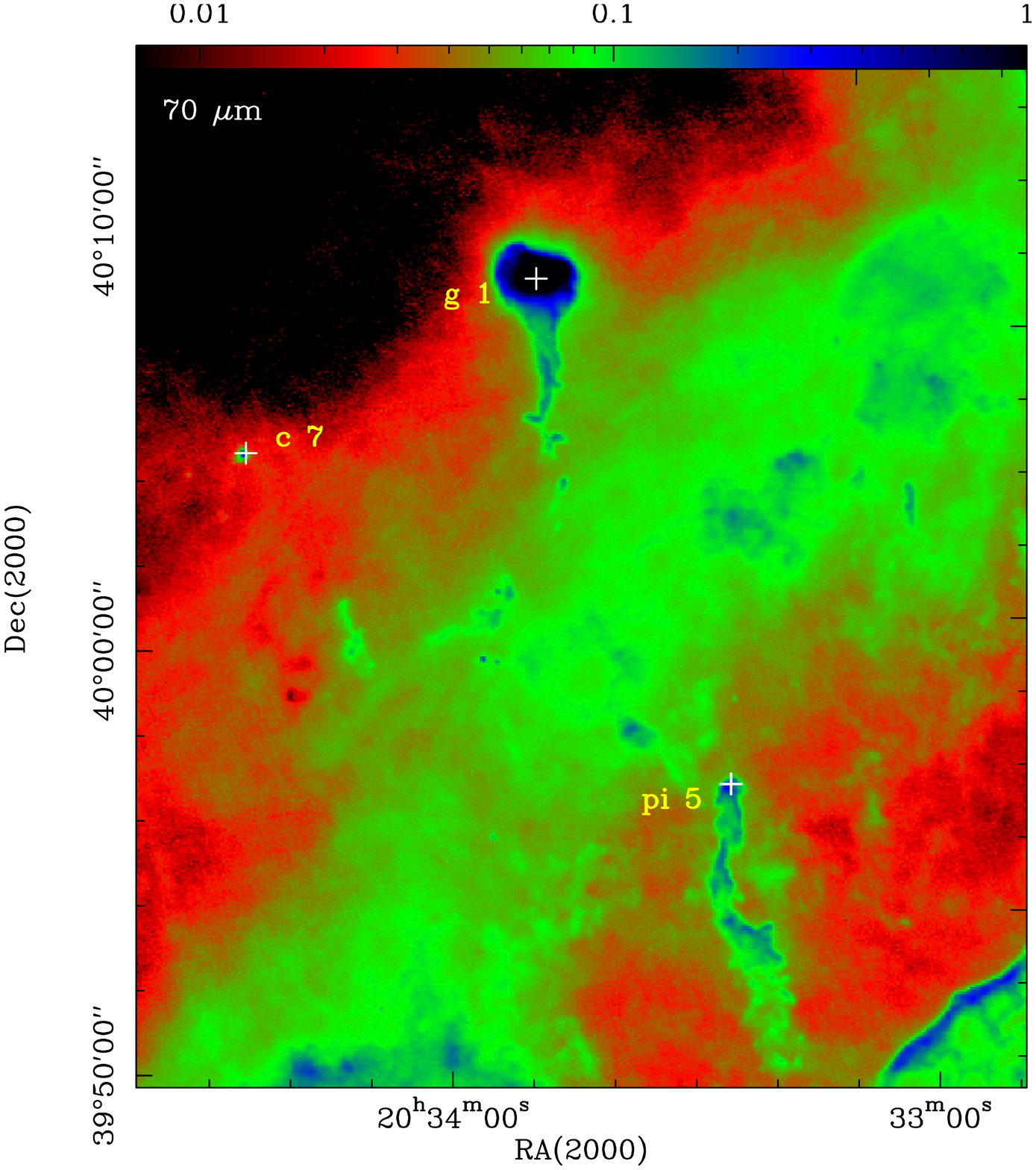} 
\includegraphics[angle=0,width=7.5cm]{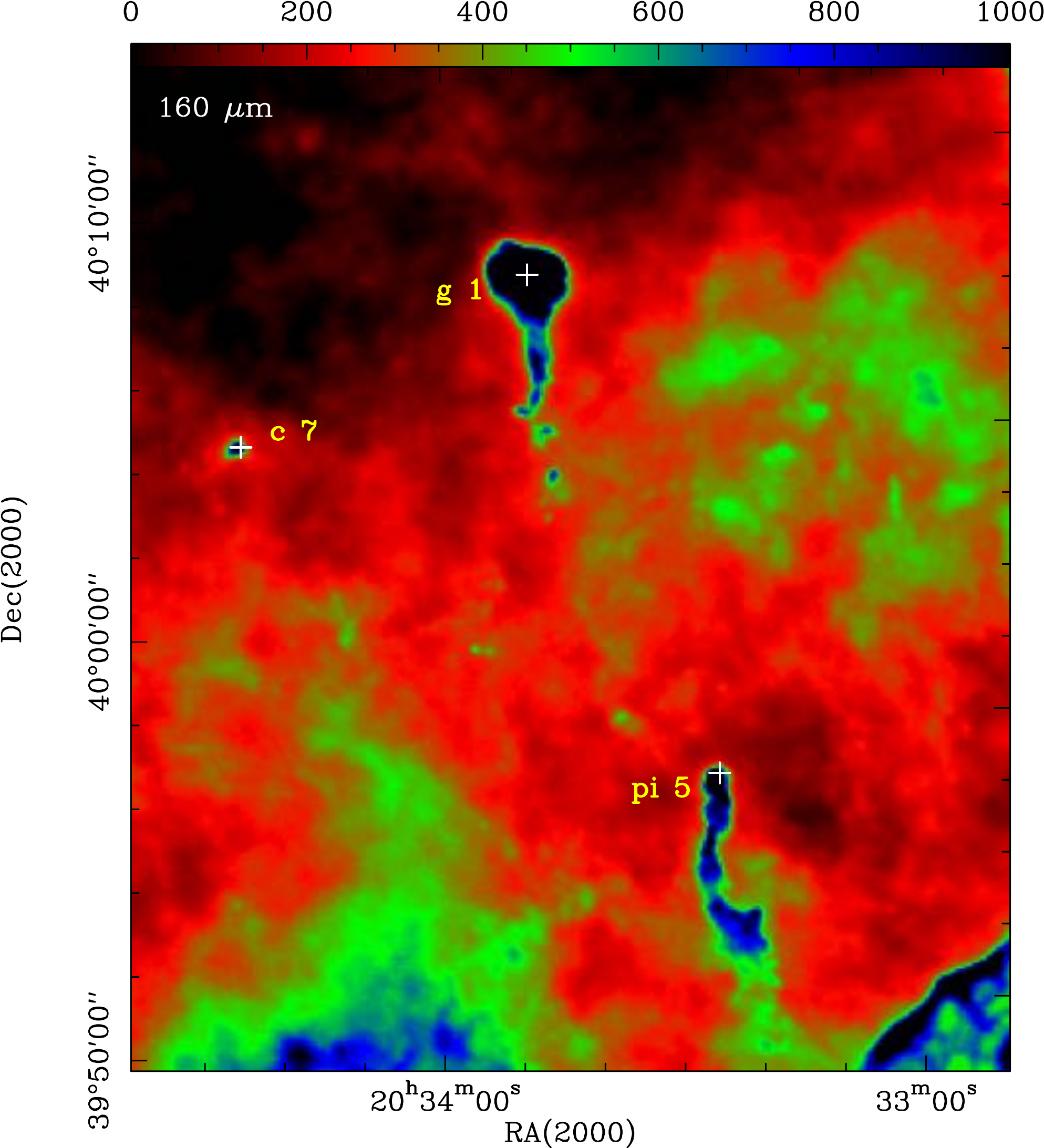} 
\includegraphics[angle=0,width=7.5cm]{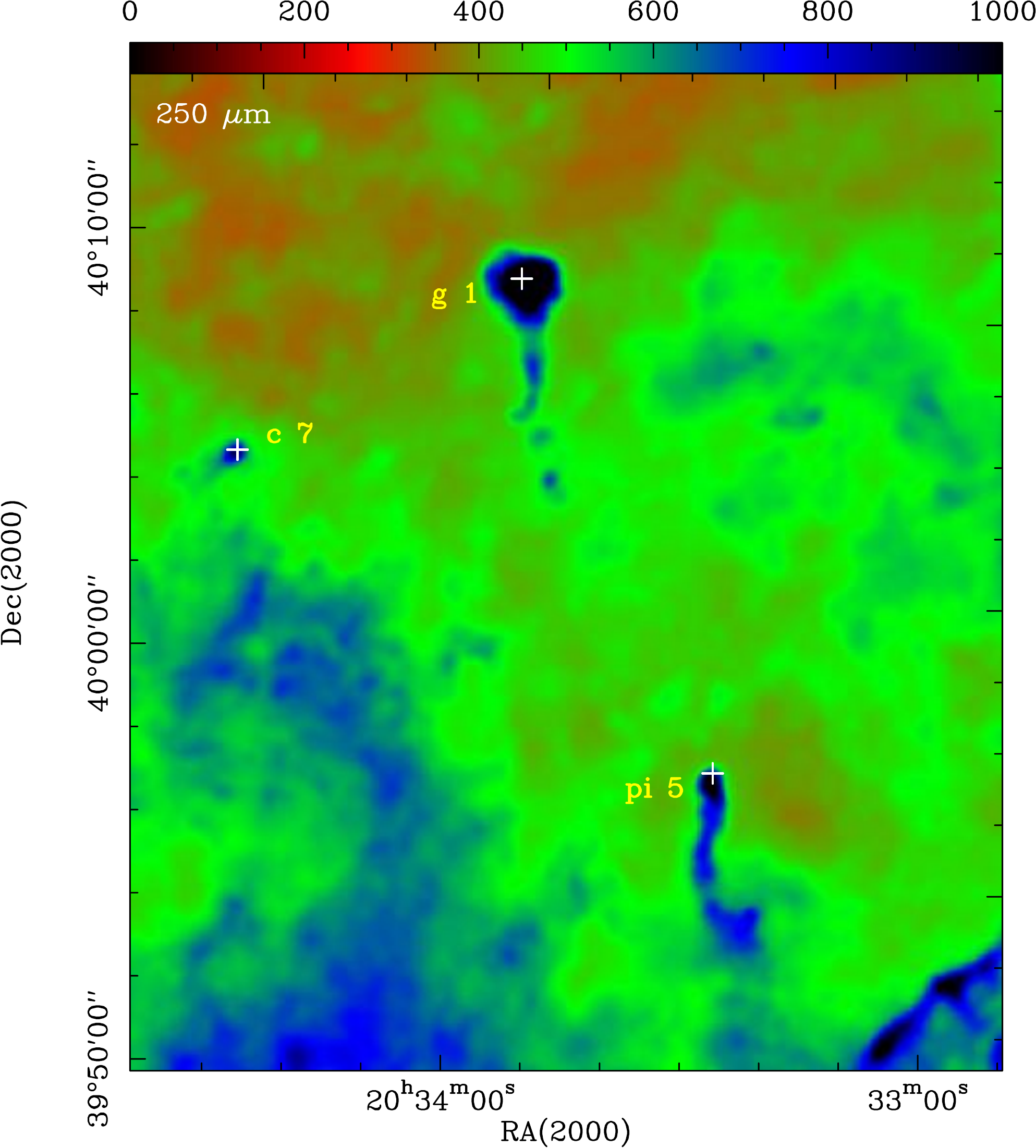} 
\includegraphics[angle=0,width=7.5cm]{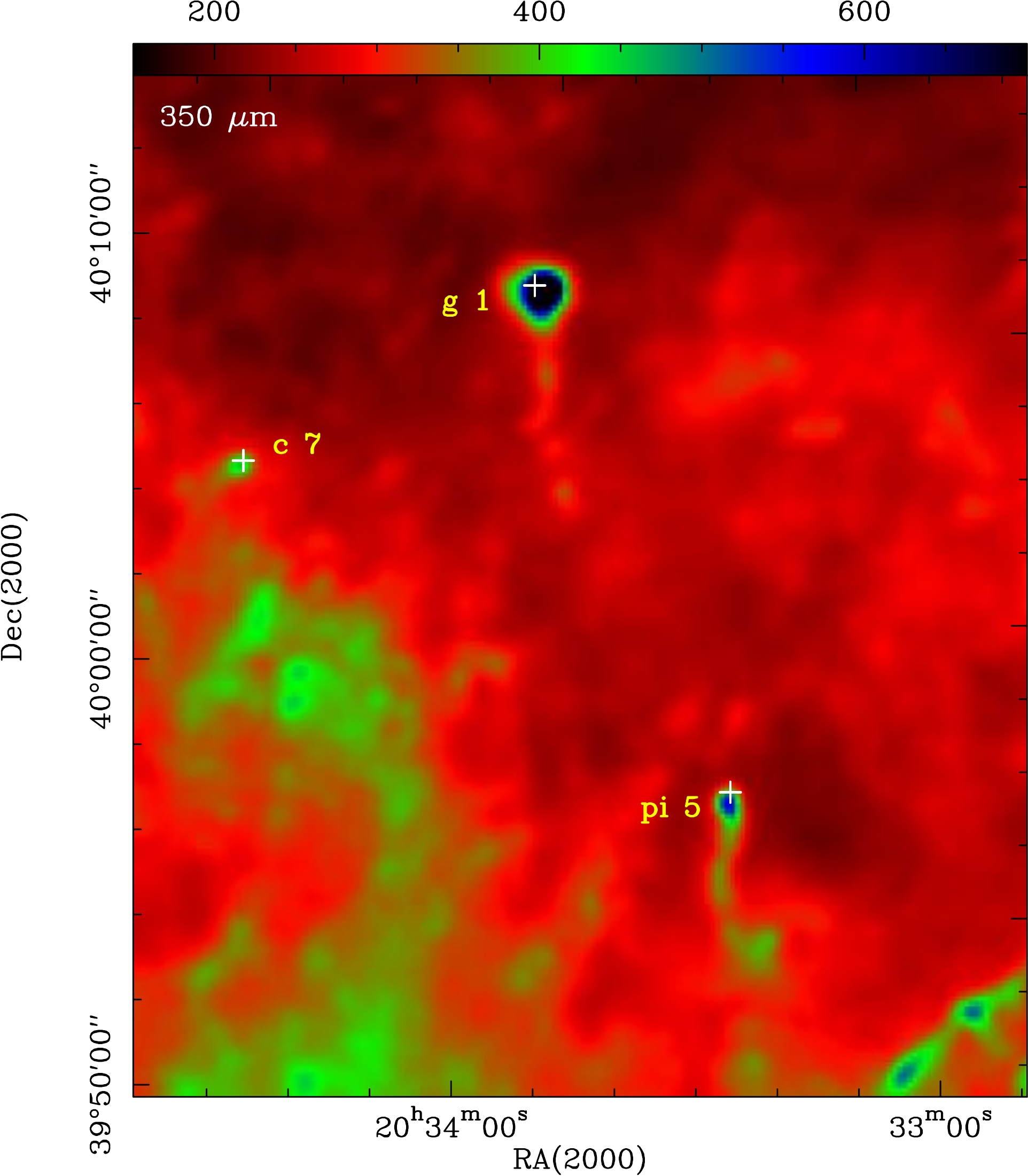} 
\includegraphics[angle=0,width=7.5cm]{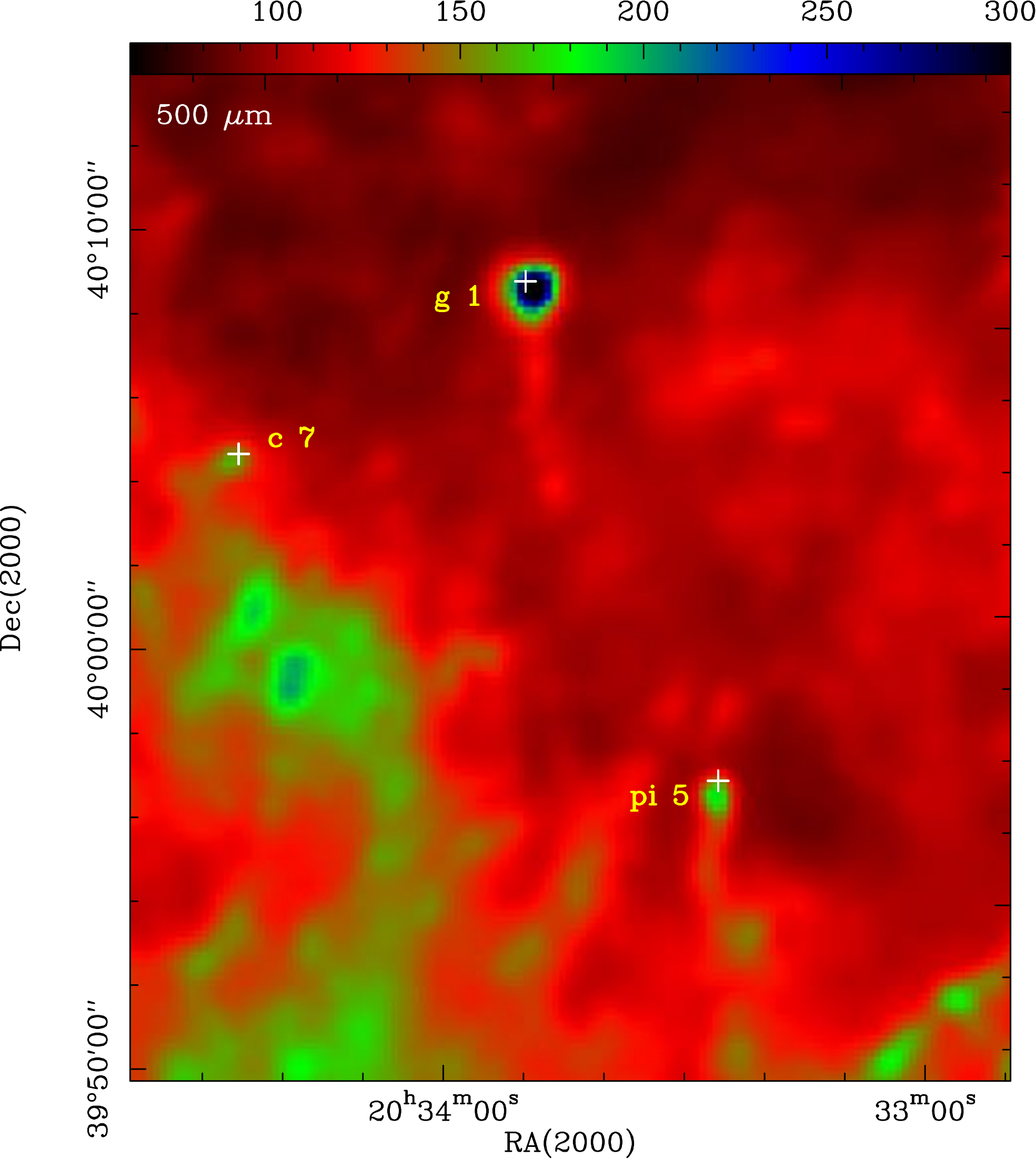} 
\end{center}  
\caption [] {PACS (70 and 160 $\mu$m) and SPIRE (250, 350, 500 $\mu$m) images 
of region 1-3.}   
\end{figure*} 

\begin{figure*}[ht]     
\begin{center}  
\includegraphics[angle=0,width=7.5cm]{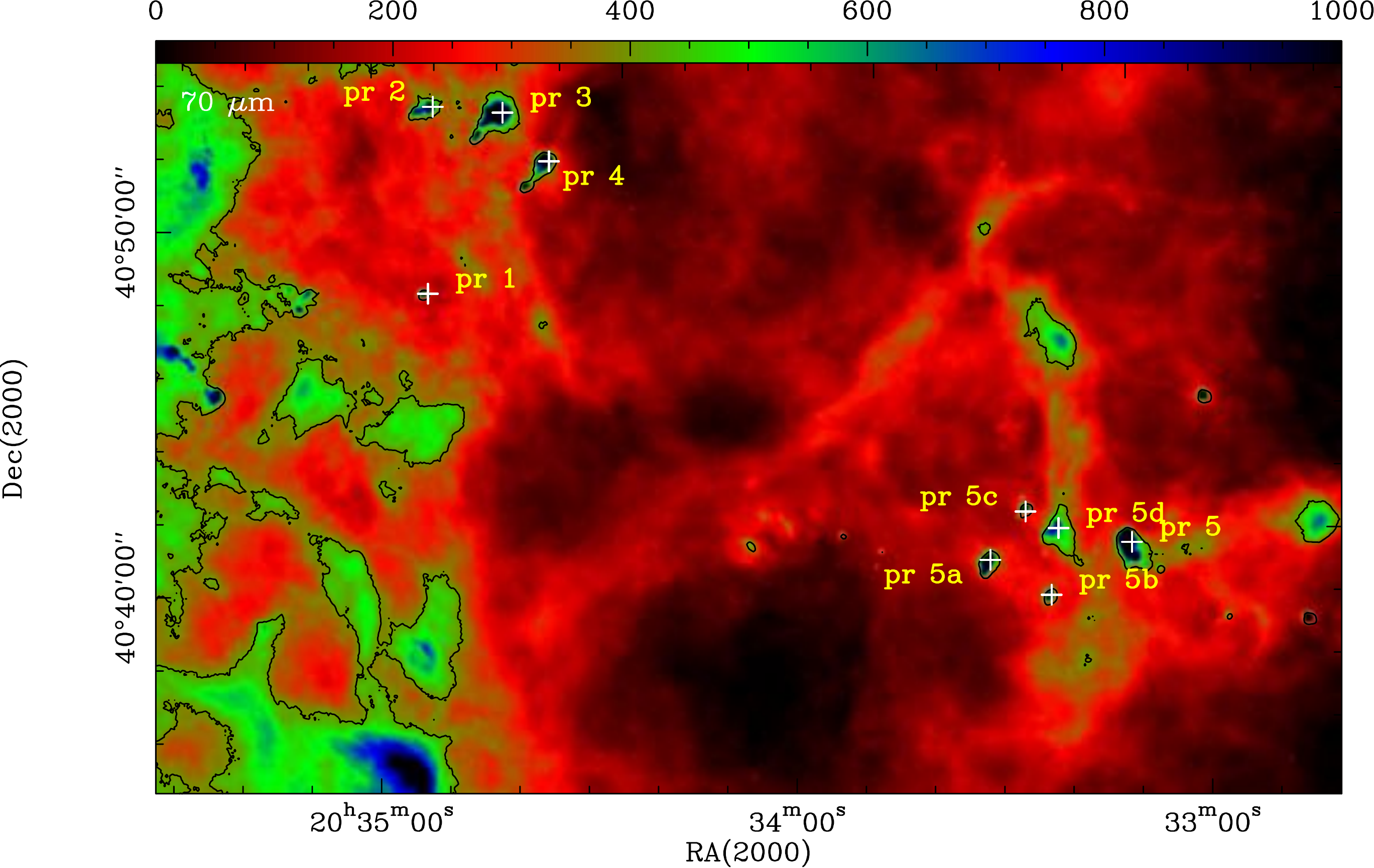} 
\includegraphics[angle=0,width=7.5cm]{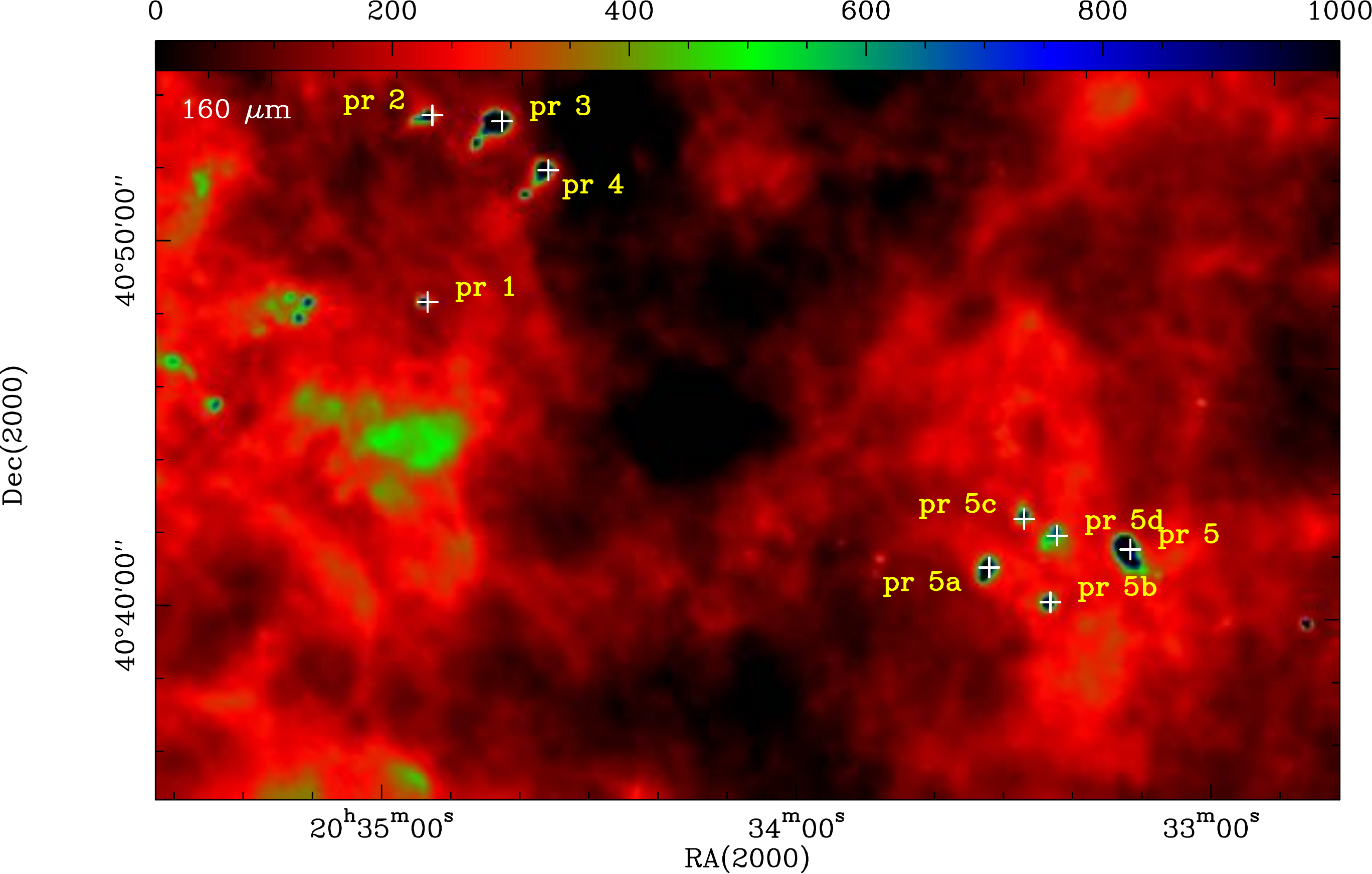} 
\includegraphics[angle=0,width=7.5cm]{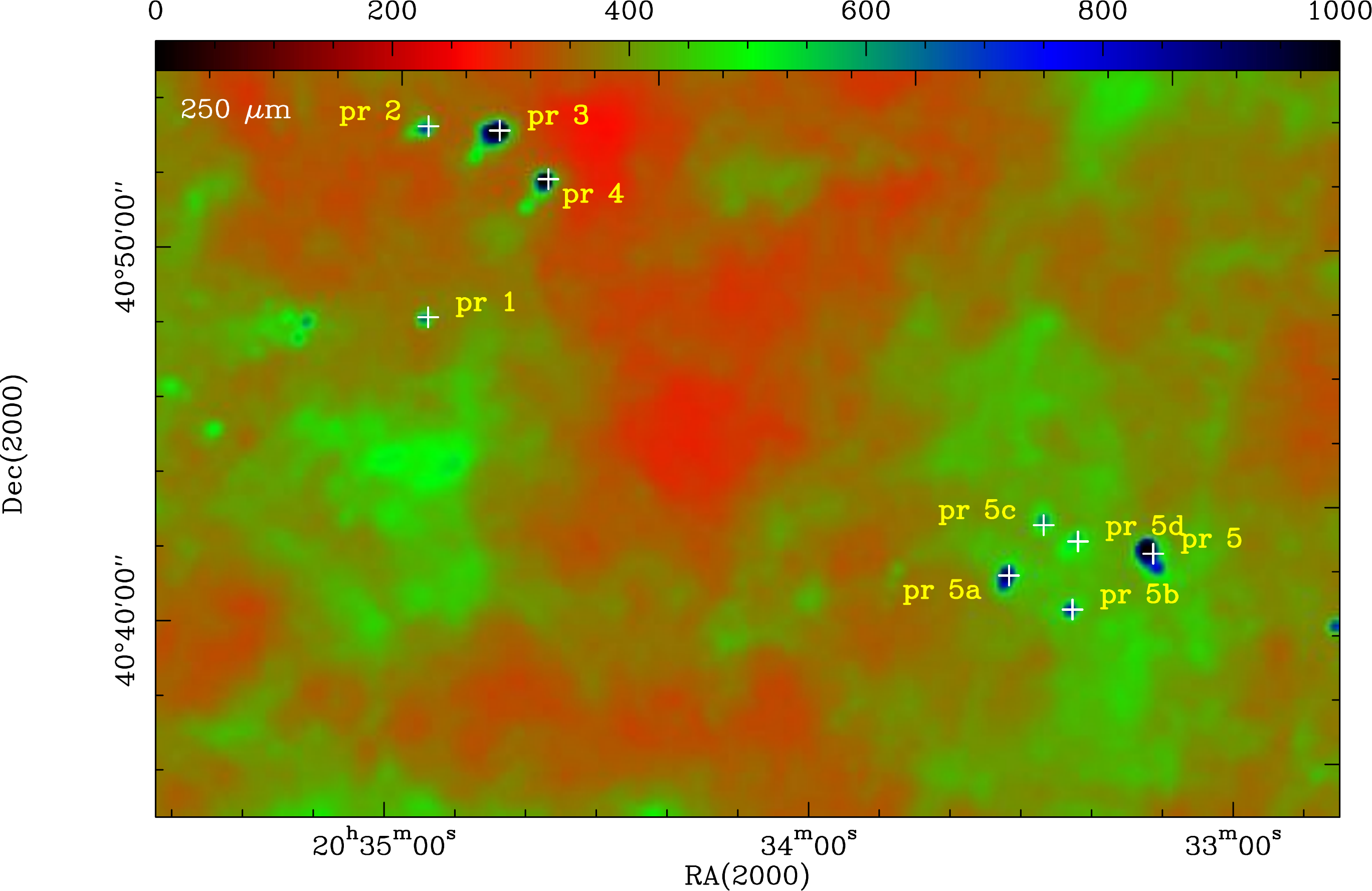} 
\includegraphics[angle=0,width=7.5cm]{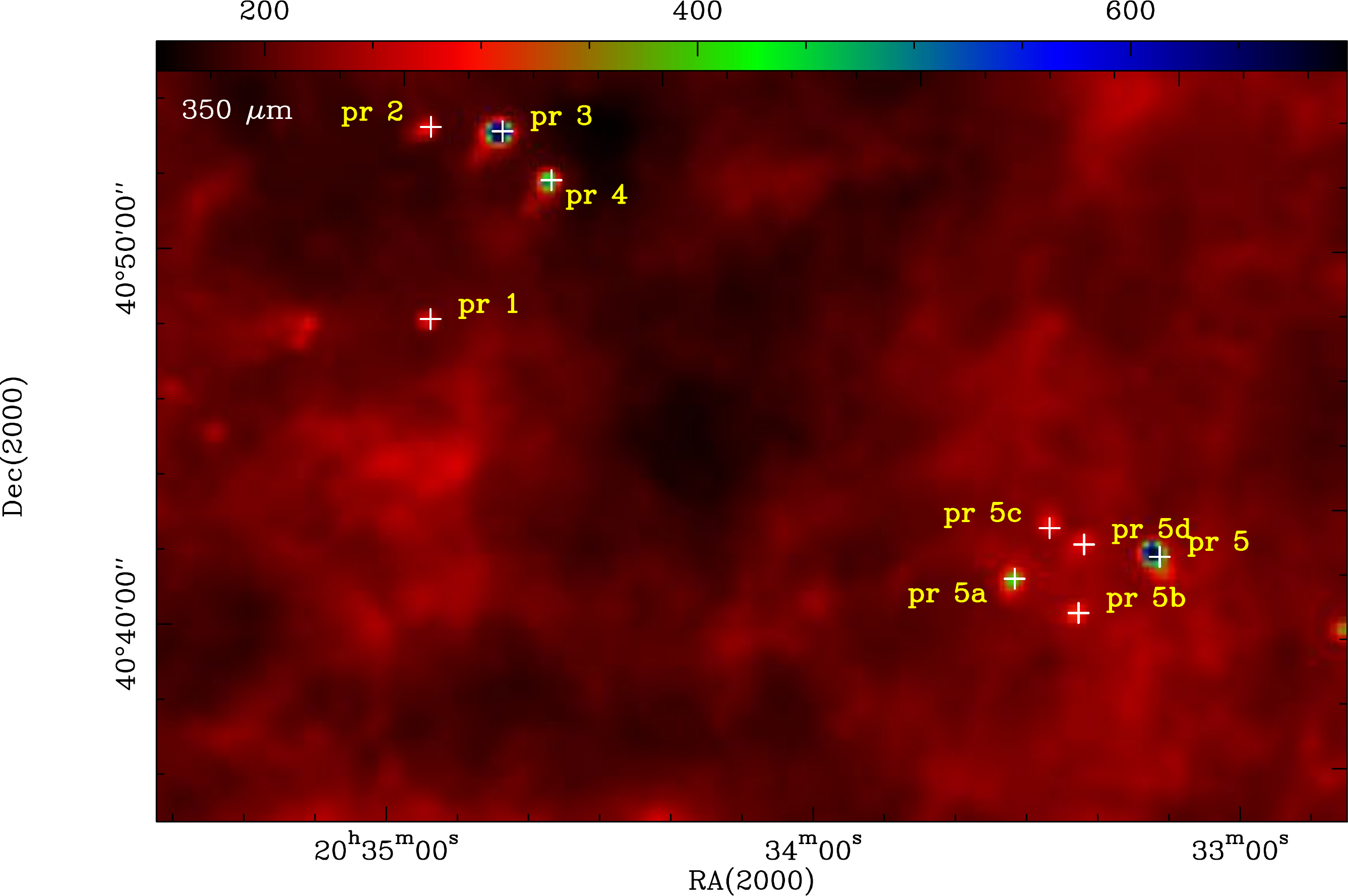} 
\includegraphics[angle=0,width=7.5cm]{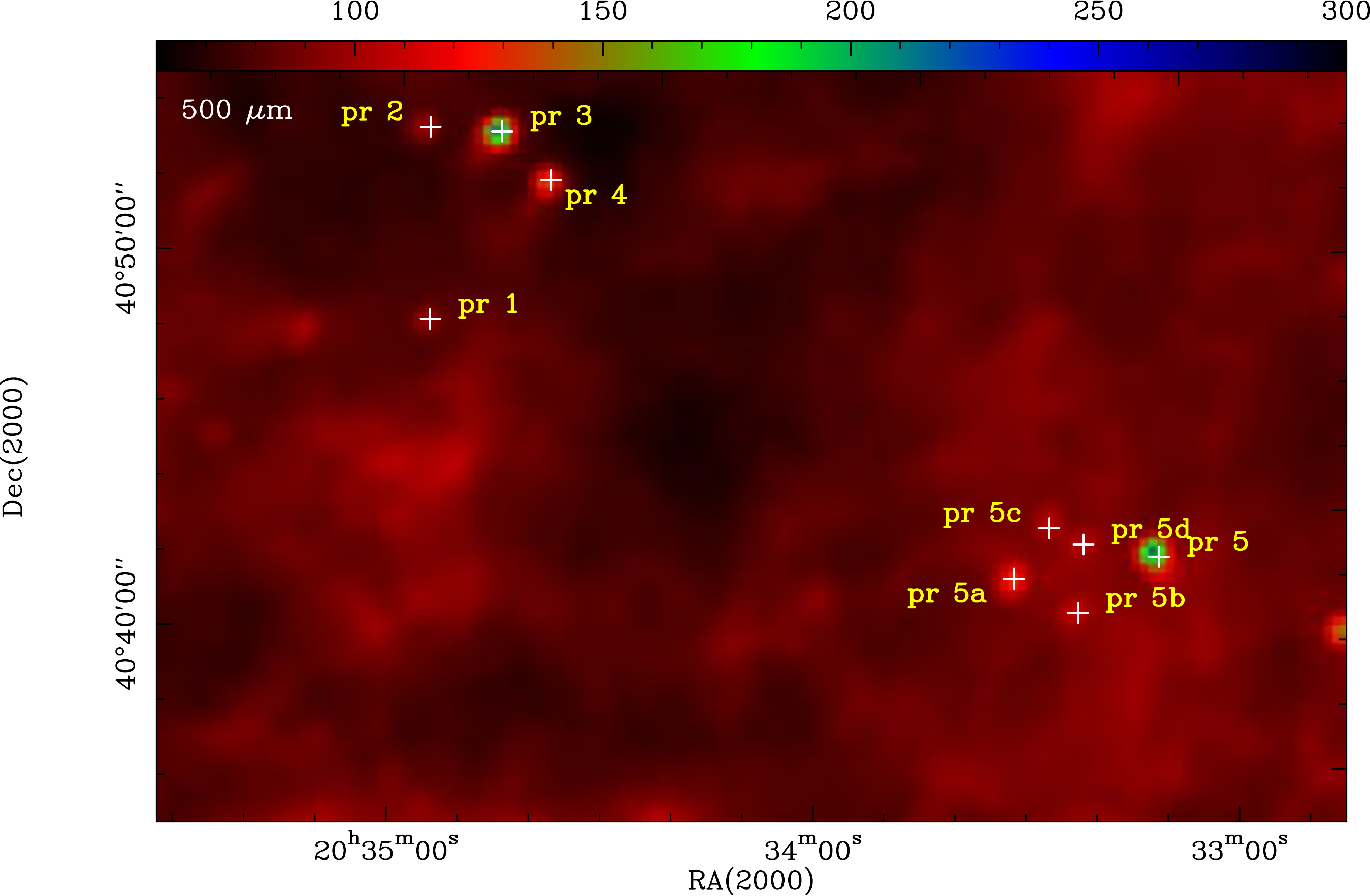} 
\end{center}  
\caption [] {PACS (70 and 160 $\mu$m) and SPIRE (250, 350, 500 $\mu$m) images 
of region 2-1.}   
\end{figure*} 

\begin{figure*}[ht]     
\begin{center}  
\includegraphics[angle=0,width=7.5cm]{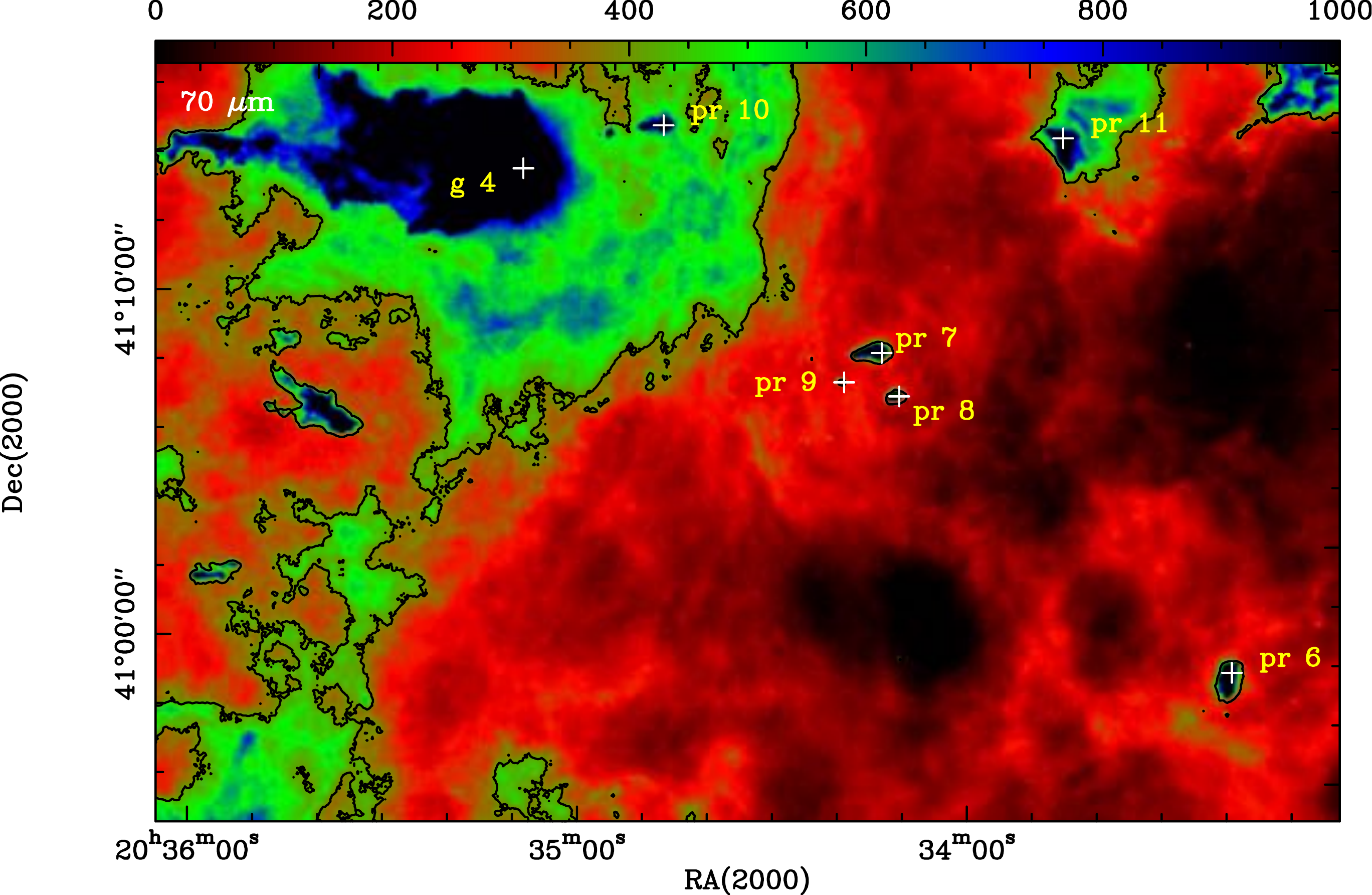} 
\includegraphics[angle=0,width=7.5cm]{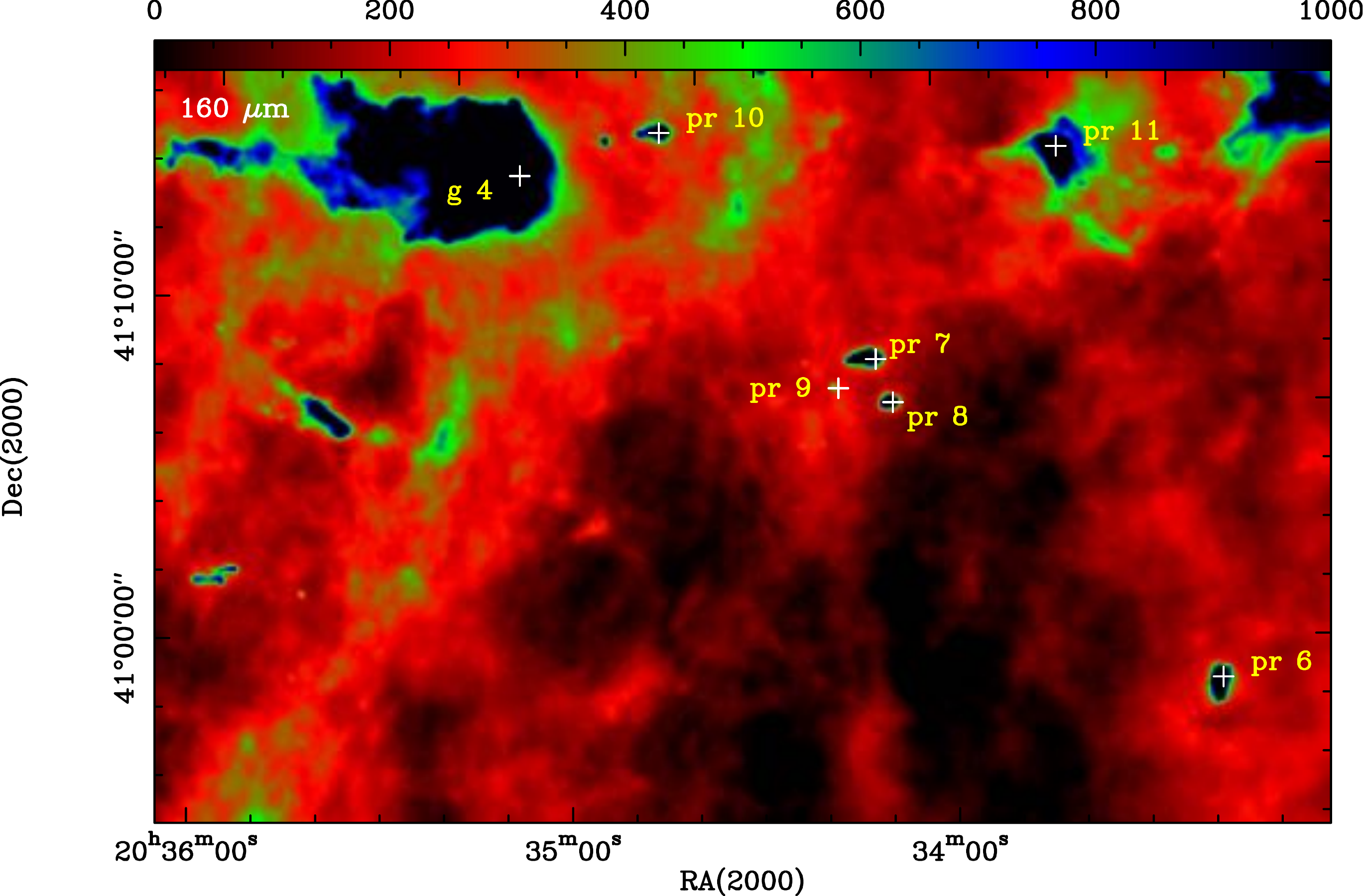} 
\includegraphics[angle=0,width=7.5cm]{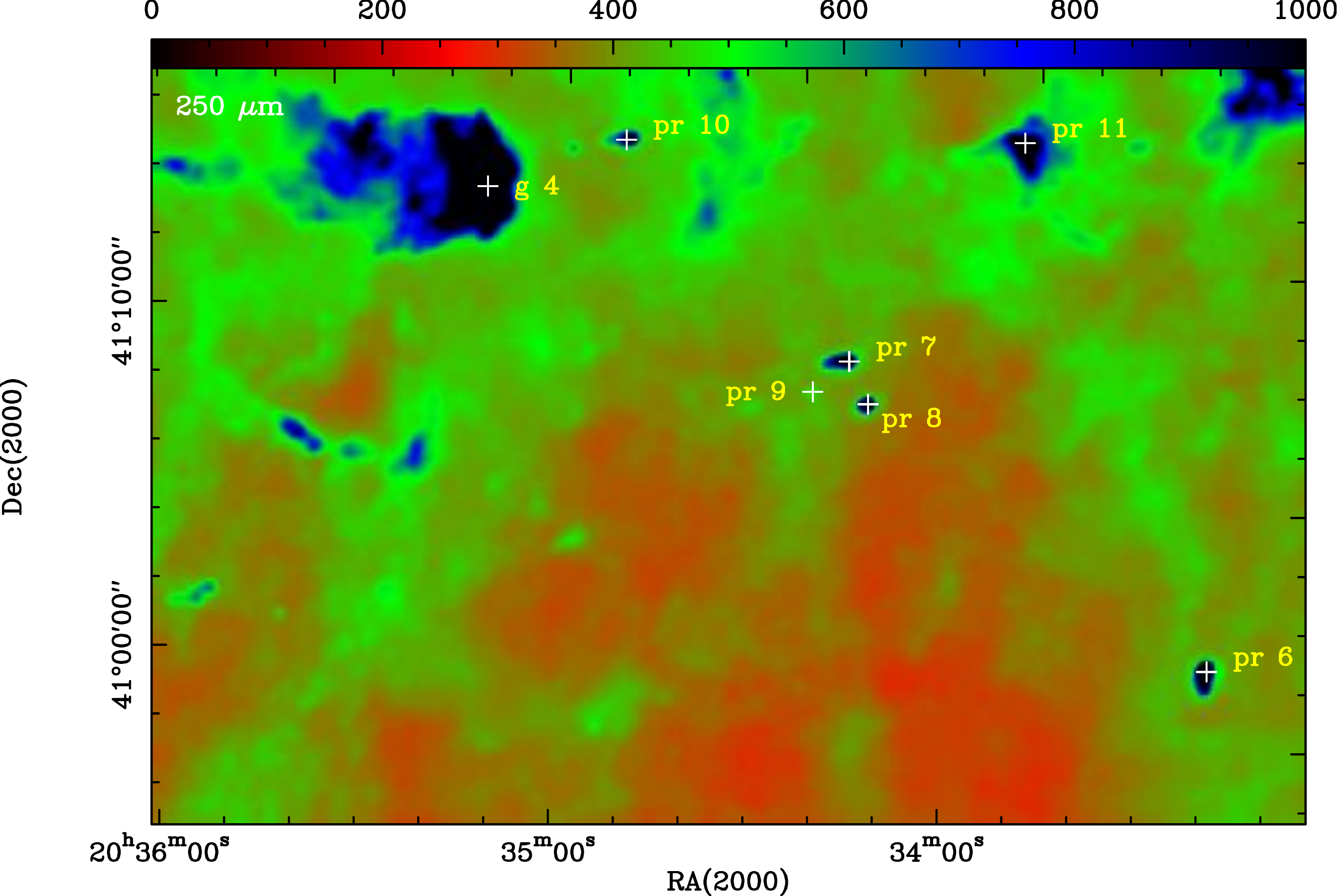} 
\includegraphics[angle=0,width=7.5cm]{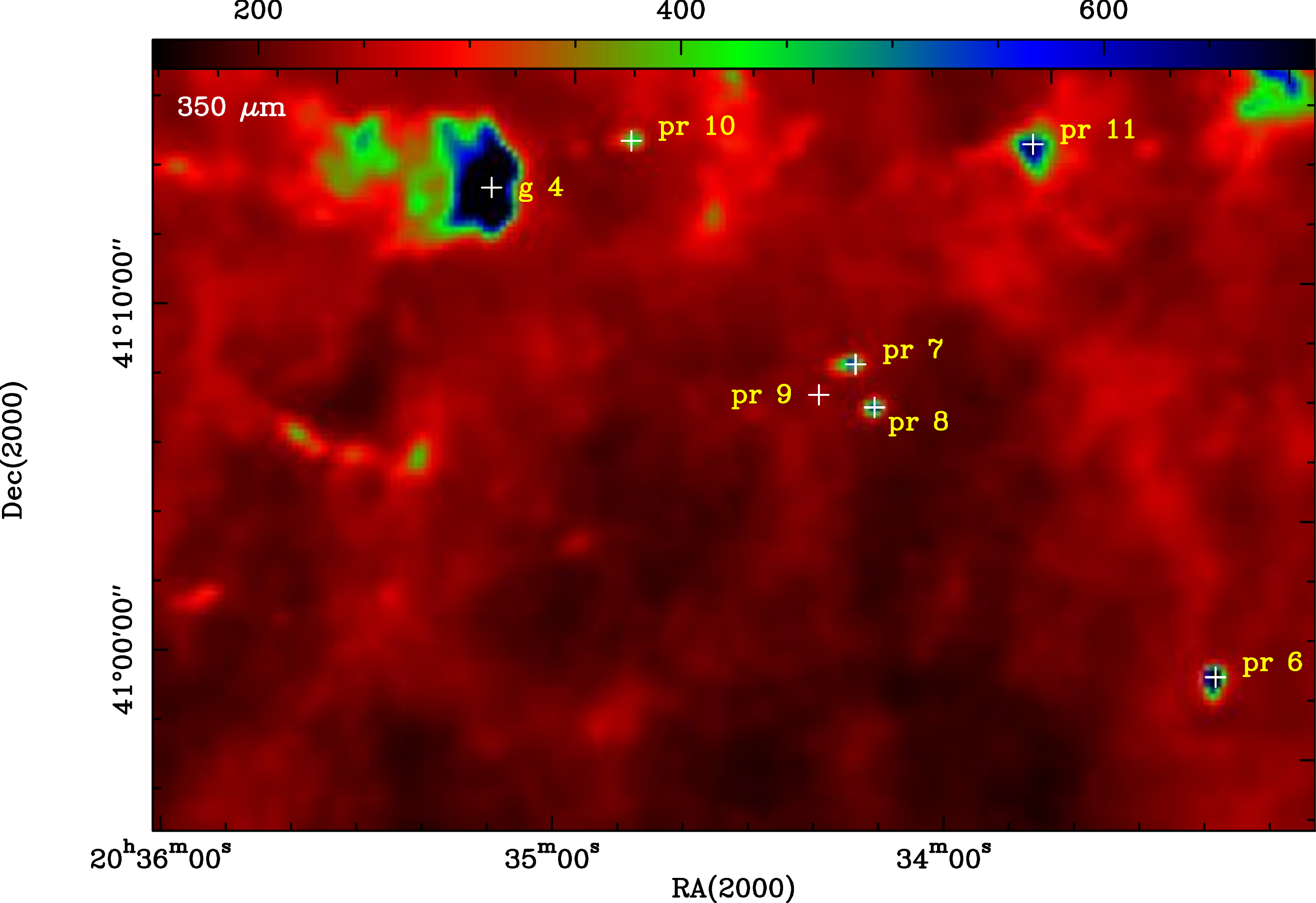} 
\includegraphics[angle=0,width=7.5cm]{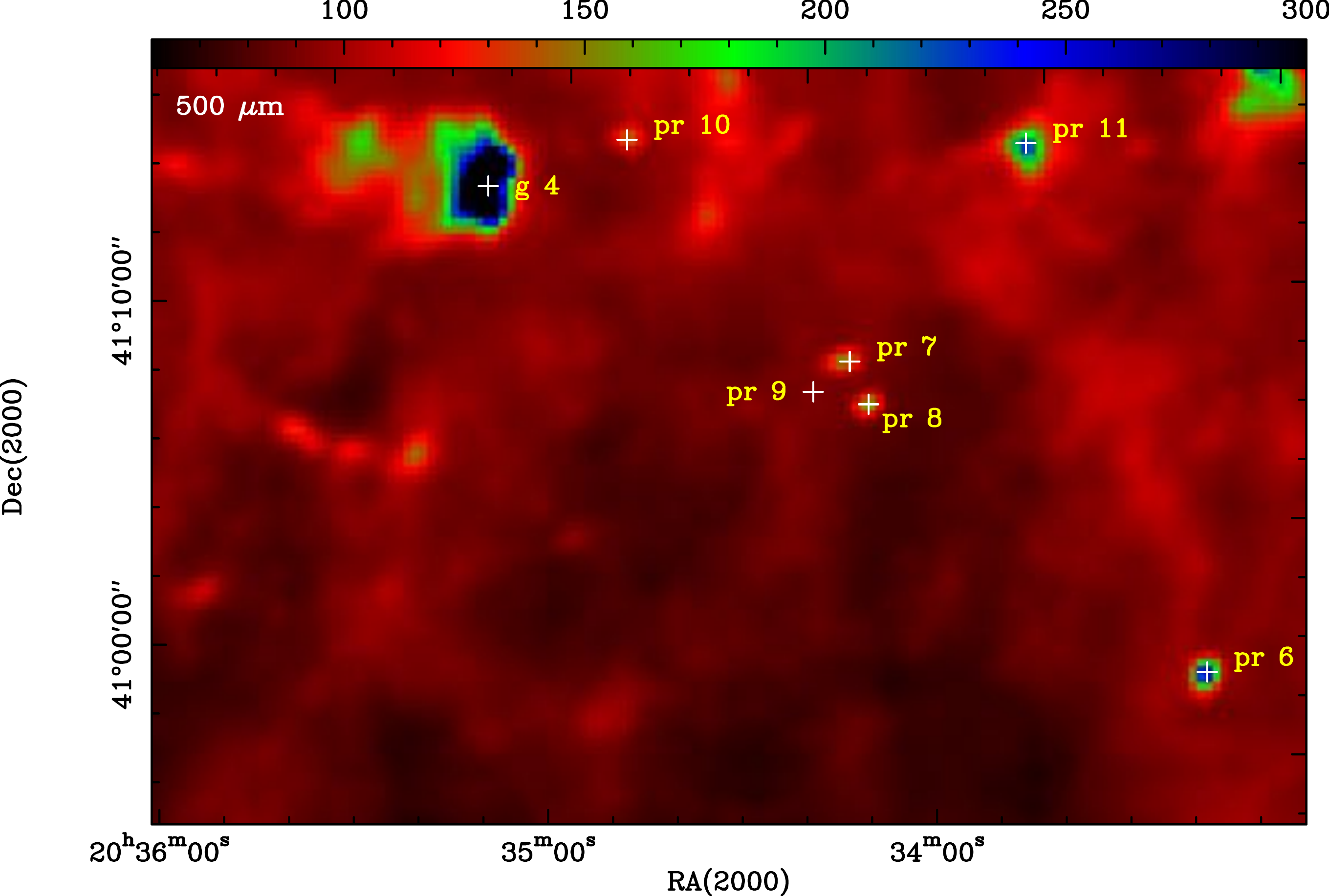} 
\end{center}  
\caption [] {PACS (70 and 160 $\mu$m) and SPIRE (250, 350, 500 $\mu$m) images 
of region 2-2.}   
\end{figure*} 

\begin{figure*}[ht]     
\begin{center}  
\includegraphics[angle=0,width=7.5cm]{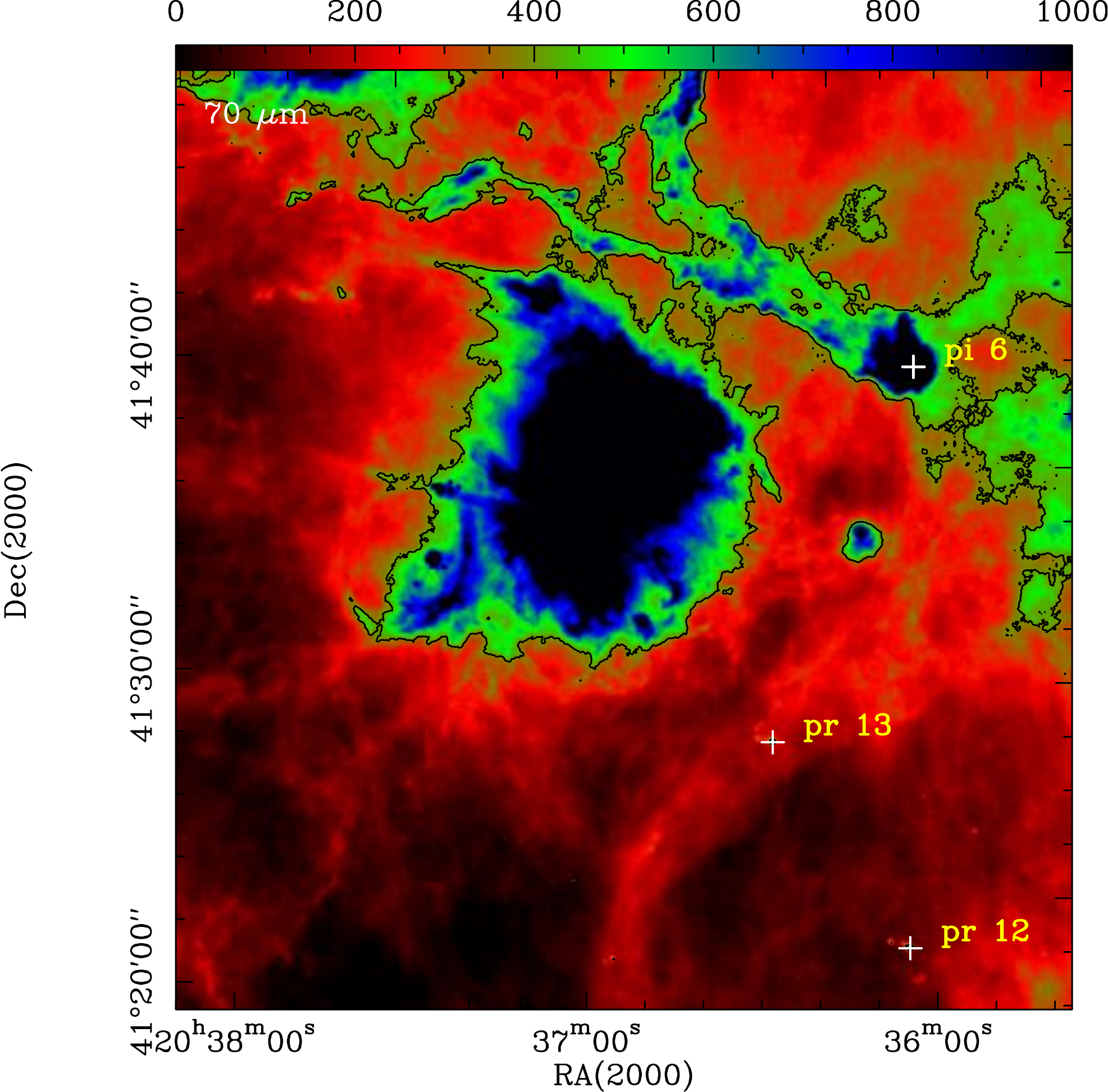} 
\includegraphics[angle=0,width=7.5cm]{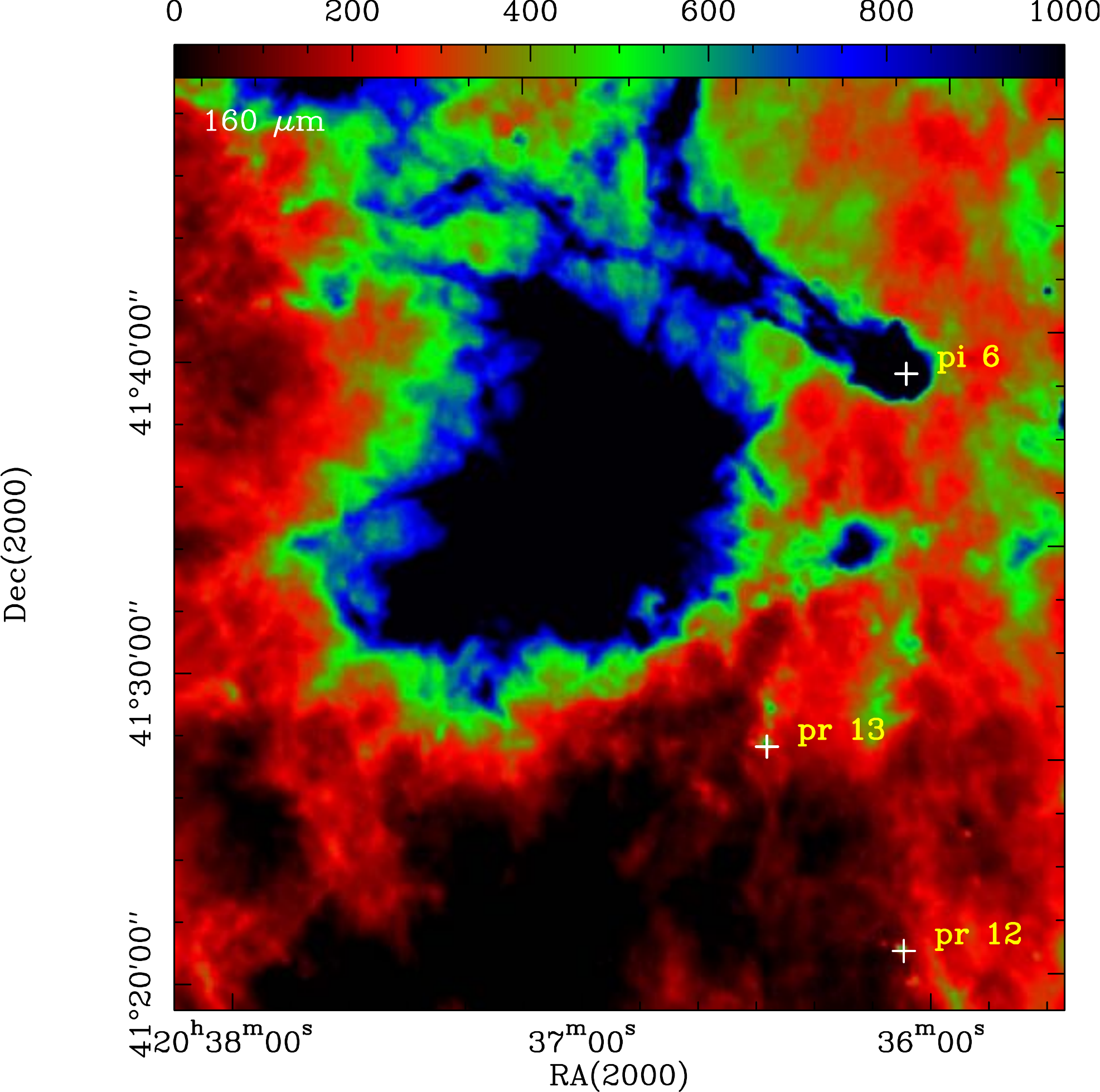} 
\includegraphics[angle=0,width=7.5cm]{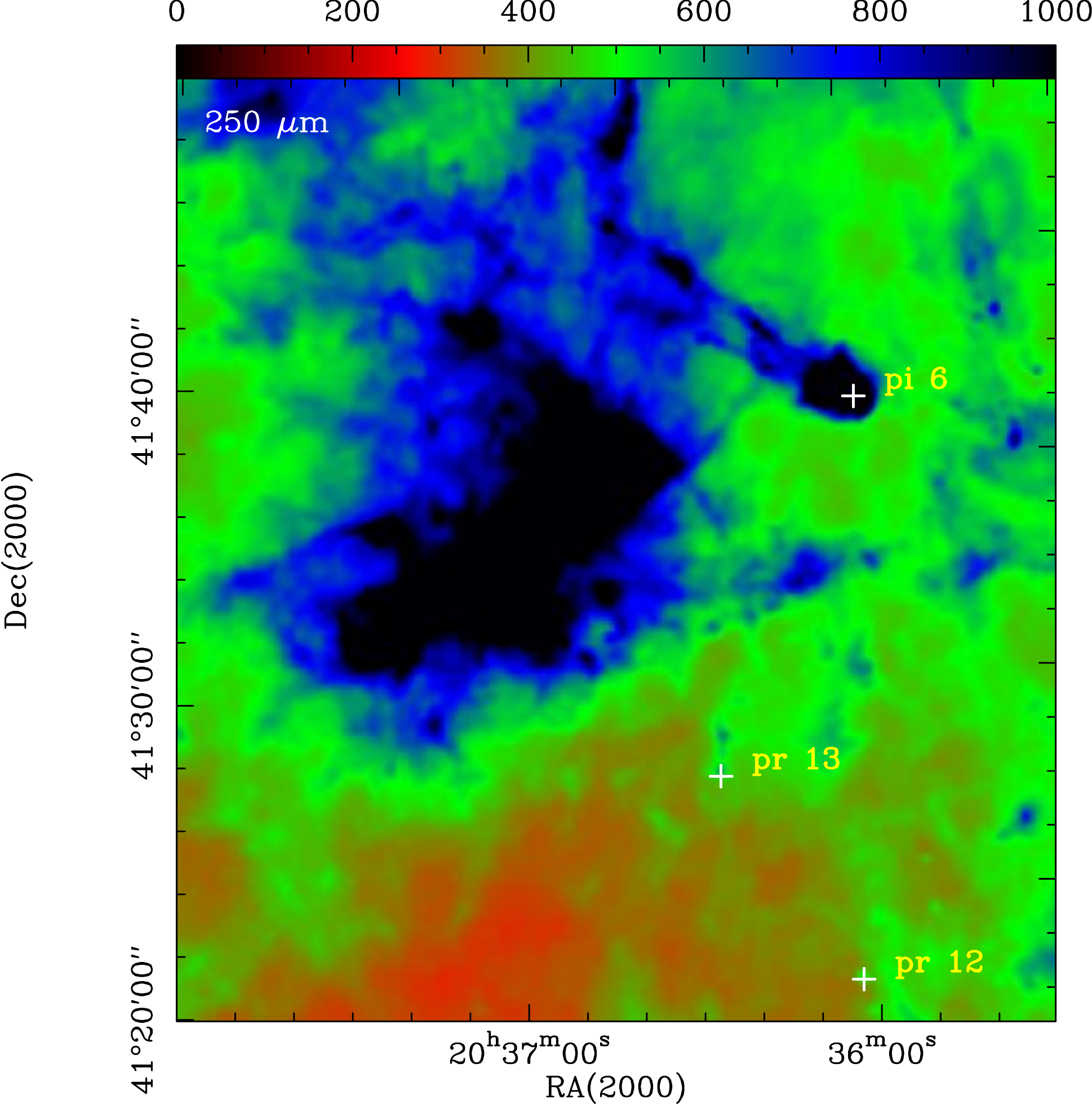} 
\includegraphics[angle=0,width=7.5cm]{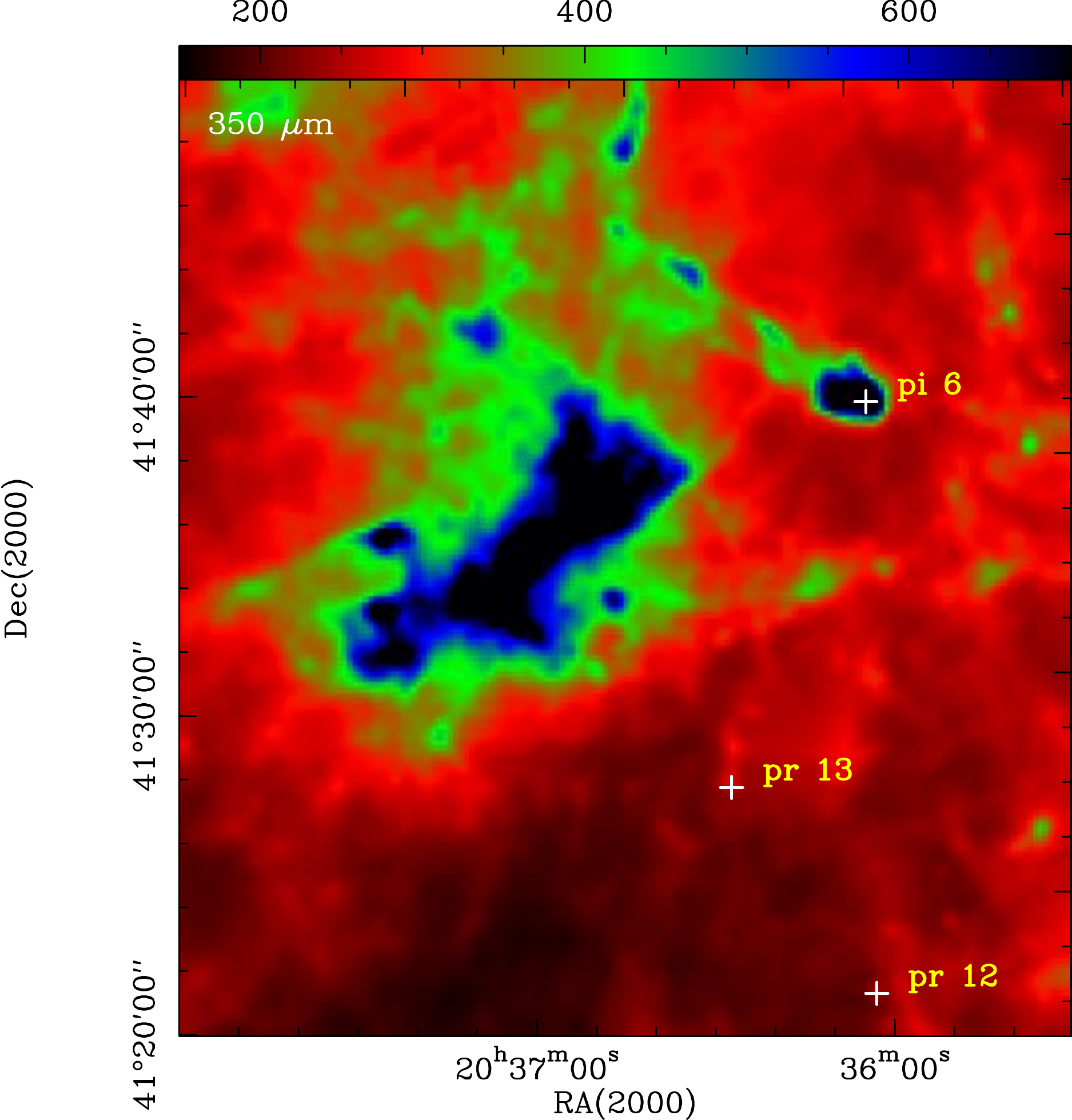} 
\includegraphics[angle=0,width=7.5cm]{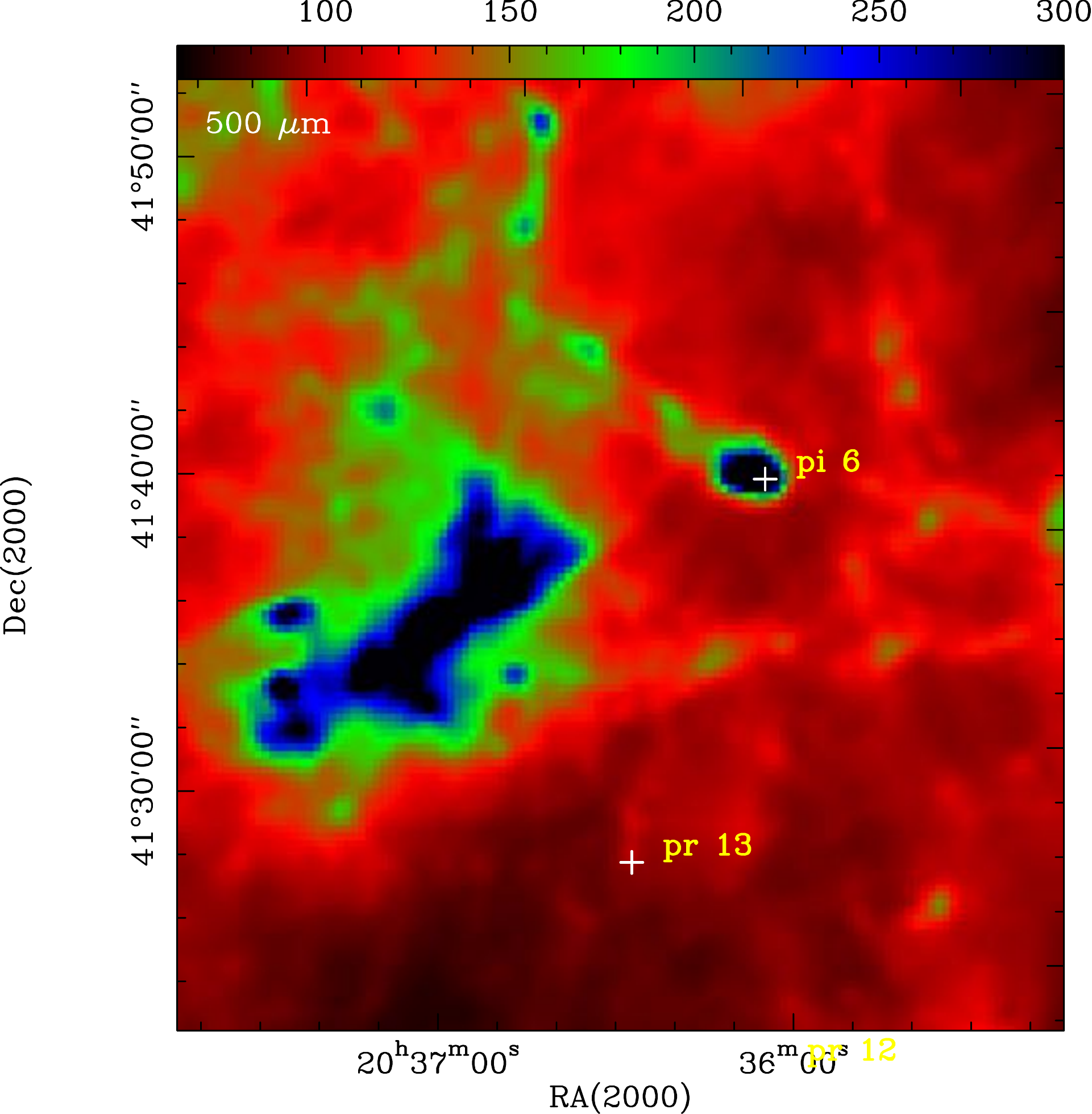} 
\end{center}  
\caption [] {PACS (70 and 160 $\mu$m) and SPIRE (250, 350, 500 $\mu$m) images 
of region 3-1.}   
\end{figure*} 
 
\begin{figure*}[ht]     
\begin{center}  
\includegraphics[angle=0,width=7.5cm]{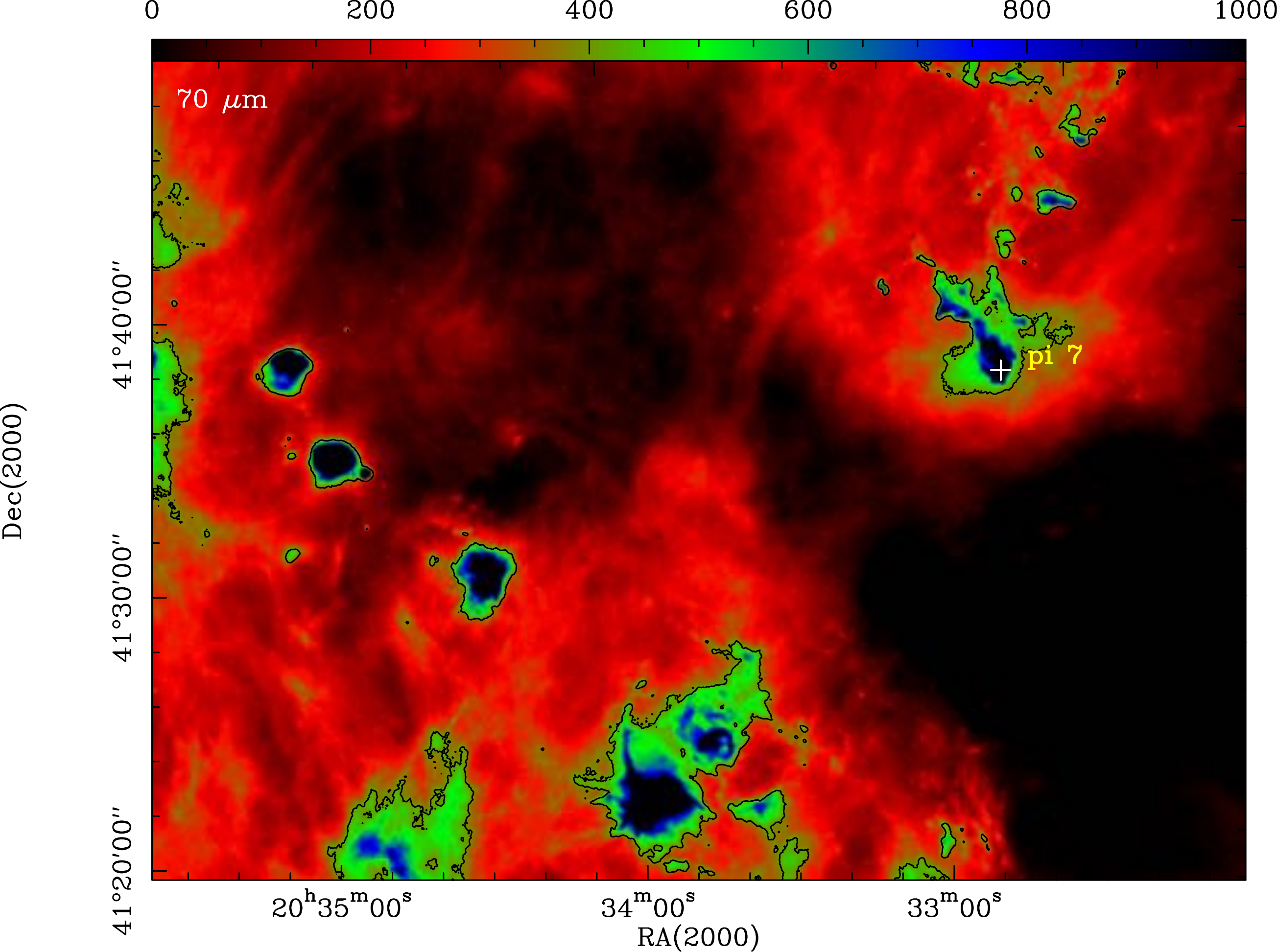} 
\includegraphics[angle=0,width=7.5cm]{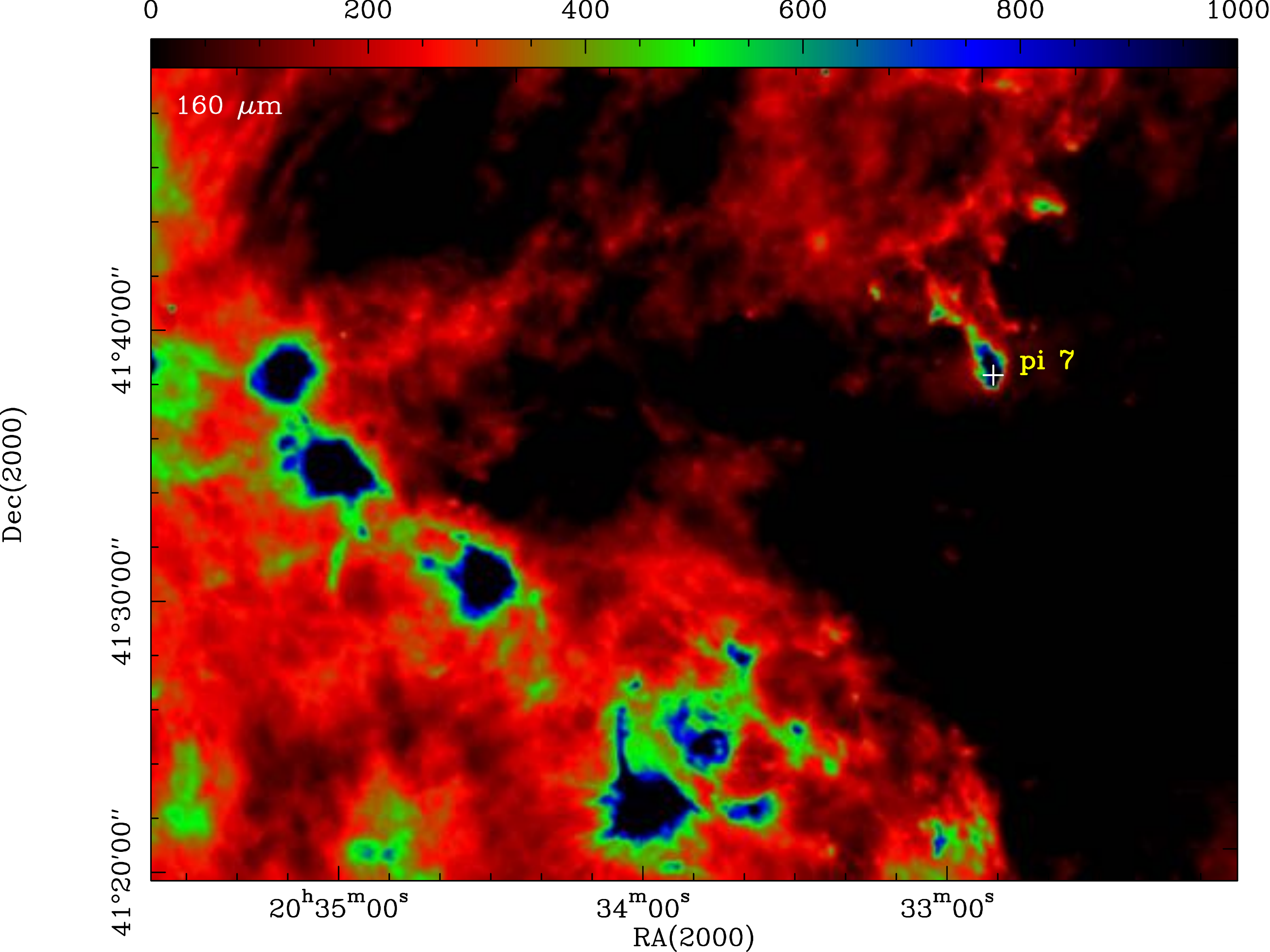} 
\includegraphics[angle=0,width=7.5cm]{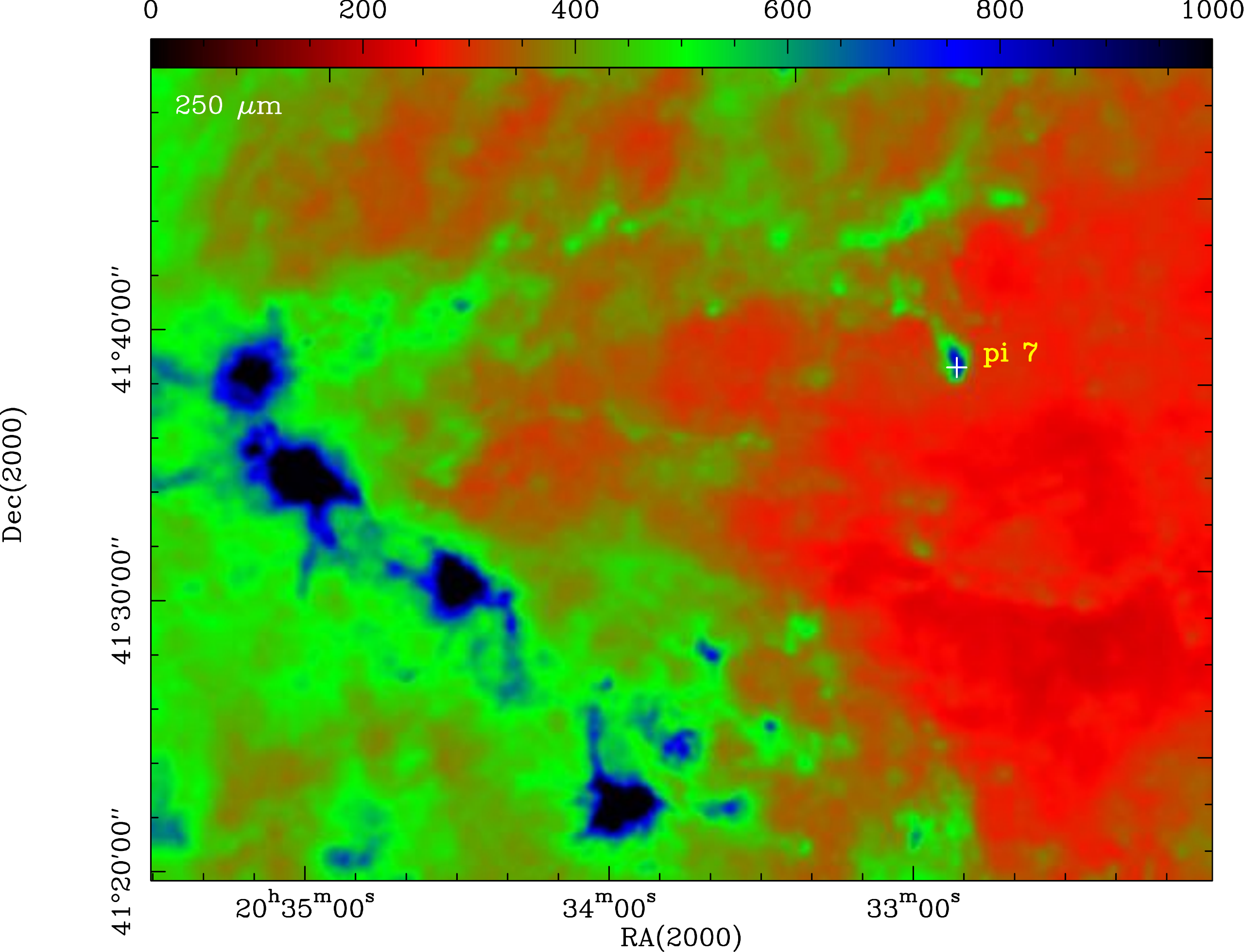} 
\includegraphics[angle=0,width=7.5cm]{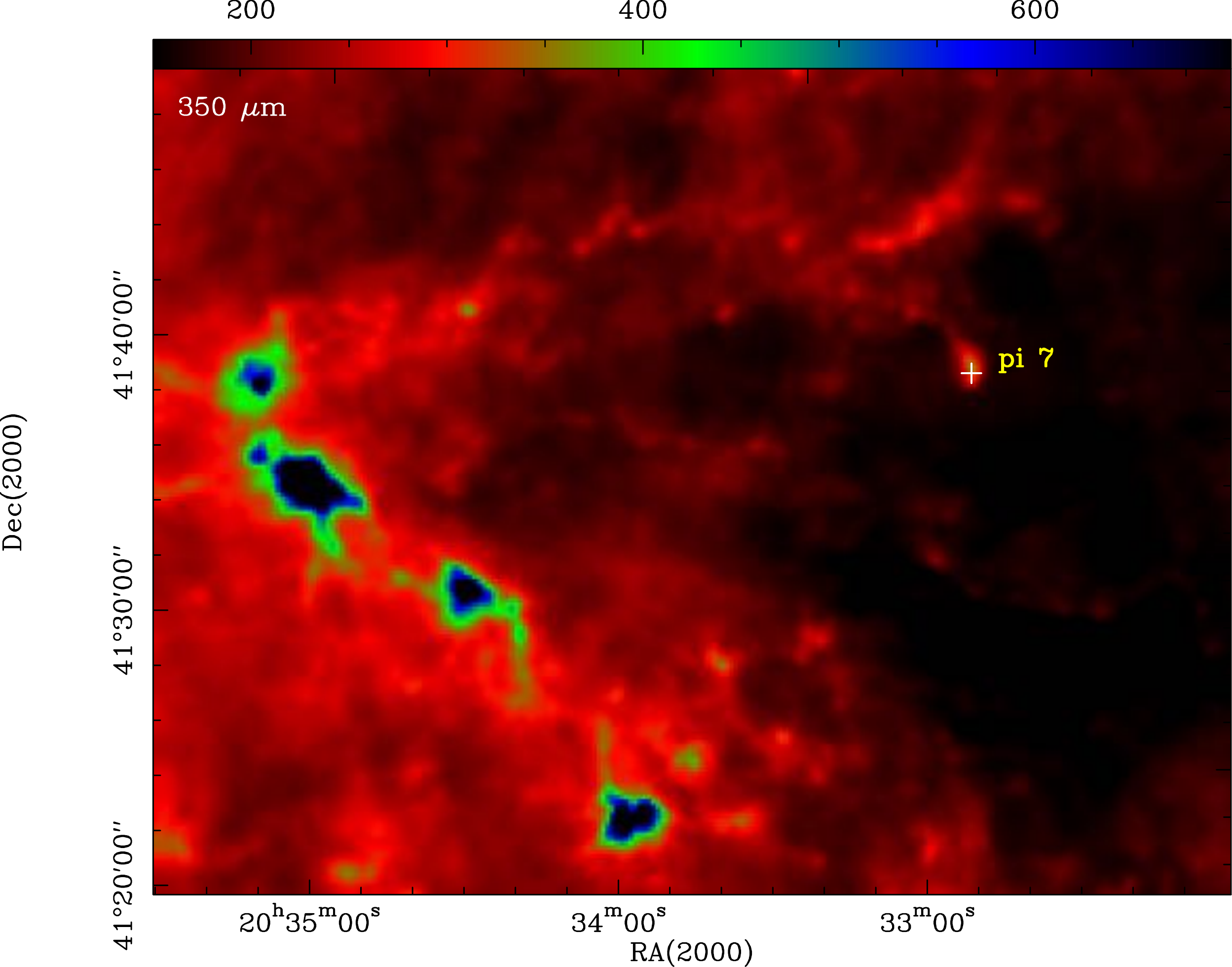} 
\includegraphics[angle=0,width=7.5cm]{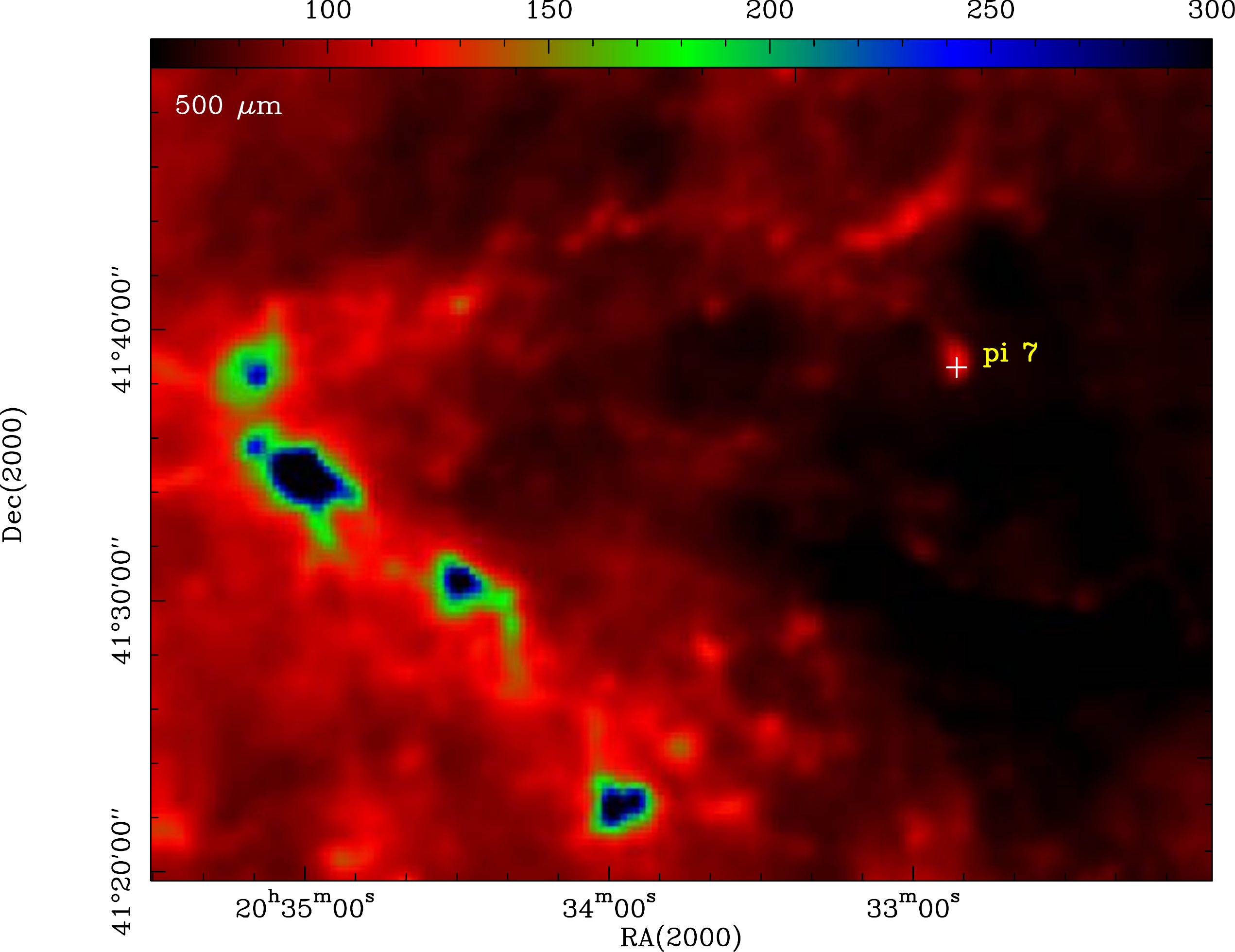} 
\end{center}  
\caption [] {PACS (70 and 160 $\mu$m) and SPIRE (250, 350, 500 $\mu$m) images 
of region 3-2.}   
\end{figure*} 

\end{appendix} 
\end{document}